\documentclass[]{jfm}

\usepackage{graphicx}
\usepackage{newtxtext}
\usepackage{newtxmath}
\usepackage{natbib}
\usepackage[hypertexnames=false]{hyperref}
\hypersetup{
    colorlinks = true,
    urlcolor   = blue,
    citecolor  = black
}

\newcommand{\RomanNumeralCaps}[1]
\linenumbers

\renewcommand{\footerflagdefns}[1]{}	
\renewcommand{\etc}{etc.\xspace}	
\renewcommand{\eg}{e.g.\xspace}	
\usepackage{calc,xspace,mparhack}
\usepackage{subfigure}
\usepackage[pdftex,pstarrows]{pict2e}
\usepackage{array,dcolumn,delarray,hhline,ltxtable,multirow,colortbl}
\usepackage{hyperxmp}
\hypersetup{%
	hypertexnames=false,%
	breaklinks,linktocpage,bookmarksnumbered,%
	colorlinks,allcolors=blue,%
	pdffitwindow,%
	pdfstartpage=1,pdfstartview={XYZ 0 \hypercalcbp{\pdfpageheight} 1},%
	pdfduplex=DuplexFlipLongEdge,pdfprintscaling=None}
\allowdisplaybreaks
\let\oldmarginpar\marginpar
\renewcommand{\marginpar}[1]{\oldmarginpar
	{\scriptsize\fbox{\color{red}\parbox[b]{\marginparwidth-2\marginparsep}{\raggedright#1}}}}
\renewcommand{\marginpar}[1]{}
\usepackage[normalem]{ulem}
\newcommand{\ie}{i.e.\xspace}
\newcommand{\bm}[1]{\boldsymbol{#1}}
\definecolor{graylll}{gray}{.95}
\definecolor{grayll}{gray}{.9}
\definecolor{grayl}{gray}{.8}
\DeclareMathOperator{\dd}{d}

\newcommand\rmH{{\rm H}}

\newcommand\tfourth{\ensuremath{{\textstyle\frac{1}{4}}}}
\newcommand\Atw{\mbox{\textit{At}}} 
\newcommand\Sch{\mbox{\textit{Sc}}} 
\graphicspath{{PoF512.fig/}}
\hypersetup{%
	pdftitle={Persistence and bimodality of large-scale turbulent structures across a Rayleigh-Taylor layer: Impact on transport and physical modelling through two-field-conditional correlations},%
	pdfauthor={R. Watteaux, A. Llor, J.A. Redford},%
	pdfsubject={},%
	pdfkeywords={Buoyancy-driven instability, Coupled diffusion and flow, Multiphase flow, Turbulence modelling}}
\shorttitle{Persistence and bimodality of large-scale turbulent structures in RT}
\shortauthor{R. Watteaux, J.A. Redford, and A. Llor}

\title{Persistence and bimodality of large-scale turbulent structures across a Rayleigh--Taylor layer:\\ Impact on transport and physical modelling\\ through two-field-conditional correlations}

\author{R. Watteaux\aff{1,2}
	J.A. Redford\aff{1},\footnote{Present address: \textsc{Eurobios}, 61 av.\ du Président Wilson, 94 235 Cachan Cedex, France}
	\and A. Llor\aff{2}\corresp{\email{antoine.llor@cea.fr}}}

\affiliation{\aff{1} École Normale Supérieure de Cachan, Centre de Mathématiques et de Leurs Applications, \\ 61 av.\ du Président Wilson, 94235 Cachan Cedex, France
\aff{2} CEA, DAM/DIF, 91297 Arpajon Cedex, France}

\begin{document}
%
\maketitle
\immediate\write18{time.cmd}
%
%
\begin{abstract}
	The distribution functions of field fluctuations of the turbulent mixing layer produced by a Rayleigh--Taylor instability (RTI) have long been hypothesized to involve bimodal effects. The present work reviews existing quantitative and qualitative evidence in support this conjecture, provides an associated theoretical framework, and measures the corresponding relevant statistical quantities on a simulation of a turbulent RTI at low Atwood number.

	The bimodal behaviour of fluctuations is readily observable close to the edges of the mixing zone, corresponding to intermittent patches of pure laminar and mixed turbulent fluid. It is less obvious within the mixing zone where indirect evidence comes from different sources, here gathered, discussed, and expanded: energy structure of buoyancy--drag equation, visual eduction from simulated RTI, two-fluid conditional analysis of energy balance, bulk non-dimensional turbulent numbers (Stokes, Knudsen, and Reynolds)\dots Notably, the last two items make appear the very significant differences of turbulent RTI compared to usual turbulent flows such as shear layers, with a higher contribution of so-called directed energy (the portion of turbulent energy due to the relative drift of fluids) and a reduced turbulent viscosity (by almost two orders of magnitude).

	In order to carry out a bimodal analysis of the statistical quantities, two complementary indicator functions are here defined. These could be globally viewed as the `light, upward moving' and `heavy downward moving' fluid zones, with some form of space and time correlation. The numerous existing structure reconstruction techniques could be considered for this purpose but none of them was found satisfactory to provide the final separation of the two-structure field.s An approximate reconstruction is thus introduced here, based on the thresholding of a space- and time-filtered field (here the vertical velocity).

	The prescription for the two-structure-field segmentation was applied to a direct numerical simulation of an RTI and the corresponding structure-conditioned averages of the main quantities were obtained (concentrations, momentum, turbulent energies\dots). Among others, two notable features for turbulence understanding and modelling are observed : i)~the extension of the structure fields across the mixing layer and ii)~the significant ratio of directed to turbulent energies, above previous estimates (up to about 40\%).

	The measured conditional averages provide the first know reference data for validation and calibration of so-called `two-structure' RANS turbulence models such as Youngs' (2015 \emph{Int.\ J.~Heat Fluid Fl.} \textbf{56}, 233--250 and refs therein) and 2SFK (2003 \emph{Laser Part.\ Beams} \textbf{21} (3), 311--315).
\end{abstract}

%
\clearpage
\tableofcontents
%
\clearpage
\section{Introduction}
%
\subsection{Motivation: Modelling-oriented understanding of turbulent gravitational flows}
\label{ssec:Motivation}
	Instabilities are found in many flows involving two stratified fluids of different densities $\rho^1 < \rho^2$ submitted to a gravitational field oriented in the unstable direction---\ie such that $\bnabla p\bcdot\bnabla\rho<0$ where~$p$ is the pressure field. These instabilities (equivalently induced by interface acceleration) have received attention in a wide range of contexts, such as astrophysics, geophysics, meteorology, applied physics (for combustion and inertial confinement fusion), \etc \citep[and references therein]{Inogamov99, Zhou17a, Zhou17b}. One of the most studied of such flows is the Rayleigh--Taylor (RT) instability, an idealized but fundamental academic situation where the two fluids are incompressible, initially at rest on each side of a perturbed horizontal plane interface, and the vertical gravity field is constant. Numerous publications are available on this topic, and the reader will be referred here to just two reviews, a selective but consistent one by \citet[§~2]{Youngs13} and a more recent and comprehensive by \citet{Zhou17a, Zhou17b}.

	At late enough times and high Reynolds number these instabilities evolve into a fully turbulent regime whose modelling is required in many applications. As for any other turbulent flow, modelling is a compromise between robustness, consistency, accuracy, and on-the-field practicality, which may lead to different options depending on the retained balance between these often opposing demands. Current trends appear to favour the approaches of Direct Numerical Simulation, Large Eddies Simulation \citep[DNS and LES, \eg][]{Lesieur05, Sagaut06} and their variants \citep[ILES, MILES, spectral, \etc \eg][]{Grinstein07} but their high computational cost and the patchy physical understanding they bring do not make them practical in complex applications. The traditional approaches centred on Reynolds Averaged Navier--Stokes (RANS) equations are thus predominantly used in practice, although they still demand careful monitoring: RANS models are calibrated to retrieve some predefined features of a limited set of flows, from which excessive departure in applications can produce inaccurate to meaningless results.

	In the modelling of RT-like turbulent flows one of the main goals is to provide a functional relationship between the possibly variable Atwood number $\Atw(t)=(\rho^2-\rho^1)/(\rho^2+\rho^1)$ and acceleration field~$g(t)$ on one hand, and the mean width~$L(t)$ of the Turbulent Mixing Zone (TMZ) on the other hand---among various possibilities, the width~$L(t)$ will here be conveniently and accurately defined below in~\eqref{eq:L} by the so-called `momentum width'. For constant~$\Atw$ and~$g$, and at late times~$t$ with the proper time origin, the growth is self similar with
\begin{equation}
L(t) = \alpha \Atw g t^2 ,
\end{equation}
where coefficient~$\alpha$ depends on~$\Atw$ and on the details of the initial perturbation at the interface. Beyond~$L(t)$, proper modelling of RT-like flows also requires predicting some supplementary relevant quantities, such as turbulent energies, turbulent fluxes, dissipation, fluctuations of fluid concentrations, \etc as provided by many RANS models.

	Now, RANS modelling faces specific challenges in the case of gravitationally induced turbulent flows---some of them discussed in parts~\ref{sec:QualitativeArg} to~\ref{sec:QuantitativeArg}---which can be especially acute for variable and transient accelerations \citep{Neuvazhaev83, Shvarts95, Llor03, Llor05, Schilling10, Redford12a, Redford12b, Griffond14, Grea16}, acceleration reversal \citep{Kucherenko93b, Kucherenko97, Dimonte07, Ramaprabhu13, Aslangil16}, coupled shear--buoyancy drive \citep{Andrews90, Ptitzyna93, Denissen14}, or shocks and compressibility effects \citep{Youngs08, Boureima18}. This has reduced to just a few the number of truly operational models ranging from simple one- and two-equation \citep[$k$ and $k$--$\varepsilon$-like as in][]{Andronov76, Neuvazhaev83, Gauthier90, Dimonte06, Sinkova16, Kokkinakis19, Morgan15, Morgan18}---many of them listed and compared by \citet{VanMaele06}---to more complex higher order \citep[and refs therein]{Andronov82, Besnard87, Besnard96, Gregoire05, Braun21}, two-fluid \citep{Youngs84, Youngs89, Youngs94, Cranfill92}, and so-called `two-structure' \citep{Youngs95, LlorBailly03, LlorPoujade04, Kokkinakis15, Kokkinakis20} also designated as `two-phase' by D.L.~Youngs.
%
\subsection{Present work: Two-structure-field analysis of turbulent gravitational flows}
	Two-structure-field RANS models \citep{Youngs84, Youngs89, Youngs94, Cranfill92, Youngs95, LlorBailly03, LlorPoujade04, Kokkinakis15, Kokkinakis20} have been developed to improve the consistency and accuracy of simulations in the presence of buoyancy, with emphasis on the driving density fluctuations and energy budget in the TMZ. In this approach the flow is segmented into \emph{two large-scale regions} here designated as \emph{`structure fields'} which can be identified and followed over time in such a way that their respective \emph{global displacements} are either \emph{upward} or \emph{downward} in RT-like instabilities. \emph{Conditional} Reynolds averaging can then be performed over each of the regions in order to produce coupled two-structure-field statistical equations amenable to closure by standard procedures of turbulent modelling \citep[§~3 and refs therein]{Llor05}.

	Various authors have also advocated the use of two-structure-field-related approaches following many different rationales, sometimes without producing actual turbulence models, or for application to other than the gravitationally driven flows considered here. Beyond such differences, these works and the present are expected to cross benefit by bringing mutual insights to the theoretical basis of models (and possibly to the relevance of underlying structure definitions). Some prominent of these works are briefly reviewed in appendix~\ref{app:Previous}.

	So far however, two-structure-field models have not been \emph{explicitly} compared or calibrated with respect to measured conditionally averaged quantities: apparent conceptual obstacles (in part explored in the present work) have impaired the detection and reconstruction of flow regions associated with the relevant structures in both experiments and simulations (by DNS or LES). This is a major downside of two-structure-field approaches compared to usual RANS models. For instance, correlations for first and second-order buoyancy-driven single-fluid RANS models have been obtained from RT DNS \citep{Livescu09, Schilling10, Soulard16} or experiments \citep{Dalziel99, Ramaprabhu04, Mueschke06, Banerjee10}. Because of this lack of validation, many modellers and users appear to have some reservations about two-structure-field models despite support from a substantial number of consistent arguments reviewed below in parts~\ref{sec:QualitativeArg} and~\ref{sec:QuantitativeArg} and despite clear success on many test cases \citep{Youngs07, Youngs09, Kokkinakis15, Kokkinakis20}.

	The aim of the present work is thus to provide a first physics-grounded approach to two-field segmentation of structures---thereafter designated as `two-structure-field segmentation'---and to the associated conditional averaging for modelling of buoyancy-driven flows. For this purpose it has appeared necessary to take a broader perspective on the turbulent structure concepts (see sections~\ref{ssec:Persistence} and~\ref{ssec:Memory}). Two principles of \emph{persistence} and \emph{bimodality} have then emerged as foundational and seem generic enough to be applicable to numerous other turbulent flows. Yet, further investigations will be required for extensions beyond the present specific case of RT.

	The two-structure-field segmentation and conditional averaging are exemplified here on LES of the turbulent incompressible RT instability at constant acceleration field and low Atwood number. Although this is the simplest of the flows to be simulated, it is considered to be strongly model constraining \citep[see part~\ref{sec:QuantitativeArg} or][§~3]{Llor05} and provides a useful benchmark for two-structure-field segmentation methods before they can be applied to other more complicated situations. For this purpose, approximate \emph{two-field-segmented presence functions of large-scale turbulent structures}, thereafter denoted~$b^\pm$ and more loosely designated as` two-structure fields', are produced by a \emph{specifically-crafted passive equation} solved by a simple on-the-fly solver added to the simulation code (see part~\ref{sec:Prescription}). An extensive set of correlations \emph{conditioned} by~$b^\pm$ (see part~\ref{sec:RTStructures}) is then produced and compared with the few basic results (see section~\ref{ssec:Profiles}) which can be predicted regardless of any modelling or closure assumptions. The present restriction to simulated incompressible RT flows at vanishing Atwood number and constant acceleration represents a first exploratory approach and a proof of concept.

	It is to be stressed that \emph{the two-structure-field statistical approach examined here must not, by any means whatsoever, be considered as mere modelling}: even if modulated and optimized by some control parameters, it is produced by \emph{perfectly well defined and exact mathematical procedures} which embody \emph{sound physical concepts}. Prominent among these concepts are the so-called `\emph{directed} fluxes' and `\emph{directed} energy' effects which are extensively and quantitatively discussed in sections~\ref{ssec:Directed} and~\ref{ssec:SurrogateDirected}. In contrast, modelling comes as an over-layer \emph{to be examined in later publications} whereby algebraic relationships between different statistical correlations (possibly two-structure-field conditioned) are \emph{postulated}, even if with relevant physical arguments---the most common of such closures being the so-called `Boussinesq', `turbulent viscosity', `first-gradient', or `gradient transport' hypothesis which relates the fluctuation-induced turbulent transport fluxes to gradients of mean quantities \citep{Schmitt07}.
%
\subsection{Structure of the present work}
	Part~\ref{sec:QualitativeArg} provides the rationale and the main physical concepts for two-structure-field segmentation in a broad qualitative way, leaving the detailed mathematical developments for later sections. It is thus highly recommended to all readers, expert or not, in order to follow the general line of later developments.

	Part~\ref{sec:Theory} contains the mathematical background to define the statistical two-structure-field correlations which appear in all later sections. It is not truly original material---readers familiar with usual two-fluid models will find obvious reminiscences with classic textbooks~\citep[\eg][]{Nigmatulin67, Delhaye68, Drew71, Ishii11},---but it defines all the correlations for the analysis of simulations in part~\ref{sec:RTStructures} and it is thus provided here for reference. It also introduces the critical concept of `directed effects' which is central to the quantitative understanding provided in part~\ref{sec:QuantitativeArg}.

	Part~\ref{sec:QuantitativeArg} provides in a condensed form the so-far-available quantitative results from experiments and DNS--LES on RT flows which justify two-structure-field approaches beyond the qualitative elements of part~\ref{sec:QualitativeArg}. Although these basic results have been published up to two decades ago, their interpretation in two-structure-field terms appears to have been so far rather uncommon, incomplete, or convoluted \citep[and refs therein]{Llor05}. This discussion must be considered as the main motivation of the present work: \emph{the strength of directed effects in RT-like flows represents the most compelling quantitative result which is markedly absent in more standard turbulent flows and which by itself justifies the usefulness of two-structure-field analysis.}

	Building on the background of part~\ref{sec:QualitativeArg}, part~\ref{sec:Prescription} prescribes and justifies a two-structure-field segmentation procedure applicable to DNS--LES calculations. It is based on the Lagrangian filtering and picture segmentation of an appropriately selected field, here the vertical velocity, performed on-the-fly during simulations. The method depends on a few adjustable parameters which must be first optimized with respect to selected criteria, primarily a \emph{bimodality coefficient} introduced here.

	Part~\ref{sec:RTStructures} regroups the results obtained from a high-resolution LES of a turbulent incompressible RT flow at $\Atw=0.01$ complemented with the on-the-fly two-structure-field segmentation procedure defined in part~\ref{sec:Prescription}. Extensive two-structure-field conditional averages are computed and the most salient are discussed---full results are provided as online supplementary material. The balance between intra-structure transport and inter-structure exchange is examined more carefully: it is a critical modelling aspect which displays the peculiarity of being equally reproduced by various very different closure assumptions of the two contributions.

	The authors are well aware that two-structure-field segmentation can be somewhat off-putting as it requires becoming acquainted with what could appear as loosely defined concepts, unusual physical quantities, tedious calculations, or unintuitive results. For the sake of readability and completeness they have therefore tried to be as thorough as possible, sometimes reinterpreting well-established approaches---hence the length and structure of parts~\ref{sec:QualitativeArg} to~\ref{sec:QuantitativeArg} and their possible redundancy with material already published elsewhere in different contexts.
%
\section{Qualitative relevance of two-structure-field approaches in RT flows}
\label{sec:QualitativeArg}
%
\subsection{Empirical arguments}
\label{ssec:EmpiricalArg}
	Prior to any developments, it appears necessary to state the \emph{explicit}, \emph{qualitative}, and \emph{consistent} physical reasons---even if \emph{indirect} or \emph{empirical}---which make two-structure-field approaches preferable for understanding and modelling turbulent RT-like flows.

	As stated in section~\ref{ssec:Motivation}, the main goal of RT modelling is to describe the evolution of the overall mixing width~$L(t)$. It is thus natural to start here from the \emph{buoyancy--drag equation}
\begin{equation}
\label{eq:BD}
L'' = C_B\,\Atw\,g - C_D\,\dfrac{L'^2}{L} ,
\end{equation}
where~$L'$ and~$L''$ are the first and second time derivatives of~$L(t)$. Adapted from previous previous investigations, it was notably introduced in this form by \citet[app.~A]{Hansom90}, \citet[p.~535]{Alon95}, \citet[eq.~1]{Dimonte96}, \citet[eqs~13 \&~27]{Ramshaw98} and others as reviewed by \citet[§~12.2]{Zhou17b}. It is the simplest phenomenological model which can be built from~$\Atw(t)$, $g(t)$ and~$L(t)$, and despite the presence of only two adjustable constants $C_B\approx\thalf$ and $C_D\approx2$, it efficiently yields consistent and reasonably accurate results, even for variable (but positive)~$g(t)$ \citep{Dimonte96, Dimonte00, Youngs02, Poujade10, Redford12a, Redford12b}. It can thus be expected that the buoyancy--drag equation derives from some physically-sound underlying principles as revealed by various earlier analysis which happen to have more or less implicitly made use of structure-like concepts similar to those to be developed here. The most salient of these works must therefore be mentioned as they provide a basic background to the present approach:
\begin{itemize}
	\item Following \citet{Ramshaw98}, the nature of the buoyancy--drag equation~\eqref{eq:BD} can be interpreted in terms of an `internal mixing momentum' of the system, given by $LL'$. The non-dissipative part of the buoyancy--drag equation can thus be obtained by applying the least-action principle on a `buoyancy--drag Lagrangian' of type $\thalf L(L')^2 + \thalf C_B\Atw gL^2$. This least-action theoretical framework presumes the existence of \emph{underlying mechanical objects} which carry this generalized momentum and can be identified as \emph{structures}, with some level of persistence.
	\item As elaborated in section~\ref{ssec:LasDrift}, the two momenta of the two-structure-field approach naturally capture the \emph{relative transport} of `heavy' and `light' fluids or structures \citep{Llor03, Llor05} which is directly related to~$L'$ (see also~\eqref{eq:MassFux} and associated comments).
	\item In the buoyancy--drag equation~\eqref{eq:BD}, the characteristic relaxation rate scaled to the growth rate $L'/L$ is given by~$C_D$, not much above unity. It indicates that underlying structures do display a significant level of persistence as discussed in section~\ref{ssec:Memory}, a critical property required in order to make their momentum evolution equations relevant to transport and TMZ growth. This is confirmed by the bulk analysis of turbulence measurements in section~\ref{ssec:NonStandard}, table~\ref{tab:0DSurrogate}, where the relative rate also appears to be similar to that of shear layers.
	\item The here-designated \emph{directed} energy $\propto L(L')^2$---which is the kinetic energy in the buoyancy--drag Lagrangian above,---can be estimated from a simplified bulk `0D' energy-budget analysis of self-similar RT experiments and DNS--LES \citep[also summarized in part~\ref{sec:QuantitativeArg}]{Llor03, Llor05} and appears to be of significant magnitude compared to the \emph{turbulent} kinetic energy. This supports two-structure-field analysis and averaging as it separates directed and turbulent energies in a quantitative and straightforward manner (see section~\ref{ssec:Directed}).
	\item The late-time self-similar growth of a thin layer of heavy fluid atop a light fluid appears to be proportional to time $L\propto t$ \citep{Smeeton87, Kucherenko93a, Jacobs05}. The buoyancy--drag equation can retrieve this behaviour only by assuming an \emph{effective} Atwood number which is `diluted' as $\Atw_e\propto 1/L$---instead of a constant apparent Atwood number which would yield $L\propto t^2$ as in the case of self-similar RT. This implies that large-scale transport cannot be simply identified with the individual fluids (the Atwood number would otherwise be constant) and must be governed by structures comprising the various fluids, whose generation and evolution is mostly dictated by mixing and entrainment.
\end{itemize}

	Despite these compelling theoretical findings, actual geometrical identification of structures in RT layers does not appear to have been carried out, except at TMZ edges where in-flowing laminar fluid transitions to turbulence. The `Alpha-Group' collaboration did observe structural effects \citep[§~V.A and figs~15 to~21]{Dimonte04} and could quantify bubble radii at the edges of the TMZ. In a similar spirit, \citet{Laney06} performed a Morse--Snale structure analysis of the turbulent--laminar boundary in an RT flow. Implicit to these studies, is the concept that `bubbles' undergoing `competition' represent identifiable and dominant transport structures contributing to the growth of the RT layer \citep{Alon95}. The relationship between these findings and the above listed properties of the buoyancy--drag equation would require being re-examined.
%
\subsection[Visual eduction of large-scale structures and two-structure fields]{Visual eduction\footnote{\citet{Hussain70} appear to have been first in using words `educe' and `eduction' in their investigations on turbulent structures.} of large-scale structures and two-structure fields}
\label{ssec:Visual}
\begin{figure}
\centerline{\includegraphics[width=\textwidth]{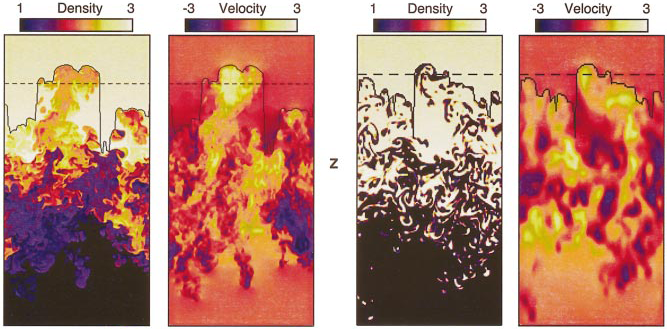}}
\caption{Colour maps in a vertical cross section of the density and vertical velocity fields in two RT DNS carried out for the `Alpha-Group' collaboration with codes \textsc{Turmoil3D} and \textsc{Alegra}, respectively left and right at $256^2\times512$ and $128^2\times256$ cells \citep[reproduced from][fig.~24]{Dimonte04}. \textsc{Turmoil3D} solved a concentration equation for the fluid mixture, hence producing numerical inter-diffusion. \textsc{Alegra} performed interface reconstruction between the fluids, hence assuming non-miscibility. Despite the sharp contrast between these density fields, the associated velocity fields display equally strong turbulent and entrainment effects, and produce overall growths of the TMZ which match within physical and numerical errors.}
\label{fig:AlphaDNS}
\end{figure}
	As a first step towards building effective two-structure-field segmentation procedures, a simple visual inspection of classical RT TMZ data reveals some basic phenomenological and qualitative principles. A relevant starting point is the set of DNS of turbulent RT flows carried out by the `Alpha-Group' collaboration \citep[five research groups, seven different codes]{Dimonte04} to extensively compare and unify findings from different available numerical approaches. Although somewhat obsolescent now due to the resolution then limited to $256^2\times512$ cells, this work still carries insightful information: figure~\ref{fig:AlphaDNS} reproduces the maps of density and vertical velocity fields in a vertical plane as obtained with codes \textsc{Turmoil3D} and \textsc{Alegra} \citep[fig.~24]{Dimonte04}. As visible from the density maps, these codes make different assumptions regarding fluid mixing: the former assumes \emph{miscible} fluids whereas the latter assumes \emph{non-miscible} fluids. One could expect the growth rate of the TMZ to be lower in the former case as the local buoyancy forces are induced by weaker density contrasts. However, the overall growth rates simulated by the two codes were found to be barely distinguishable \citep[fig.~13]{Dimonte04}. This was also observed experimentally \citep[and refs therein]{Kucherenko91, Olson09} although in more disputable ways \citep{Roberts16} as various perturbations are brought about by physical side effects (surface tension, mixture ideality, initial conditions, \etc).

	The weak dependence of the overall growth of the TMZ on small-scale mixing shows that~$L'(t)$ is controlled by turbulent transport at and above \emph{large turbulent scales}. Such a result suggests that fluids are entrained by \emph{turbulent structures} (at energy containing sizes) within which velocity fluctuations are strong enough to neutralize the weaker buoyancy-driven small-scale counter flows of heavy and light fluid elements. In other words, the effective turbulent viscosity generates a form of cohesive behaviour of large scale turbulent structures and makes small-scale inclusions experience a low Stokes number motion. It is supported by visual inspection of figure~\ref{fig:AlphaDNS} where the vertical velocity fields appear to be weakly correlated with small-scale fluctuations of density, and it has also been predicted theoretically by a scaling and anisotropy analysis of the various terms in the turbulent energy equation \citep[and refs therein]{Poujade06, Griffond12}.\marginpar{Add more recent references by these authors and Gréa?} \emph{This qualitative understanding is the fundamental background of all the forthcoming considerations to sort the various eductive approaches to the identification of relevant turbulent structures.}

	Quite strikingly and readily visible in both the miscible and non-miscible simulations in figure~\ref{fig:AlphaDNS}, the large-scale upward- or downward-moving turbulent structures contain just a \emph{small} fraction of the respectively light or heavy driving fluids \citep[about 20\%, see density scales in figure~\ref{fig:AlphaDNS} and][§~V.B]{Dimonte04}. As a further surprising consequence, the `light' upward going structures at the top edge are actually \emph{denser} than the `heavy' downward going structures at the bottom edge!
%
\subsection{Conditional averaging over two-structure fields}
\label{ssec:Averaging}
\begin{figure}
\centerline{\begin{tabular}{cccc}
\includegraphics[width=\textwidth/4-1em]{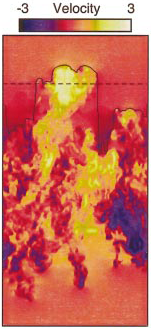}
&	\includegraphics[width=\textwidth/4-1em]{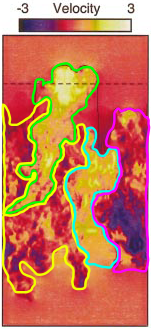}
&	\includegraphics[width=\textwidth/4-1em]{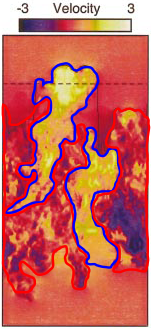}
&	\includegraphics[width=\textwidth/4-1em]{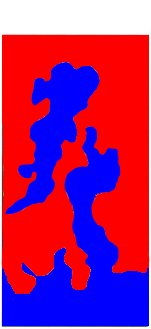}
\\
(a) & (b) & (c) & (d)
\end{tabular}}
\caption{Qualitative visual segmentation of two-structure fields from the vertical velocity in the vertical plane of an RT DNS carried out for the `Alpha-Group' collaboration with code \textsc{Turmoil3D}, see figure~\ref{fig:AlphaDNS} \citep[fig.~24]{Dimonte04}:
(a) original velocity map,
(b) map with boundaries of individual turbulent structures,
(c) two classes of associated turbulent structures, downward and upward moving, respectively bounded in red and blue,
(d) connection of turbulent structures with the laminar fluids from which they originate, giving structure fields~$b^+$ (red) and~$b^-$ (blue).}
\label{fig:structures_turb}
\end{figure}
	The comments above suggest that the vertical velocity component~$u_z$ has relevance as an indicator of structures in RT flows. As illustrated by hand-drawn lines (see figures~\ref{fig:structures_turb}a and~b), individual structures could be broadly defined by connected points in the flow where~$u_z$ is of same sign, possibly matching the sharp density and turbulence contrasts at the edges of the TMZ. These structures can then be regrouped (see figure~\ref{fig:structures_turb}c) and associated with the respectively heavy and light laminar fluids on each side to produce two complementary \emph{structure presence fields} labelled `$+$' and `$-$': $b^+(t,\bm{r})$ (downward moving) and $b^-(t,\bm{r})$ (upward moving), such that $b^\pm=0$ or~1 with $b^+\!+b^-=1$ (see figure~\ref{fig:structures_turb}d). It must be noticed that the loosely defined concept of `structure' is here applied to two related but distinct objects: `structures' (presumably separated and turbulent) and `structure fields'~$b^\pm$ (made of many structures and of laminar fluid outside of the TMZ).

	In the spirit of section~\ref{ssec:EmpiricalArg}, it now appears natural to build models starting from \emph{structure-conditioned} averages of relevant quantities and their fluctuations. For any given per volume quantity~$a$, the usual single-fluid average is then replaced by two per-structure averages
\begin{equation}
\label{eq:TSA}
\overline{a} \qquad\longrightarrow\qquad \overline{b^\pm a},
\end{equation}
with the usual density weighting (or so called `Favre' averaging) for per mass quantities. This comes of course at the expense of i)~dealing with twice as many and more complex evolution equations for per-structure averages, and ii)~providing a proper and relevant definition for the fields~$b^\pm$. The first point involves convoluted yet rigorous and univocal calculations carried out in part~\ref{sec:Theory}. In contrast, the second point is a somewhat ad hoc recipe, here provided in part~\ref{sec:Prescription}, whose relevance must be estimated and optimized from the principles exposed in the rest of this part~\ref{sec:QualitativeArg}.
%
\subsection{Picture segmentation for two-structure fields}
\label{ssec:Segmentation}
	The visually-guided construction of~$b^\pm$ presented in section~\ref{ssec:Averaging} and figure~\ref{fig:structures_turb} actually reduces to a special case of \emph{picture segmentation} by threshold selection~\citep{Otsu79}. Starting from some proper \emph{separator} field~$\beta$ with some level of \emph{contrast} and choosing some \emph{threshold} value~$\beta^\circ$ the structure fields are then defined by
\marginpar{Mention Lumley (which?) for segmenting (or is it filtering?) as in part~\ref{sec:Prescription}?}%
\begin{equation}
\label{eq:bpm_PM}
b^\pm(t,\bm{r}) = \rmH\big[
	\mp \big( \beta(t,\bm{r}) - \beta^\circ(t,\bm{r}) \big) \big] ,
\end{equation}
where~$\rmH$ is the Heaviside step function. In principle, picture segmentation could be considered and optimized in \emph{any} type of flow provided that the ensemble Probability Density Function (PDF) of~$\beta$ be known at each time~$t$ and position~$\bm{r}$ from which $\beta^\circ(t,\bm{r})$ could be prescribed. This is a tedious task in general but it is here simplified by the homogeneity of the RT flow along the transverse coordinates: the ensemble PDF can then be replaced by the in-plane PDF and~$\beta^\circ$ depends on~$t$ and height~$z$ only.

\begin{figure}
\centerline{\begin{tabular}{@{}ccccc@{}}
\includegraphics[width=\textwidth/5]{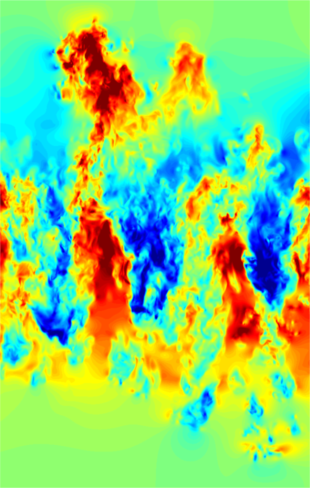}
&	\includegraphics[width=\textwidth/5]{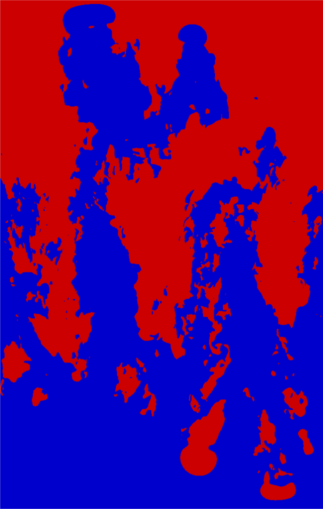}
&	\raisebox{\textwidth*3/2/5-\height}{
		\includegraphics[width=\textwidth/8]{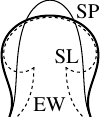}}
&	\includegraphics[width=\textwidth/5]{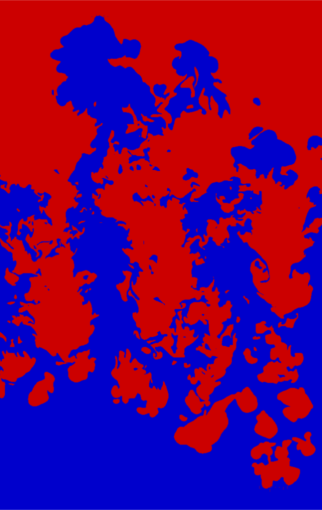}
&	\includegraphics[width=\textwidth/5]{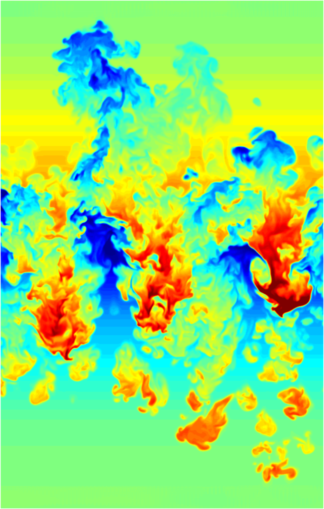}
\\
(a) & (b) & (c) & (d) & (e)
\end{tabular}}
\caption{Illustration of two-structure fields produced in a simulation by segmentation of vertical velocity fluctuations~$u_z'=u_z-\overline{u_z}(z)$ (a and~b, poor man's instantaneous approach) or density fluctuations~$\rho'=\rho-\overline{\rho}(z)$ (e and~d); and schematic representation of distortions expected and observed with the two methods on a `bubble' or `mushroom' like structure at the edge of an RT TMZ (c): an expected two-structure field boundary (thick line) would follow either~$u_z'$ or~$\rho'$ contrasts (solid and dotted thin lines) at respectively the entrained wakes (EW) or the stagnation point (SP) and shear layers (SL).}
\label{fig:RTruc}
\end{figure}
\marginpar{Figure~\ref{fig:RTruc}: Switch color palette on~$\rho$. Give~$\beta^\circ$ in text. Optimized poor man's at $q=\thalf$? Redo figures on next 1024 simulation? Match with 4:1 figure? Romain please check and approve.}%
	An illustration of~\eqref{eq:bpm_PM} on a presently simulated turbulent RT flow with $\beta=u_z$ is provided in figures~\ref{fig:RTruc}a and~b---with~$\beta^\circ$ coarsely defined here as $10^{-2}(z/L)L'\approx0$ to clip residual noise in laminar regions. In all the following, this segmentation based on~$u_z$ will be designated as the `poor man's instantaneous' approach---as will be justified in section~\ref{ssec:Persistence}.

	Many other choices of separator fields~$\beta$ can be considered in principle---for instance density $\beta=\rho'$ in figures~\ref{fig:RTruc}d and~e (or equivalently, fluid mass fractions $c^2=1-c^1$)---but their relevance and efficiency for two-structure-field segmentation requires careful examination.\marginpar{Give~$\beta^\circ$.} They generally suffer from three major limitations related to the PDF properties of~$\beta$: i)~weak \emph{persistence}, ii)~weak \emph{bimodality} as confirmed in section~\ref{ssec:Coefficients} and figures~\ref{fig:bimod_sensi}a to~c, and as a consequence iii)~fractality of the two-structure-field boundaries as visible in figure~\ref{fig:2DMaps}. These properties are related to the large (noisy) turbulent fluctuations on~$\beta$.

	Despite introducing these and other distortions to actual two-structure presence fields as discussed in section~\ref{ssec:Distortions}, the poor man's instantaneous approach is a sensible starting point for crafting more evolved segmentation strategies in RT flows (see section~\ref{ssec:Persistence}). Furthermore, it is a convenient approximation whenever the persistence-retrieving approaches (see section~\ref{ssec:Memory}) put insuperable technical burdens, as for instance and prominently in experiments.
%
\subsection{Distortions induced by instantaneous field-segmentation approaches}
\label{ssec:Distortions}
	As all direct field-segmentation approaches, the instantaneous poor man's introduces many distortions which are revealed by a closer educated analysis of the visual eduction process illustrated in figure~\ref{fig:RTruc}.

	One of our simulations provided an insightful event shown in figure~\ref{fig:RTruc}: at the top edge of the TMZ the segmented maps of velocity and density do not match and both distort what would be the expected two-structure presence fields (compare the horse-head and slender shapes in respective figures~\ref{fig:RTruc}d and~b). Another similar but less striking example is found at the bottom middle edge of the TMZ (compare triangle and slender shapes in same figures).

	The observed distortions can be qualitatively traced to the main features of velocity and density fields in the vicinity of a turbulent structure at the edge of the TMZ. Around a typical `mushroom-like' density-segmented structure (see figure~\ref{fig:RTruc}c, dashed line), the velocity field displays three important peculiarities: i)~a stagnation point at the top (SP), ii)~shear layers at the sides (SL), and iii)~entrainment zones around the wake (EW). These features displace the threshold surface of~$u_z$ thus distorting the associated structure boundaries towards slenderer shapes (see figure~\ref{fig:RTruc}c, thin line).

	Intuitively however, one would expect a sensible structure detection to produce a somewhat interpolated result in between the density and velocity fields (see figure~\ref{fig:RTruc}c, thick line). Similar effects are also present within the TMZ but in a much more convoluted way because of the fractal character of the boundary and the weaker contrast (or bimodality) of turbulent velocity between structures. A possible correction could be brought by considering $\mathcal{P}(\rho',u_z)$, the joint PDF of $\rho'$ and $u_z$, but extensive exploratory work not reported here showed scant improvement of neither bimodality nor persistence.

	As quantified in section~\ref{ssec:PMLimits}, the distortions and the fractal character produced by field-segmentation affect \emph{mass exchange} between structures and \emph{mass fluxes} within structures, two central quantities in the two-structure-field statistical framework. It thus appears necessary to re-examine the visual eduction of two-structure presence fields in order to correct the poor man's instantaneous approach.

\subsection{Two-structure field segmentation from persistent individual turbulent structures}
\label{ssec:Persistence}
	The instantaneous field-segmentation approaches such as the poor man's (see sections~\ref{ssec:Averaging}, \ref{ssec:Segmentation}, and~\ref{ssec:Distortions}) may appear as a reasonable embodiment of the visual eduction of two-structure presence fields introduced in section~\ref{ssec:Visual}---of course, after some optimized selection of~$\beta$ and~$\beta^\circ$. In principle as illustrated in figure~\ref{fig:structures_turb}, a $\beta$-field contrast would, \emph{all at once}, delineate, sort, and merge the \emph{individual} large-scale turbulent structures to produce~$b^\pm$ (see figures~\ref{fig:structures_turb}b and~d)---notice that laminar areas constitute individual structures on their own. In fact these $\beta$-segmentation approaches are approximate: applied to~$u_z$ and~$\rho'$, as respectively visible for instance in figures~\ref{fig:RTruc}a and~b, and d and~e, they provide clearly different but equally distorted and fractal two-structure fields.

\begin{figure}
\centerline{\begin{tabular}{cc}
\includegraphics{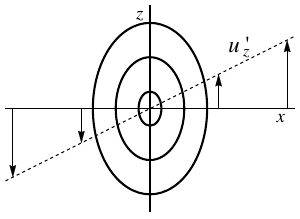}
&	\includegraphics{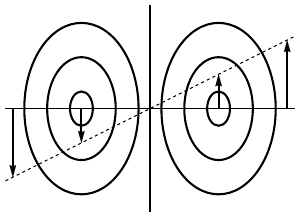}
\\
(a) & (b)
\end{tabular}}
\caption{2D schematic representation of segmentation by a locally uniform vertical velocity gradient~$u_{z,x}$ with threshold~${u_z}^\circ=0$ in two different but compatible configurations of large-scale vortex-like structures (here idealized as laminar). The ensuing two-structure-field boundary would in (a)~erroneously intersect a structure with about-neutral motion, but in (b)~properly separate two crossing structures with opposite motions.}
\label{fig:Gradient}
\end{figure}
	More generally, individual structures cannot be properly delineated by \emph{local} segmentations of \emph{still} images. A trivial example illustrated in figure~\ref{fig:Gradient} is provided by the segmentation at threshold ${u_z}^\circ=0$ of a locally uniform velocity gradient~$u_{z,x}$: two different flows can be considered which are consistent with~$u_{z,x}$, but only one is consistent with structure-field boundaries given by ${u_z}^\circ=0$. A proper delineation of large-scale structures thus demands that fluid elements be typically correlated over integral (energy containing) space and time scales. This is actually implicit to an intuitive visual eduction and represents a form of `\emph{persistence}' of turbulent structures---also related to their `cohesiveness'.

	\emph{Persistence is here defined as the property of a fluid domain whose elements share a common mean collective motion over space and time scales of the order of the turbulent energy containing scales.} While long thought as intrinsic to most existing structure detection schemes (briefly reviewed in appendix~\ref{app:Structures}), observation of this collective motion actually requires an adapted processing of relevant fluid fields.

	In the present RT analysis, unveiling persistence requires reducing both space and time fluctuations on~$\beta$ to make the PDF bimodal and the boundaries non fractal (see figures~\ref{fig:RTruc}b and~d). Recirculating zones are then properly segmented (see figures~\ref{fig:RTruc}c and~\ref{fig:Gradient}). The persistence concept is already embedded into the buoyancy--drag equation as discussed in section~\ref{ssec:EmpiricalArg} and must therefore be somehow incorporated into any two-structure-field detection approach.

	In principle, incorporating persistence into a two-structure-field segmentation would start by identifying large-scale persistent structures labelled~$s$ (turbulent or possibly laminar) and reconstructing their presence functions~$b^s$ such that $\sum b^s = 1$ (somewhat similarly to the honeycomb or space-filling tessellation illustrated in figure~\ref{fig:structures_turb}). Final two-structure presence fields would then be given by
\begin{equation}
\label{eq:Regroup}
b^\pm = {\textstyle\sum}_{\langle u_z\rangle_s
		\lessgtr \langle u_z\rangle^\circ} b^s ,
\end{equation}
where~$\langle u_z\rangle_s$ is the average vertical velocity over structure~$s$, $\langle u_z\rangle^\circ$ is some optimized velocity threshold, and order tests~$<$ or~$>$ are associated with respective superscripts~$+$ or~$-$. Segmentation would thus inherit the regularity, persistence, and bimodality of the PDF of~$\langle u_z\rangle_s$---instead of the PDF of~$u_z$.

\begin{figure}
\centerline{\includegraphics[width=\textwidth]{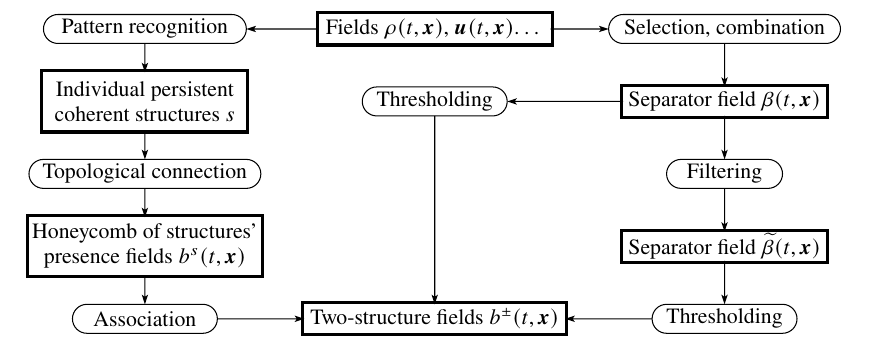}}
\caption{Sketch of three possible strategies implicit to the visual eduction of two-structure presence fields introduced in section~\ref{ssec:Visual}. The path on the left is the approximate but simple `poor man's' approach of instantaneous field thresholding---\eqref{eq:bpm_PM} in section~\ref{ssec:Segmentation}. The path on the right represents the theoretically rigorous yet very tedious approach whereby individual turbulent (persistent) structures are identified and regrouped---\eqref{eq:Regroup} in section~\ref{ssec:Persistence}. The path in the middle represents a more approximate yet tractable approach whereby a relevant combination of fields is selected, filtered, and segmented---section~\ref{ssec:Memory} and~\eqref{eq:Filtering} in part~\ref{sec:Prescription}.}
\label{fig:StructureApproaches}
\end{figure}
	The two-structure segmentation approaches through thresholding of instantaneous fields and through association of persistent individual turbulent structures are summarized by the respective paths at the left and the right of figure~\ref{fig:StructureApproaches}.
%
\subsection{Final prescription: persistence and bimodality through space--time filtering}
\label{ssec:Memory}
\marginpar{Romain please check section!}
	As elaborated in the previous sections and summarized in figure~\ref{fig:StructureApproaches}, the rich, yet intuitive, visual eduction of two-structure fields in section~\ref{ssec:Visual} would be best embodied by a reconstruction and association of the individual volume-filling structures as sketched in section~\ref{ssec:Persistence}. The daunting complexity of this approach (see appendix~\ref{app:Structures}) has here motivated the development of a simpler approximation based on an extension of the instantaneous poor man's thresholding. It is defined as \emph{some} segmentation of a separator field~$\widetilde{\beta}$ produced by \emph{some} space--time low-pass filtering of \emph{some} relevant raw field~$\beta$---as summarized by the path in the middle of figure~\ref{fig:StructureApproaches}.

	The above prescription involves three occurrences of `\emph{some}' which apply to the main objects and tools under investigators' control in order to optimize the two-structure field detection. In the same spirit of simplicity which leads to the segmentation of the vertical velocity field in section~\ref{ssec:Segmentation}, a basic approach to space--time filtering is retained here: an adapted convection--diffusion--relaxation equation~\eqref{eq:Filtering} detailed in part~\ref{sec:Prescription}. It is to be noticed that the usual poor man's instantaneous approach does displays some level of filtering: a standard thresholding field~$\beta$ is generally produced through an evolution equation which necessarily introduces some level of time correlation. However, the ensuing persistence and bimodality of~$\beta$ cannot be controlled.

	As will appear in part~\ref{sec:RTStructures} this simplified prescription eventually yields \emph{statistically appropriate} results with enhanced bimodality and reduced fractality when optimizing the filtering and segmentation conditions so as to best mimic the expected behaviour of individual turbulent structures. The trade-off is that, despite statistical appropriateness, the two-structure-field boundary can still run across (instead of around) some of the individual large-scale turbulent structures---somewhat as the segmented unfiltered fields do in figures~\ref{fig:RTruc}c and~\ref{fig:Gradient}. As a consequence, exchange terms may experience enhanced high-frequency noise coming from higher small-scale fluctuations at the structure interface.

	The relevance of convection--diffusion--relaxation equations in turbulent systems has been often reported and it is just worth quoting here \citet[p.~204, eq.~1 \& fig.~2]{Lumley92} who provided a clear justification in modelling contexts: \emph{`This idea that a non-local model is needed comes from the realization that conditions at a point in a turbulent flow depend on the history of the material elements that arrive at this point, and hence should depend on some sort of weighted integral with a fading memory back over the mean path through the point in question, with a progressively broadening domain of integration, corresponding to the backward turbulent diffusion, perhaps in first approximation a Gaussian'.} Such \emph{Lagrangian filtering} techniques were also pioneered notably by \citet[§~12 and refs therein]{Pope00} for modelling turbulent fluctuations and intermittency. In the same spirit of these works, the separation of two-structure-fields will also appear to capture the large-scale turbulent intermittency near the edges of the TMZ where fluctuations display an obvious bimodal behaviour.

	Within this general framework of convection--diffusion--relaxation equations, there is still a wide range of possible adjustments of options and parameters which have significant impact on final two-structure fields. In preliminary explorations to the present study, many such options and parameters were thus tested essentially on a trial and error basis---a short commented list is provided in \citet[app.~B]{Watteaux11}. Here, we shall only describe and justify the most efficient method found so far according to optimization criteria which are elaborated in part~\ref{sec:Prescription}.

	\emph{Filtering is thus a critical ingredient in the present work} as will be elaborated in part~\ref{sec:Prescription}.
%
\section{Theoretical framework of the two-structure-field approach}
\label{sec:Theory}
%
\subsection{Two-structure-field conditionally averaged equations}
\label{ssec:TSCAE}
	The formal procedure to produce conditionally averaged two-field statistical equations has already been shown many times in the slightly different contexts of intermittency and transition \citep{Libby75, Libby76}, energy balance analysis in RT type flows \citep{Llor03, Llor05}, combustion \citep{Spalding86, Spalding87},\marginpar{Find more references. See Veynante.} or two-phase and two-fluid flows \citep{Nigmatulin67, Delhaye68, Drew71, Ishii11}---although the latter generally involve space, time or section averages for respectively homogeneous, stationary, or pipe flows, instead of an ensemble average as in RANS approaches (for more modern discussions and references see for instance \citet[§~3]{Worner03}; \citet[§~3]{Morel05}; \citet[§~1]{Brennen05}). It is here adapted to the two-structure fields~$b^\pm$ with mass transfer.

	Whatever the procedure adopted by the modeller for defining structures, here labelled~$+$ and~$-$, these will be fully defined for any given flow realization and at any given time and position by two presence functions $b^\pm(t,\bm{r})$. For approaches such as the poor man's instantaneous segmentation~\eqref{eq:bpm_PM} with $\beta=u_z$, $b^\pm$~is constrained to take values~0 or~1, but all the following equations are also valid if~$b^\pm$ takes intermediate values---this can be important in numerical applications where the interface between structures is necessarily spread over at least one mesh cell, or also in the surrogate two--fluid approach of section~\ref{ssec:Surrogate}. In the initial state, the fluids are fully separated and $b^- = c^1$ and $b^+ = c^2$ (where~$c^1$ and~$c^2$ are the mass fractions of the two fluids). The evolution of the structure fields, \ie the evolution of~$b^\pm$, also follows from the particular choices adopted by the modeller, presumably tailored to best capture a given relevant contrast.

	Independent of the definition and behaviour of~$b^\pm$, the flow is described by the usual Navier--Stokes equation and the complementary conservation laws. Any generic per-mass quantity~$a$ (fluid mass fractions~$c^2$ and~$c^1$, momentum~$\bm{u}$, internal energy~$e$, \etc) evolves according to
\begin{equation}
\label{eq:a}
\partial_t ( \rho a ) + ( \rho a u_j )_{,j} = - \vartheta^a_{j,j} + s^a ,
\end{equation}
where~$\bm{u}$, $\bm{\vartheta}^a$, and~$s^a$ are respectively the local fluid velocity, flux of~$a$, and source (or dissipation) of~$a$---Einstein's notation of implicit summations on repeated indices will be used throughout the following. The corresponding two-structure-field RANS equations are readily obtained by expanding the \emph{$b^\pm$-conditional ensemble average} of~\eqref{eq:a}, $\overline{ b^\pm \times \eqref{eq:a} }$, and give
\begin{subequations}
\begin{align}
\label{eq:asta0}
\partial_t (\overline{b^\pm \rho a})
	&+ \big(\overline{b^\pm \rho a u_j}\big)_{,j}
		- \overline{(\dd_t b^\pm) \rho a}
	= \overline{b^\pm_{,j}\vartheta^a_j}
		- (\overline{b^\pm\vartheta^a_j})_{,j}
		+ \overline{b^\pm s^a} ,
\\*
\label{eq:asta}
	&= - \overline{b^\pm} \, \big(\overline{\vartheta^a_j}\big)_{,j}
		+ \overline{b^\pm_{,j}
			(\vartheta^a_j-\overline{\vartheta^a_j})}
		- \big(\overline{b^\mp}\,\overline{b^\pm\vartheta^a_j}
		- \overline{b^\pm}\,\overline{b^\mp\vartheta^a_j}\big)_{,j}
		+ \overline{b^\pm s^a} ,
\end{align}
\end{subequations}
where $\overline{\rule{0em}{.66em}\cdots\rule{0em}{.66em}}$ is the ensemble averaging operator and $\dd_t b^\pm = \partial_t b^\pm\!+b^\pm_{,j} u_j$ is the Lagrangian derivative of~$b^\pm$ (non vanishing in general as~$b^\pm$ is not transported by the flow). The integration by parts included here reveals the \emph{mean per-structure quantities} $\overline{b^\pm\cdots}$ and \emph{mean exchange terms} $\overline{(\dd_t b^\pm)\cdots}$ and $\overline{b^\pm_{,j}\cdots}$, whereas the decomposition of~$\bm{\vartheta}^a$ terms highlights the exchanges produced by fluctuations within and imbalances between the structures. The exchange terms couple the~$+$ and~$-$ equations in a conservative way since $b^+_{,j}+b^-_{,j}=0$. Form~\eqref{eq:asta0} is equivalent and is privileged for high contrasts of fluxes between structures. 

	Introducing the per-structure averages of
\begin{subequations}
\label{eq:all}
\begin{align}
&\text{volume fractions} & \alpha^\pm &= \overline{b^\pm} ,
	&&\text{densities} & \rho^\pm &= \overline{b^\pm\rho}/\overline{b^\pm} ,
\\
&\text{quantities~$a$} & A^\pm
			&= \overline{b^\pm\rho a}/\overline{b^\pm\rho} ,
	&&\text{velocities} & \bm{U}^\pm
			&= \overline{b^\pm\rho\bm{u}}/\overline{b^\pm\rho} ,
\\\label{eq:Phia}
&\text{turbulent transport} & \bm{\varPhi}^{a\pm}
			&= \overline{b^\pm\rho a\bm{u}^\pm} ,
	&&\text{with fluctuations} & \bm{u}^\pm
			&= \bm{u} - \bm{U}^\pm ,
\\
&\text{fluxes} & \bm{\varTheta}^{a\pm}
			&= \overline{b^\pm\bm{\vartheta}^a}/\overline{b^\pm} ,
	&&\text{and sources} & S^{a\pm} &= \overline{b^\pm s^a} ,
\end{align}
the mean flux
\begin{equation}
\bm{\varTheta}^a = \overline{\bm{\vartheta}^a}
	= \alpha^+\bm{\varTheta}^{a+} \!+ \alpha^-\bm{\varTheta}^{a-} ,
\end{equation}
and the mean exchange terms by
\begin{align}
\label{eq:Psia}
&\text{interfacial fluid transport} & \varPsi^a
	&= \overline{\mp (\dd_t b^\pm) \rho a} ,
\\
&\text{interfacial flux fluctuations} & D^a
	&= \overline{\mp b^\pm_{,j} (\vartheta^a_j-\overline{\vartheta^a_j})} ,
\\
&\text{and inter-structure flux imbalance} & \bm{\varDelta}^a
	&= \alpha^+\alpha^- (\bm{\varTheta}^{a+}\!-\bm{\varTheta}^{a-}) ,
\end{align}
\end{subequations}
the statistical equations~\eqref{eq:asta} are finally written as
\begin{equation}
\label{eq:amean}
\partial_t (\alpha^\pm \rho^\pm A^\pm)
	+ (\alpha^\pm \rho^\pm A^\pm U^\pm_j)_{,j}
		= - \varPhi^{a\pm}_{j,j} \mp \varPsi^a
		- \alpha^\pm\varTheta^a_{j,j} \mp D^a \mp \varDelta^a_{j,j}
		+ S^{a\pm} .
\end{equation}
\emph{This is the basic equation of two-structure-field analysis and modelling}. It displays the same structure as usual single-fluid averaged equations with transport, fluxes, and sources, but with new additional exchange terms.

	The full set of two-structure-field statistical equations, which will be analysed in the present work, is obtained when substituting~$a$ in~\eqref{eq:all} and~\eqref{eq:amean} by $1/\rho$ for structure volume fractions, $1$~for densities, $c^2$ and~$c^1$ for fluid mass fractions, $\bm{u}$ for momenta, and $\thalf(\bm{u}^\pm)^2$ for turbulent kinetic energies. For compactness, the corresponding explicit equations are provided in~\eqref{eq:D2SF} appendix~\ref{app:2SF}, and only their most critical features will be examined in all the following.
%
\subsection{Alternative expressions of exchange terms}
\label{ssec:Interface}
	Exchange terms in~\eqref{eq:amean} are critical ingredients for two-structure-field description and analysis. Although their general expression in~\eqref{eq:Psia} appears canonical, they can be given other forms which, although equivalent, may lead to alternative physical interpretations and modelling options. Three such aspects are examined below.
%
\subsubsection{Exchange as volume transport across interface between structures}
\label{ssec:Psiab}
	Expression of the exchange terms~\eqref{eq:Psia} involves the Lagrangian derivatives of the two-structure fields $\dd_t b^\pm$, which combine their time and space derivatives. Those are well defined if~$b^\pm$ are smooth in time and space, and an effective (finite) velocity field $\bm{w}(t,\bm{r})$ can then be found to describe their evolution
\begin{equation}
\label{eq:b}
\partial_t b^\pm + w_j \, b^\pm_{,j} = 0 .
\end{equation}
Substituting~\eqref{eq:b} into~\eqref{eq:Psia} then yields \citep[eqs~27 \&~28]{Kataoka86, Drew83}
\begin{equation}
\label{eq:Psiab}
\varPsi^a = \overline{\mp b^\pm_{,j} (u_j\!-\!w_j) \rho a} ,
\end{equation}
which shows that exchange between structures is carried by the volume transfer $b^\pm_{,j} (u_j\!-\!w_j)$ due to the material velocity relative to the interface $\bm{u}-\bm{w}$.

	If~$b^\pm$ only take values~0 and~1, both $\partial_t b^\pm$ and $\bm{\nabla}b^\pm$ become singular. However, if the (sharp) interface between structures is smooth and evolves smoothly---as defined for instance by level-set approaches based on continuous Hamilton--Jacobi evolution equations,---then~\eqref{eq:b} and~\eqref{eq:Psiab} still hold with Dirac-like scalar and vector functions $\partial_t b^\pm$ and $\bm{\nabla}b^\pm$ at the interface. This situation is expected for the filtered two-structure-field segmentation presented in part~\ref{sec:Prescription}. Of course, only the normal component of~$\bm{w}$ at the interface remains relevant.

	Now, important situations may be considered where the interface does not evolve continuously, notably: i)~in discretized simulations, where thresholding depends on number representations and can make packets of neighbouring cells switch in one time step from one structure to the other; ii)~for small-scale noisy and fractal separator fields; and iii)~for segmentation based on the separation--association of individual large-scale turbulent structures presented in section~\ref{ssec:Persistence}, where large space domains transition suddenly whenever a $\langle u_z\rangle_s$ switches sign in~\eqref{eq:Regroup}. In such cases neither~\eqref{eq:b} nor~\eqref{eq:Psiab} can hold any more, but the ensemble averaging of these singular events in~\eqref{eq:Psia} still produces smooth quantities.

	Therefore, in all the present work, despite the useful and intuitive theoretical understanding brought by~\eqref{eq:Psiab}, only~\eqref{eq:Psia} will be retained to compute actual exchange terms in simulations.
%
\subsubsection{Upwind decomposition of exchange}
\label{ssec:Psiau}
	Understanding and modelling of exchange terms~\eqref{eq:Psiab} inspires an upwind decomposition of transport at the interface (invariant through $b^-\rightarrow-b^+$ substitution)
\begin{align}
\varPsi^a &= \varPsi^{a\oplus} - \varPsi^{a\ominus}
\nonumber\\
	&= \thalf \overline{\big[ \dd_t b^- \!+ |\dd_t b^-| \big] \rho a}
	+ \thalf \overline{\big[ \dd_t b^- \!- |\dd_t b^-| \big] \rho a}
\nonumber\\
	&= \varPsi^{\alpha\oplus}\!\rho^\oplus\!A^\oplus
		\!- \varPsi^{\alpha\ominus}\!\rho^\ominus\!A^\ominus,
\end{align}
where effective (positive) \textcircled{$\pm$} to \textcircled{$\mp$} volume transfer rates and effective (upwind) \textcircled{$\pm$} quantities are given by
\begin{subequations}
\begin{align}
\label{eq:omegaOMean}
\varPsi^{\alpha\text{\textcircled{$\pm$}}}
	&= \thalf \overline{\big[ \dd_t b^- \!\pm |\dd_t b^-| \big]} ,
\\\label{eq:rhoaOMean}
\rho^\text{\textcircled{$\pm$}}\!A^\text{\textcircled{$\pm$}}
	&= \thalf \overline{\big[ \dd_t b^- \!\pm |\dd_t b^-| \big] \rho a}
		\Big/ \varPsi^{\alpha\text{\textcircled{$\pm$}}} .
\end{align}
\end{subequations}
A basic and sensible modelling closure commonly consists in approximating $\rho^\text{\textcircled{$\pm$}} \! A^\text{\textcircled{$\pm$}} \approx \rho^\pm A^\pm$ in~\eqref{eq:rhoaOMean}. The closure of \citet[§~2]{Youngs95} is retrieved by further setting $\varPsi^{\alpha\text{\textcircled{$\pm$}}} \propto \alpha^\pm$, a more questionable assumption as will be discussed in section~\ref{ssec:Exchange}.\marginpar{Check section label.}

	According to~\eqref{eq:omegaOMean}, the volume exchange rates~$\varPsi^{\alpha\text{\textcircled{$\pm$}}}$ appear to scale as the area density $\overline{\|\bm{\nabla}b^\pm\|}$. Therefore, when the interface is highly corrugated or even fractal, the exchange rates become asymptotically infinite and the corresponding upwind quantities $\rho^\text{\textcircled{$\pm$}}\!A^\text{\textcircled{$\pm$}}$ must become asymptotically equal. Proper large scale averaging of~$b^\pm$ must be introduced to make exchange terms insensitive to such small-scale details.
%
\subsubsection{Intertwining of intra-structure fluxes and inter-structure exchanges}
\label{sssec:Intertwin}
	All the different flux and exchange terms on the right~hand side of~\eqref{eq:amean} may appear to have unique well defined definitions in~\eqref{eq:all} which would let carry term-to-term comparisons between different simulated and modelled results.

	This is actually misleading because flux and exchange terms are intimately coupled: only their \emph{combined} effect is relevant in~\eqref{eq:amean}. For instance, if $\bm{F}(t,\bm{r})$ is an arbitrary vector field then the modified fluid-transport flux and exchange terms
\begin{align}
\label{eq:PsiPhi}
\bm{\varPhi}^{a\pm*} = \bm{\varPhi}^{a\pm} \pm \bm{F}
&&
\text{and}
&&
\varPsi^{a*} = \varPsi^a - F_{j,j} ~,
\end{align}
leave as \emph{invariant} the evolution equations of mean quantities~$A^\pm$
\begin{equation}
\partial_t (\alpha^\pm \rho^\pm A^\pm) + \cdots
	= - \varPhi^{a\pm}_{j,j} \mp \varPsi^a + \cdots
	= - \varPhi^{a\pm*}_{j,j} \mp \varPsi^{a*} + \cdots ~.
\end{equation}
This invariance holds whatever the couple of flux and exchange terms considered in the equation and whatever their definition or source (simulation, modelling, \etc). Thus, provided that all other terms of the balance equations are identical term to term, simulations and models yielding different flux and exchange terms must be accepted as consistent if related by~\eqref{eq:PsiPhi}.

	A particularly striking example is provided by the flux and exchange terms of volume as obtained for $a=1/\rho$ in the case of an RT flow at vanishing $\Atw$. In this case, the Boussinesq approximation makes $\varPsi^\alpha(t,z)$ odd around the initial interface position and thus $\int_{-\infty}^{+\infty}\varPsi^\alpha \dd z = 0$. Accordingly, defining $\bm{F}$ by
\begin{equation}
F_z(z) = \int_{-\infty}^z \varPsi^\alpha(z') \dd z' ~.
\end{equation}
would yield $\varPsi^{\alpha*}=0$ in~\eqref{eq:PsiPhi} with $F_z(\pm\infty)=0$ as expected in pure laminar domains outside the TMZ. Volume fraction evolution in RT at vanishing $\Atw$ can thus be described with a vanishing volume exchange.

	Expressions~\eqref{eq:all} will be retained here as canonical for their algebraic simplicity and intuitive meaning, but comparisons between different simulations or with models should preferably be carried out on the combinations $\varPhi^{a\pm}_{j,j} \pm \varPsi^a$.
%
\subsection{Growth rate of mixing layer as inter-structure drift velocity}
\label{ssec:LasDrift}
	As a follow up to the empirical arguments in section~\ref{ssec:EmpiricalArg}, it is now possible to rigorously relate the two-structure-field approach and~$L(t)$, defined in the present section as the total width of the TMZ between points where \emph{structure} volume fractions vanish, $\alpha^\pm\rightarrow0$. For practical reasons given in section~\ref{ssec:Bulk} a different but close definition of~$L$~\eqref{eq:L} with points where \emph{fluid} volume fractions vanish will be retained in all the other sections of this work. For a given structure-segmentation method in 1D self-similar conditions these definitions match up to a constant factor close to unity.

	In the 1D-symmetric case, the two edges of the TMZ---assumed to be of compact support---are given by the two points~$z^\pm(t)$ where $\overline{b^\pm}(t,z(t))$ vanish, \ie
\begin{equation}
\overline{b^\pm}(t,z^\pm(t)) = \alpha^\pm(t,z^\pm(t)) = 0
	\qquad\text{or}\qquad \overline{b^\pm\rho}(t,z^\pm(t))
		= (\alpha^\pm\rho^\pm)(t,z^\pm(t)) = 0 ,
\end{equation}
The time derivative of this last equation immediately yields the time derivative of~$z^\pm(t)$
\begin{align}
\dd_t z^\pm &= \lim_{z \rightarrow z^\pm}
	\frac{-\,\partial_t(\alpha^\pm\rho^\pm)}{(\alpha^\pm\rho^\pm)_{,z}}
	= \lim_{z \rightarrow z^\pm}
		\frac{(\alpha^\pm\rho^\pm U_z^\pm)_{,z}\pm\varPsi^\rho}
			{(\alpha^\pm\rho^\pm)_{,z}}
\nonumber\\
	&= \lim_{z \rightarrow z^\pm}
		\frac{\alpha^\pm\rho^\pm U_z^\pm}{\alpha^\pm\rho^\pm}
	= U_z^\pm(t,z^\pm) ,
\end{align}
The first line is obtained by substituting $\partial_t(\alpha^\pm\rho^\pm)$ from the two-structure-field mass-conservation equations~\eqref{eq:DR_pm}. The two ensuing contributions are simplified separately to obtain the last expression: i)~as the behaviour of structures is not singular at the edges, the mass exchange term must scale as $\varPsi^\rho\sim\alpha^\pm$ and the ratio $\varPsi^\rho/(\alpha^\pm\rho^\pm)_{,z}$ vanishes for any $\alpha^\pm(z)\sim(z-z^\pm)^n$ and ii)~l'Hospital's rule is applied at~$z^\pm$ where $\alpha^\pm\rho^\pm=0$.

	The growth of the TMZ width is then obtained as
\begin{align}
\label{eq:LasDrift}
L'(t) = \dd_t z^- \!- \dd_t z^+ = U_z^-(z^-) - U_z^+(z^+)
	&= - \delta U_z(z^-) - \delta U_z(z^+)
\nonumber\\*&
	\stackrel{\Atw\,\rightarrow\,0}{\longrightarrow}\quad
		-2\delta U_z(z^\pm) ,
\end{align}
where $\delta\bm{U}=\bm{U}^+\!-\bm{U}^-$ is the \emph{drift velocity} between structures and bearing in mind that $U_z^-(z^+)=U_z^+(z^-)=U_z(\pm\infty)$. The last expression is valid for vanishing Atwood number where the TMZ is symmetrical. This extends to the general two-structure-field case the specific two-fluid result of \citet[eq~4.4]{Llor05}. For non-vanishing Atwood number, $\delta U_z$~is still expected to provide a reasonable estimate of~$L'$.

	The buoyancy--drag equation~\eqref{eq:BD} is thus an equation for the drift velocity and constrains the consistency of modelled equations for two-structure-field momenta $\bm{U}^\pm$~\eqref{eq:DU_pm}.
%
\subsection{Relationship with usual single-fluid approach, directed effects}
	Because $b^+\!+b^-=1$, the single-fluid and two-structure-field quantities are trivially connected by decomposing the so-called `Favre' average $A=\overline{\rho a}/\overline{\rho}$ as the sum of the conditional averages $\overline{b^\pm\rho a}$
\begin{equation}
\overline{\rho} A = \alpha^+\rho^+A^+ \!+ \alpha^-\rho^-A^- .
\end{equation}
The single-fluid statistical equation can thus be obtained by a usual ensemble average of the~$a$ equation in~\eqref{eq:a}, or by adding the~$A^\pm$ equations in~\eqref{eq:amean}, giving either
\begin{subequations}
\label{eq:1F2S}
\begin{align}
\partial_t (\overline{\rho} A) + (\rho A U_j)_{,j}
	&= - \varPhi^a_{j,j} - \varTheta^a_{j,j} + S^a ,
\\*
	&= - ( \varPhi^{a+} \!+ \varPhi^{a-}
	\!+ \alpha^+\rho^+A^+U^+_j + \alpha^-\rho^-A^-U^-_j - \rho A U_j )_{,j}
\nonumber\\*&\hspace{12em}
			- \varTheta^a_{j,j} + S^{a+} \!+ S^{a-} ,
\end{align}
\end{subequations}
with single-fluid averages defined by
\begin{subequations}
\begin{align}
&\text{quantity~$a$} & A
			&= \overline{\rho a}/\overline{\rho} ,
	&&\text{velocity} & \bm{U}
			&= \overline{\rho\bm{u}}/\overline{\rho} ,
\\
&\text{turbulent flux} & \bm{\varPhi}^{a}
			&= \overline{\rho a\bm{u}''}
	&&\text{with~$\bm{u}$ fluctuation} & \bm{u}''
			&= \bm{u} - \bm{U} ,
\\
&\text{flux} & \bm{\varTheta}^a &= \overline{\bm{\vartheta}^a} ,
	&&\text{and source} & S^a &= \overline{s^a} .
\end{align}
\end{subequations}
Identifying the turbulent fluxes in the right hand sides of~\eqref{eq:1F2S}, elementary rearrangements eventually yield the \emph{fundamental decomposition}
\begin{equation}
\label{eq:Phi1F2S}
\bm{\varPhi}^a = \bm{\varPhi}^{a+}\!+\bm{\varPhi}^{a-}
		+ \alpha^+\alpha^-\frac{\rho^+\rho^-}{\overline{\rho}}
					(A^+\!-A^-)(\bm{U}^+\!-\bm{U}^-) ,
\end{equation}
which has been reported numerous times under different or particular forms for instance by \citet[eq.~19]{Spiegel72}; \citet[eq.~16]{Libby81}; \citet[eq.~1.2--6]{Spalding85}; \citet[eq.~7.5]{Llor05}.

	This relationship shows that the two-structure-field approach separates each single-fluid turbulent flux into three contributions: two per-structure turbulent fluxes of similar nature, and a complementary inter-structure term, designated as `directed flux' in the present and previous works \citep{Llor03, Llor05}---also known as `ordered' \citep[§~I.2.d]{Mangin77}, `sifting' \citep[§~1.2]{Spalding85}, or `ordered convective' \citep[§~3]{Cranfill92}. The directed flux combines the \emph{contrast} between structures $\delta A=A^+\!-\!A^-$ and the \emph{drift velocity} between structures~$\delta\bm{U}$. Its importance is further discussed in section~\ref{ssec:Directed} below.
%
\subsection{Critical importance of directed effects, directed energy, energy paths}
\label{ssec:Directed}
	Beyond the very important but obvious ability to capture strong contrasts of equations of state or constitutive laws between components, the major advantage of two-structure-field approaches lies in their ability to also capture the relative strengths of the directed and per-structure fluxes \emph{even for the degenerate case of mixing between identical fluids}. If the combination of contrast and drift between structures is large enough in~\eqref{eq:Phi1F2S}, the directed flux can be dominant, making most classical diffusive-like closures of the single-fluid flux inconvenient at best, and inconsistent at worst \citep{Llor03, Llor05}. Moreover, a two-structure-field approach provides evolution equations for~$A^\pm$ and~$\bm{U}^\pm$, thus closing the directed flux with higher-order exchange terms, such as drag in the case of momentum. It is thus worth examining the relationship between the two approaches.

	In the presently considered flows the transport of heavy and light structures is the central characteristic quantity. In the single-fluid framework, this is accounted for by the mean transport velocity~$\bm{U}$ and by the turbulent transport fluxes of~$b^\pm$ which can be rewritten in the two-structure-field framework as
\begin{equation}
\label{eq:MassFux}
\overline{b^\pm \rho \bm{u}''} = \overline{b^\pm \rho (\bm{u}-\bm{U})}
	= \alpha^\pm\rho^\pm(\bm{U}^\pm\!-\bm{U})
	= \pm\alpha^+\alpha^-\frac{\rho^+\rho^-}{\overline{\rho}}\delta\bm{U} .
\end{equation}
Therefore, by the very definition of~$\bm{U}^\pm$, these particular turbulent fluxes reduce to their directed contributions, without any per-structure fluxes. Modelling of this flux in the single-fluid approach is then replaced by closures of the~$\bm{U}^\pm$ momentum equations which are severely constrained by their symmetric and conservative character.

	Application of~\eqref{eq:Phi1F2S} to the Reynolds stress tensor---that is the turbulent flux of momentum as defined in~\eqref{eq:app:URP}---yields
\begin{equation}
\label{eq:Rij}
\overline{\rho u''_i u''_j} = R_{ij} = R^+_{ij} + R^-_{ij}
		+ \alpha^+\alpha^-\frac{\rho^+\rho^-}{\overline{\rho}}
						\delta U_i \delta U_j .
\end{equation}
In contrast to the per-structure terms~$R^\pm_{ij}$, the directed term is purely axial (no components perpendicular to~$\delta\bm{U}$) and can thus contribute significantly to the total anisotropy of the Reynolds stress tensor \citep[see section~\ref{ssec:SurrogateDirected} and table~\ref{tab:0DSurrogate}, or][eq.~7.6]{Llor03, Llor05}---notice however, that the reverse is \emph{not} true: a large Reynolds stress anisotropy does not necessarily imply a large directed component.

	The half-traces of the tensors in~\eqref{eq:Rij} are related to the mean energies according to
\begin{subequations}
\label{eq:kdec}
\begin{align}
\overline{\rho} \, K
	&= \alpha^+ \rho^+ K^+ \!+ \alpha^- \rho^- K^-
		+ \alpha^+\alpha^-\frac{\rho^+\rho^-}{\overline{\rho}} \, K_d ,
\\
\thalf \, \overline{\rho} \, \bm{U}^2
	&= \thalf \, \alpha^+ \rho^+ (\bm{U}^+)^2
		+ \thalf \, \alpha^- \rho^- (\bm{U}^-)^2
	- \alpha^+\alpha^-\frac{\rho^+\rho^-}{\overline{\rho}} \, K_d ,
\end{align}
\end{subequations}
where the total turbulent, per-structure turbulent, and directed per-mass energies are given by
\begin{align}
\label{eq:kkk}
K &= \overline{\thalf\rho(\bm{u}'')^2} \,/\, \overline{\rho} ,
&
K^\pm &= \overline{b^\pm \thalf \rho (\bm{u}^\pm)^2}
		\,/\, \overline{b^\pm \rho} ,
&
K_d &= \thalf\,\delta\bm{U}^2 ,
\end{align}
---notice that~$K_d$ is defined here as per-mass instead of per-volume in previous works \citep[eq.~3.13]{Llor03, Llor05}. The statistical description of turbulence is thus very significantly enriched in the two-structure-field framework by the separation of the three energy contributions.

\begin{figure}
\centerline{\includegraphics[width=\textwidth]{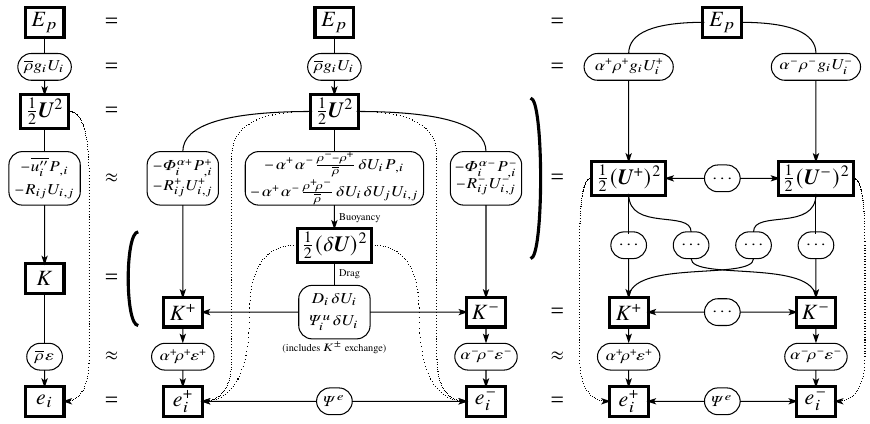}}
\caption{Flow chart of energy reservoirs and dominant transfers (respectively per-mass and per-volume, in respectively square and rounded boxes) from potential to internal energies (respectively~$E_p$ and~$e_i$, compressibility related terms not represented) in the single-fluid and two-structure-field approaches (respectively left and centre--right) according to~\eqref{eq:DKm}, \eqref{eq:DKd}, \eqref{eq:DK}, and~\eqref{eq:DK_pm}. The two-structure-field approach separates the turbulent kinetic energy of the single-fluid approach~$K$ into three contributions, two per-structure kinetic energies~$K^\pm$ and the directed kinetic energy $K_d=\delta\bm{U}^2/2$, see~\eqref{eq:kkk}. The chart on the right is rigorously equivalent to that on the centre but involves the mean kinetic energies of the structures $(\bm{U}^\pm)^2/2$ instead of mean and directed energies $\bm{U}^2/2$ and~$K_d$---expressions for the associated transfer terms, more complex, are not given here. Dotted arrow lines represent transfers controlled by fluid viscosities, here considered negligible at high Reynolds number. The source and sink terms of directed energy are identified with buoyancy and drag terms between structures.}
\label{fig:EnergyPaths}
\end{figure}
\marginpar{Figure~\ref{fig:EnergyPaths}: Reduce figure width by squeezing third column.}%
	Statistical equations for the various kinetic energy reservoirs can also be derived from the Navier--Stokes equations as summarized in appendix~\ref{app:2SF}: single-fluid mean~\eqref{eq:DK} and turbulent~\eqref{eq:DKm}, and two-structure-field directed~\eqref{eq:DKd} and per-structure turbulent~\eqref{eq:DK_pm}. Various energy transfer terms can then be identified, of which the most important are related to the directed energy balance as represented in figure~\ref{fig:EnergyPaths}. The measurement of these and other transfer terms for model calibration is one of the major goals of the present work. It must be noticed in particular that, as visible on the evolution equation of~$K_d$ in~\eqref{eq:DKd} and on the energy paths in figure~\ref{fig:EnergyPaths}: i)~potential energy is only \emph{indirectly} transferred into directed energy through mean kinetic energy, and ii)~at high Reynolds number, only the two per-structure turbulence energies dissipate at small scales into internal energy through~$\varepsilon^\pm$ given in~\eqref{eq:epsilon}. Kolmogorov cascades are thus expected to be embedded in the~$K^\pm$ reservoirs within the structures.

	The approach of \citet{Ramshaw98} mentioned in section~\ref{ssec:EmpiricalArg} considers $\thalf L(L')^2$ to represent an effective `kinetic energy' in the `buoyancy--drag Lagrangian' associated to the `internal momentum' embedded in~\eqref{eq:BD}. Bearing in mind the relationships between~$L'$ and $\delta U_z$~\eqref{eq:LasDrift} and between~$\delta U_z$ and~$K_d$~\eqref{eq:kkk}, the `buoyancy--drag kinetic energy density' can thus be identified with the 0D directed energy~$\langle K_d\rangle$ (up to the factor~$L$). The production and destruction of directed energy are then identified with the respective buoyancy and drag terms as visible in figure~\ref{fig:EnergyPaths}.

	The importance of an energy-balance analysis in RT flows has also been emphasized using different single-fluid energy decomposition procedures, such as the poloidal--toroidal \citep{Cambon13} and available mixing--energy approaches \citep{Winters95}.
%
\subsection{Effective mixing fraction and Atwood number; impact on RT growth coefficient}
\label{ssec:Theta}
	The level of mixing or entrainment is usually quantified by the `molecular mixing' fraction~$\theta$ \citep[eq.~14]{Youngs91} related to the `segregation' intensity $1-\theta$ \citep[eq.~14]{Danckwerts52} and defined as the normalized cross correlation
\begin{equation}
\label{eq:ThetaA}
\theta(t,\bm{r}) = \frac{\overline{v^2v^1}}{\overline{v^2}\,\overline{v^1}} ,
\end{equation}
where $v^m=c^m\rho/\overline{\rho^m}$ is the local volume fraction of fluid~$m$, $\overline{\rho^m}$ being the mean density of fluid~$m$ in the fluid mixture assumed to be ideal ($\overline{\rho^m}=\rho^m$ is thus constant, uniform and fluctuation-free in the incompressible case). Similarly to the separation of directed effects in~\eqref{eq:Phi1F2S}, the molecular mixing fraction can be decomposed into two per-structure molecular segregation intensities and an \emph{effective} or inter-structure mixing fraction according to
\begin{subequations}
\label{eq:Theta}
\begin{align}
\theta &= \tfrac{1}{\alpha^2\,\alpha^1}
	\Big( \overline{b^+\big(v^2\!-\tfrac{\alpha^{2+}}{\alpha^+}\big)
			\big(v^1\!-\tfrac{\alpha^{1+}}{\alpha^+}\big)}
		+ \overline{b^-\big(v^2\!-\tfrac{\alpha^{2-}}{\alpha^-}\big)
			\big(v^1\!-\tfrac{\alpha^{1-}}{\alpha^-}\big)}
		\Big) + \theta_e ,
\\
\label{eq:ThetaE}
\theta_e &= \tfrac{1}{\alpha^2\,\alpha^1} ( \alpha^{2+}\alpha^{1+}/\alpha^+
		+ \alpha^{2-}\alpha^{1-}/\alpha^- ) ,
\end{align}
\end{subequations}
where $\alpha^{m\pm}=\overline{b^\pm v^m}$ is the mean volume fraction of fluid~$m$ of structure~$\pm$ and $\alpha^m=\overline{v^m}$ (thus $\alpha^{2\pm}+\alpha^{1\pm}=\alpha^\pm$ and $\alpha^{m+}\!+\alpha^{m-}=\alpha^m$). $\theta_e$~characterizes the composition contrast between the structure fields as it coincides with~$\theta$ when structures are homogeneously mixed, \ie when $b^\pm(v^m-\alpha^{m\pm}/\alpha^\pm)=0$: in contrast with~$\theta$, its value is thus independent on the miscible or non-miscible character of the fluids (being for instance nearly identical in the two cases of figure~\ref{fig:AlphaDNS}).

	The per-structure segregation intensities are necessarily negative and thus $\theta\leq\theta_e$. As will be observed from the simulations in part~\ref{sec:RTStructures} the respective inter- and intra-structure contributions to the segregation intensities, $1-\theta_e$ and $\theta_e-\theta$, are of similar magnitude, further justifying the relevance of the two-structure-field analysis to describe buoyancy effects.

	Another mixing parameter of relevance in the presence of buoyancy forces is the \emph{effective} or inter-structure Atwood number
\begin{equation}
\Atw_e(t,\bm{r}) = (\rho^+\!-\rho^-)/(\rho^2\!+\rho^1) ,
\end{equation}
where the denominator $\rho^2\!+\rho^1$ acts as a common density to scale all buoyancy effects---in contrast with a $\rho^+\!-\rho^-$ denominator which would merely provide a \emph{local} scaling. As $\alpha^\pm\rho^\pm = \alpha^{2\pm}\rho^2 + \alpha^{1\pm}\rho^1$, the effective-to-apparent Atwood-number ratio can be rewritten as
\begin{equation}
\label{eq:AtE}
\frac{\Atw_e}{\Atw}
	= \frac{\rho^+\!-\rho^-}{\rho^2\!-\rho^1}
	= 1 - \frac{\alpha^{1+}}{\alpha^+} - \frac{\alpha^{2-}}{\alpha^-} .
\end{equation}
The simplicity of this formula comes from the definition of~$\Atw_e$ and would be lost with the \emph{local} Atwood number $(\rho^+\!-\rho^-)/(\rho^+\!+\rho^-)$.

	The resemblance of terms in~\eqref{eq:ThetaE} and~\eqref{eq:AtE} lets derive exact algebraic relationships in two limiting cases: i)~at the TMZ edges (here at $z\rightarrow L/2$ where $\alpha^+\rightarrow1$) and ii)~at the TMZ centre for vanishing Atwood number (here at $z=0$ or $\alpha^\pm=\thalf$), where it is respectively found
\begin{subequations}
\label{eq:AtTheta}
\begin{align}
\label{eq:AtThetaE}
& \text{at } z \rightarrow L/2 ,
	\text{ then } \alpha^2 \rightarrow 1 ,~
	\alpha^{1+}/\alpha^{1-} \rightarrow 0 ,
& \theta_e &\rightarrow 1 - \Atw_e/\Atw ,
\\
\label{eq:AtThetaC}
& \text{at } z = 0
	\text{ with } \Atw \rightarrow 0 ,
	\text{ then } \alpha^2 = \alpha^1 = \thalf ,~ \alpha^{1+} = \alpha^{2-} ,
& \theta_e &= 1 - (\Atw_e/\Atw)^2 .
\end{align}
\end{subequations}
It must be stressed that the first expression---which also applies at the other TMZ edge through the $+\leftrightarrow-$ and $2\leftrightarrow1$ permutation---holds because the inflowing structure~$+$ being laminar, it cannot entrain any of the light fluid~$1$ carried by the outflowing turbulent structure~$-$: hence $\alpha^{1+}/\alpha^{1-}\rightarrow0$.

	In the case of turbulent RT flows, the effective Atwood number controls the driving buoyancy force. It was thus conjectured that coefficient~$\alpha_b$---which defines the growth of the self-similar bubbles as $L_b=\alpha_b\Atw gt^2$---should be related to~$\theta$. More accurately it is related to the effective~$\theta_e$ (to correct the dependence to fluid miscibility) similarly to~\eqref{eq:AtTheta} as large-scale buoyancy is proportional to $\Atw_e$. The single-fluid analysis of \citet[eq.~18]{Ramaprabhu04} at the TMZ centre hinted at $\alpha_b^2 \propto 1 - \theta$, consistently with~\eqref{eq:AtThetaC}. However, simulations and theoretical analysis of \citet[eq.~53 \&~fig.~9]{Grea13}; \citet[fig.~4b]{Youngs13}; \citet[fig.~5a]{Soulard16} later contradicted this relationship and hinted instead at $\alpha_b \propto (1 - \theta)^2$. This discrepancy can be traced to the fact that changes in turbulent intensity simultaneously impact both~$\theta$ (buoyancy) and the dissipation processes (drag): the $\alpha_b$--$\theta_e$ connection is thus indirect as can already be expected from the coexistence in the same TMZ of two significantly different $\Atw_e$--$\theta_e$ relationships~\eqref{eq:AtTheta}. In any case, all the $\alpha_b$--$\theta$ relationships reported previously are actually proxies of more natural $\alpha_b$--$\theta_e$ relationships embedded in~\eqref{eq:AtTheta}.
%
\subsection{Bulk `0D' averages of energies and mixing in planar TMZs}
\label{ssec:Bulk}
	For many purposes \citep{Llor03} it is useful to have global estimates of the various energy related quantities in a plane TMZ (\ie with two homogeneous dimensions). These are conveniently obtained from the ensemble averages by a (1D) \emph{averaging over the axis} perpendicular to the TMZ: the ensuing quantities will be here designated as `bulk' or `0D' averages. The most relevant of these are the mean per mass
\begin{subequations}
\label{eq:0DEB}
\begin{align}
\label{eq:0DEp}
&\text{change in potential energy} &
\langle E_p\rangle(t)
	&= \frac{1}{\langle\rho\rangle L} \int_{-\infty}^{\infty} \!\!
		\!\! \big(\, \overline{\rho}(t,z)-\overline{\rho}(0,z) \,\big)
			\, g_z z \dd z ,
\\
\label{eq:0DU2}
&\text{mean kinetic energy} &
\langle \thalf \bm{U}^2 \rangle(t)
	&= \frac{1}{\langle\rho\rangle L} \int_{-\infty}^{\infty}
		\!\! \thalf \overline{\rho} \bm{U}^2 \dd z ,
\\
\label{eq:0DKd}
&\text{directed energy} &
\langle K_d\rangle(t)
	&= \frac{1}{\langle\rho\rangle L} \int_{-\infty}^{\infty}
		\!\! \alpha^+\alpha^-\frac{\rho^+\rho^-}{\overline{\rho}} K_d \dd z ,
\\
\label{eq:0DKpar}
&\text{longitudinal turb.\ energies} &
\langle K_z^\pm\rangle(t)
	&= \frac{1}{\langle\rho\rangle L} \int_{-\infty}^{\infty}
		\!\! \thalf \overline{b^\pm \rho (u_z-U_z^\pm)^2} \dd z ,
\\
\label{eq:0DKperp}
&\text{transverse turb.\ energies} &
\langle K_{xy}^\pm\rangle(t)
	&= \frac{1}{\langle\rho\rangle L} \int_{-\infty}^{\infty}
		\thalf \overline{b^\pm \rho (u_x^2+u_y^2)} \dd z ,
\\
\label{eq:0DKpm}
&\text{turbulent energies} &
\langle K^\pm\rangle(t)
	&= \langle K_z^\pm\rangle + \langle K_{xy}^\pm\rangle ,
\\[\halflineskip]
\label{eq:0DKz}
&\text{total longit.\ turb.\ energy} &
\langle K_z\rangle(t)
	&= \langle K_z^+\rangle + \langle K_z^-\rangle + \langle K_d\rangle ,
\\[\halflineskip]
\label{eq:0DKxy}
&\text{total transv.\ turb.\ energy} &
\langle K_{xy}\rangle(t)
	&= \langle K_{xy}^+\rangle + \langle K_{xy}^-\rangle ,
\\[\halflineskip]
\label{eq:0DK}
&\text{total turbulent energy} &
\langle K\rangle(t)
	&= \langle K_z\rangle + \langle K_{xy}\rangle ,
\end{align}
\end{subequations}
where the mean mixing width and mass of the TMZ \citep[eq.~3]{Andrews90}---extending the `momentum width' first introduced for shear layers by von Kármán---are here taken as
\begin{subequations}
\label{eq:LrhoL}
\begin{align}
\label{eq:L}
L(t)
	&= C_L \int_{-\infty}^{\infty}
		\!\! \alpha^2 \alpha^1 \dd z ,
\\
\langle\rho\rangle(t) L(t)
	& = C_L \int_{-\infty}^{\infty} \!\! \overline{\rho}(t,z)
		\, \alpha^2 \alpha^1 \dd z ,
\end{align}
\end{subequations}
$\alpha^m$ introduced after~\eqref{eq:Theta} being the ensemble averaged volume fractions of fluids $m=1$ and~2---not of structures~$\pm$. Because of homogeneity along transverse directions, all ensemble averages such as~$\alpha^m$ are functions of height~$z$ only and coincide with averages over the $xy$ planes for asymptotically wide domains. The domain under consideration can extend to $z\rightarrow\pm\infty$ and the~$z$ origin is arbitrary. For vanishing~$\Atw$, the Boussinesq limit also provides $\langle\rho\rangle = \thalf(\rho^2+\rho^1)$.

	Coefficient~$C_L$ corrects~$L$ as given by~\eqref{eq:L} so as to match the actual width of the TMZ as defined by the zero and unit values on the fluids' volume-fraction profiles: at $C_L=6$ or 70/9 for instance, $L$~is the exact width of the TMZ if~$\alpha^m$ display respectively linear or cubic profiles. Since the observed profiles in the present simulations are close to cubic (see part~\ref{sec:RTStructures} and figure~\ref{fig:1Dprofiles}a) $C_L=70/9$ is retained in all the following.
\marginpar{Check section and figure labels.}

	In principle, consistency with the two-structure-field quantities in~\eqref{eq:0DKd} to~\eqref{eq:0DKpm} could have led to defining $L\propto\int\alpha^+\alpha^- \dd z$. However, such a definition would have been sensitive to the different structure-field-segmentation methods, with correspondingly varying coefficients $C_L$, and would not facilitate comparisons with the large body of well documented measurements of~$L$.

	From~\eqref{eq:0DEB}, the 0D energy budget in the TMZ also provides the missing contributions
\begin{subequations}
\begin{align}
\label{eq:0DEi}
&\text{internal energy} &
\langle E_i\rangle(t)
	&= - \big( \langle E_p\rangle
		+ \langle \thalf \bm{U}^2 \rangle + \langle K\rangle \big) ,
\\
\label{eq:0DDissipation}
&\text{kinetic-energy dissipation} &
\langle\varepsilon\rangle(t)
	&= \frac{1}{\langle\rho\rangle L} \; \frac{\dd}{\dd t}
		\big( \langle\rho\rangle L \langle E_i\rangle \big) ,
\end{align}
\end{subequations}
---notice that $\langle E_p\rangle<0$ from its definition in~\eqref{eq:0DEp}. The major advantage of these formulæ is that they do not require the knowledge of the \emph{local} dissipation rate which is seldom accessible in experiments or can be strongly affected by numerical and sub-grid scale dissipation in DNS and LES. For the same reason, the separation of the~$E_i^\pm$ and~$\varepsilon^\pm$ per-structure contributions was not attempted---but in the present work anyhow, their importance on two-structure-field segmentation and averaging at large scales is marginal.

	Bulk apparent and effective mixing fractions are also obtained by extending their point values~\eqref{eq:ThetaA} and~\eqref{eq:Theta} according to
\begin{subequations}
\begin{align}
\langle\theta\rangle(t)
	&= \frac{C_L}{L} \int_{-\infty}^{\infty} \!\! \overline{v^2v^1} \dd z ,
\\
\langle\theta_e\rangle(t)
	&= \frac{C_L}{L} \int_{-\infty}^{\infty} \!\!
		( \alpha^{2+}\alpha^{1+}/\alpha^+ \!+ \alpha^{2-}\alpha^{1-}/\alpha^- )
			\dd z ,
\end{align}
\end{subequations}
as previously given by \eg~\citet[eq.~21]{Youngs91, Dimonte04}.
%
\subsection{Expected profiles of structure volume fractions in a TMZ}
\label{ssec:Profiles}
	Although two-structure fields in the sense discussed here have not been produced in experimental or simulated flows so far, it is desirable to estimate the general profiles of the average volume fractions of two-structure fields and fluids in a TMZ, $\alpha^\pm(z)$ and~$\alpha^m(z)$, from known and expected properties of turbulent RT flows \citep{Llor04}.

	For this purpose three assumptions are required:
\begin{itemize}
	\item linear profiles of average fluid volume fractions~$\alpha^m$, as approximately but consistently observed by \citet[fig.~3]{Anuchina78}; \citet[figs~3 \&~7]{Andrews90}; \citet[fig.~11]{Kucherenko91}; \citet[fig.~9]{Youngs91}; \citet[figs~8 \&~15]{Linden94}; \citet[fig.~5]{Youngs94}; \citet[figs~6 \&~7]{Kucherenko97}; \citet[fig.~2]{Schneider98}; \citet[fig.~5]{Kucherenko00}; \citet[fig.~11]{Cook01}; \citet[fig.~2]{Wilson02}; \citet[fig.~17]{Cook04}; \citet[fig.~7]{Dimonte04}; \citet[fig.~10]{Banerjee06}; \citet[fig.~6]{Livescu09}; \citet[fig.~5]{Mueschke09}; \citet[fig.~7b]{Livescu10}; \citet[fig.~9]{Olson09}; \citet[fig.~8]{Banerjee10}; \citet[fig.~2]{Boffetta10}; \citet[fig.~7]{Livescu10}; \citet[fig.~3]{Schilling10}; \citet[fig.~3a]{Livescu13}; \citet[fig.~6]{Youngs13};
	\item uniform effective mixing fraction~$\theta_e$ in~\eqref{eq:ThetaE}, paralleling the uniform apparent molecular mixing fraction~$\theta$ in~\eqref{eq:ThetaA} as approximately but consistently observed by \citet[fig.~9]{Youngs91}; \citet[figs~5 \&~15]{Linden94}; \citet[fig.~5]{Youngs94}; \citet[fig.~4]{Wilson02}; \citet[fig.~26]{Dimonte04}; \citet[fig.~7b]{Livescu10}; \citet[fig.~3a]{Livescu13}; \citet[fig.~6]{Youngs13};
	\item and monotonic profiles of per-structure densities~$\rho^\pm$, with continuous first derivatives.
\end{itemize}

	While the first two are good approximations, well supported by experimental and simulation results, the last is just a weak constraint inferred from the qualitative observations and phenomenological analysis provided in section~\ref{ssec:Visual} and illustrated in figure~\ref{fig:AlphaDNS}---it is also equivalent to the condition $\alpha^{1+}/\alpha^{1-}\rightarrow0$ when $\alpha^+\rightarrow1$ in~\eqref{eq:AtThetaE}. Remarkably, these assumptions do not require prior specific knowledge about the phenomenon which drives mixing---gravity, shear, or other---and the ensuing results significantly constrain both the possible outputs from two-structure-field segmentation schemes, and the transport and exchange coefficients in models \citep{Llor04}.

	From now on in this section, an incompressible self-similar flow at vanishing Atwood number will be considered---results will thus apply to an RT flow but not exclusively. All fields become functions of the reduced coordinate $\zeta=z/L\in[-\thalf,+\thalf]$, and because of the Boussinesq approximation at $\Atw\rightarrow0$, the structure density profiles are symmetric and can be scaled to the fluid densities through a single function~$r(\zeta)$
\begin{align}
\rho^-(\zeta) = \rho^1 + r(\zeta) \times (\rho^2-\rho^1) ,
&&
\rho^+(\zeta) = \rho^2 - r(-\zeta) \times (\rho^2-\rho^1) .
\end{align}
The fluid-of-structure mean volume fractions $\alpha^{m\pm}(\zeta)$---as defined after~\eqref{eq:ThetaE}---are thus constrained by the experimentally observed linear profiles of fluid volume fractions, and by the above profiles of structure densities according to
\begin{subequations}
\label{eq:AlphaCond}
\begin{align}
\alpha^{2+}\!+\alpha^{2-}
	&= \thalf+\zeta ,
&
\alpha^{2-}\big/(\alpha^{2-}\!+\alpha^{1-})
	&= r(\zeta) ,
\\
\alpha^{1+}\!+\alpha^{1-}
	&= \thalf-\zeta ,
&
\alpha^{1+}\big/(\alpha^{2+}\!+\alpha^{1+})
	&= r(-\zeta) .
\end{align}
\end{subequations}
This closed linear system yields explicit expressions for $\alpha^{m\pm}(\zeta)$ as functions of~$r(\zeta)$, from which $\theta_e(\zeta)$ and $\Atw_e(\zeta)/\Atw$ can be obtained according to~\eqref{eq:ThetaE} and~\eqref{eq:AtE}.

	To fully define the $\alpha^{m\pm}(\zeta)$ profiles, it is now necessary to provide a reasonable function~$r(\zeta)$. Visual inspection of figure~\ref{fig:AlphaDNS} and general understanding of the TMZ leads to two basic constraints: i)~$r(\zeta)$ is monotonic because structures entrain more and more of the opposite fluid as they advance across the TMZ, and ii)~both $r(\zeta)$ and its slope cancel at $\zeta=-\thalf$ because structures are initially made of \emph{pure non-turbulent} fluid which is therefore unable to entrain the opposite fluid. The \emph{simplest} rational function following these constraints must have a \emph{double zero} at $\zeta=-\thalf$ and \emph{one adjustable pole} at $\zeta<-\thalf$ as given by
\begin{equation}
\label{eq:RhoProf}
r(\zeta) = \frac{2r_0\,(\thalf+\zeta)^2}{\thalf+\varkappa\zeta} .
\end{equation}
This form depends on the two adjustable parameters~$\varkappa$ and~$r_0$ such that $0\leq\varkappa<1$ and $0\leq r_0<(1+\varkappa)/4$ in order to have $0\leq r(\zeta)\leq1$---a third order polynomial for~$r(\zeta)$ is not advisable as it can produce large unphysical oscillations for acceptable values of the parameters. Elementary algebraic transformations of~\eqref{eq:AlphaCond} with~\eqref{eq:RhoProf} yield explicit expressions for $\alpha^{m\pm}(\zeta)$ and $\theta_e(\zeta)$ as rational fractions of degree four in~$\zeta$---these are only given below in~\eqref{eq:Prof} for a special value of~$r_0$.

\begin{figure}
\centerline{\includegraphics{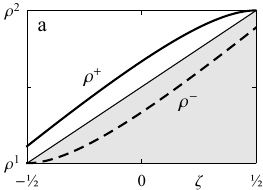}\includegraphics{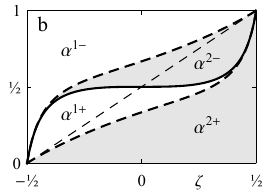}\includegraphics{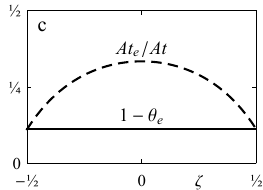}}
\caption{Expected profiles as functions of reduced height $\zeta=z/L$ for: (a)~structure densities~$\rho^+$ and~$\rho^-$~\eqref{eq:RhoProf}, in respectively plain and dashed lines, thin line being total mean density~$\overline{\rho}$; (b)~stacked volume fractions of structures~$\alpha^{m\pm}$ in~\eqref{eq:Prof}, plain line representing $\alpha^+=\alpha^{2+}\!+\alpha^{1+}$ with gray zones for~$\alpha^{2\pm}$; and (c)~effective or inter-structure segregation fraction $1-\theta_e$ and effective Atwood number reduction factor $\Atw_e/\Atw$ in~\eqref{eq:Prof}, in respectively plain and dashed lines. Shaded regions represent the `heavy' fluid~2 spread in~(b) over the `heavy' or `light' structures~$\pm$. All profiles here assume $\varkappa=\thalf$ in~\eqref{eq:RhoProf} and~\eqref{eq:Prof}.}
\label{fig:AlphaProfiles}
\end{figure}
	Selection of~$r_0$ and~$\varkappa$ values must now be performed so as to match some desired observations, mainly on an effective mixing fraction level $\theta_e(\zeta)$ estimated from a measured apparent mixing level $\theta(\zeta)$. \citet[p.~5]{Llor04} assumed $r_0=\varkappa$ which reduced to three the degree of $\alpha^{m\pm}(\zeta)$, but yielded a non uniform $\theta_e(\zeta)$ whose average value across the layer was adjusted to the observed mean~$\langle\theta\rangle$. However, simulated and experimental profiles of $\theta(\zeta)$ in RT flows are linear at moderate~$\Atw$ \citep{Dimonte04} and become uniform at vanishing~$\Atw$ \citep[and refs therein]{Youngs91, Mueschke09}. Assuming that $\theta_e(\zeta)$ should parallel $\theta(\zeta)$ and anticipating results below, it is here chosen $r_0=\varkappa/(1+\varkappa)$ which makes~$\theta_e$ uniform and eventually yields
\begin{subequations}
\label{eq:Prof}
\begin{align}
\alpha^{2+}(\zeta)
	&= \thalf+\zeta - \alpha^{2-}(\zeta) ,
&
\alpha^{2-}(\zeta)
	&= \frac{2\varkappa\,(\thalf+\varkappa\zeta)\,(\thalf+\zeta)\,(\squart-\zeta^2)}
		{(1+\varkappa)\,\big(\squart-\varkappa(2-\varkappa)\zeta^2\big)} ,
\\
\alpha^{1-}(\zeta)
	&= \thalf-\zeta - \alpha^{1+}(\zeta),
&
\alpha^{1+}(\zeta)
	&= \frac{2\varkappa\,(\thalf-\varkappa\zeta)\,(\thalf-\zeta)\,(\squart-\zeta^2)}
		{(1+\varkappa)\,\big(\squart-\varkappa(2-\varkappa)\zeta^2\big)} ,
\\
\theta_e(\zeta)
	&= \frac{4\varkappa}{(1+\varkappa)^2} ,
&
\frac{\Atw_e(\zeta)}{\Atw}
	&= \frac{(1-\varkappa)\,\big(\squart-\varkappa(2-\varkappa)\zeta^2\big)}
		{(1+\varkappa)\,(\squart-\varkappa^2\zeta^2)} .
\end{align}
\end{subequations}
The parameter~$\varkappa$ can then be adjusted to retrieve the value of the effective mixing fraction or the effective-to-apparent Atwood number ratio. As expected, these forms comply with the general relationship~\eqref{eq:AtTheta} linking~$\theta_e$ and~$\Atw_e$ at edges and centre of the TMZ. The case $\varkappa=\thalf$, chosen so that $\dd\alpha^\pm/\dd\zeta=0$ at $\zeta=0$ and yielding $\theta_e=\ensuremath{{\textstyle\frac{8}{9}}}\approx0.89$, is typical of expected profiles as illustrated in figure~\ref{fig:AlphaProfiles}. This specific value appears in section~\ref{ssec:Mixing} in light of results from experiments and DNS--LES of RT flows.
%
\section{Quantitative relevance of two-structure-field approaches in RT flows}
\label{sec:QuantitativeArg}
%
\subsection{Surrogate two-fluid approach to estimate two-structure-field quantities}
\label{ssec:Surrogate}
	The arguments given in section~\ref{ssec:EmpiricalArg} in favour of two-structure-field approaches in RT modelling must now be quantitatively substantiated in terms of the directed effects presented in section~\ref{ssec:Directed}. Although two-structure-field data on RT flows are unavailable at the moment, the magnitude of directed effects can still be \emph{estimated} from existing experimental or simulated measurements.

	Most of the measurements available to date monitor fluid fields---sometimes with correlations---and the simplest approach therefore consists in substituting two-structure-field approaches by the \emph{two-fluid} conditional analysis. This amounts to applying all the formulæ of section~\ref{ssec:TSCAE} with structure fields replaced by fluid mass fractions
\begin{align}
b^+(t,\bm{r}) \rightarrow c^2(t,\bm{r}) ,
	&& b^-(t,\bm{r}) \rightarrow c^1(t,\bm{r}) .
\end{align}
All the ensuing concepts explored in part~\ref{sec:Theory} still hold even if~$b^\pm$ can possibly take intermediate values between 0 and 1. Instead of `well delineated', the corresponding `structures' are actually `fractal like' for immiscible fluids and `smoothed' for miscible fluids. This approximation, here designated as the `surrogate two-fluid' approach, was extensively analysed elsewhere \citep{Llor05}. The analysis of \citet{Soulard16} appears very similar to this approach although it was formally carried out on single-fluid concentration fluxes.

	The surrogate two-fluid approach is a relevant approximation because relationship~\eqref{eq:LasDrift} still holds and relates the growth of the TMZ and the drift velocity~$\delta\bm{U}$---the critical directed parameter introduced in section~\ref{ssec:LasDrift}, appearing in~\eqref{eq:Phi1F2S}, and discussed in section~\ref{ssec:Directed}. As discussed below in section~\ref{ssec:SurrogateDirected}, the simple measurement of~$L(t)$ thus provides valuable information on the strength of directed effects.

	The surrogate two-fluid approach, however crude it may appear and despite the distortions it introduces, can also be applied to any type of mixing layer. \citet[§~4]{Llor05} provided estimates of the main bulk 0D structure-characteristic features defined in section~\ref{ssec:Bulk} for RT as well as Richtmyer--Meshkov (RM) and Kelvin--Helmholtz (KH)---respectively free decaying and shear turbulent layers---which then revealed significant differences in their energy balance and turbulence structure. This retrospectively justified or eventually motivated the development of original two-structure-field models \citep{Spalding85, Youngs84, Youngs89, Cranfill92, Youngs94, LlorBailly03, LlorPoujade04, Kokkinakis15, Kokkinakis20} where directed and standard turbulent effects are closed separately from very different assumptions.

	Now, these early estimates were based on the then available experimental and simulated results, sometimes inconsistent, patchy, noisy, or poorly resolved. Updated reinterpretations are thus provided in table~\ref{tab:0DSurrogate} for RT, RM, and KH, each based on a single recent high-resolution simulation in order to ensure self-consistency.
\begin{table}
\centerline{\begin{tabular}{ll*{2}{D{.}{.}{1.3}}*{6}{D{.}{.}{1.2}}}
Flow & Source
	& \multicolumn{1}{c}{$\tfrac{\langle K_d\rangle}{\langle K\rangle}$}
	& \multicolumn{1}{c}{$\tfrac{\langle K_{xy}\rangle}{\langle K\rangle}$}
	& \multicolumn{1}{c}{$\tau$}
	& \multicolumn{1}{c}{$\tau_{xy}$}
	& \multicolumn{1}{c}{$\kappa$}
	& \multicolumn{1}{c}{$\kappa_{xy}$}
	& \multicolumn{1}{c}{$\sigma$}
	& \multicolumn{1}{c}{$\sigma_{xy}$}
\\*\hline\rule{0pt}{10pt}%
Rayleigh--Taylor & \citet[$\epsilon=0$]{Soulard16} & 0.085 & 0.56 & 0.39 & 0.22 & 0.19 & 0.08 & 0.10 & 0.03
\\*
Richtmyer--Meshkov & \citet{Thornber17} & 0.005 & 0.58 & 0.16 & 0.14 & 0.34 & 0.28 & 0.73 & 0.56
\\*
Kelvin--Helmholtz & \citet{Takamure18} & 0.015 & 0.78 & 0.35 & 0.41 & 0.42 & 0.53 & 0.49 & 0.68
\\*\hline
\end{tabular}}
\caption{Main 0D ratios characterizing the RT, RM, and KH self-similar TMZ as estimated from the surrogate two-fluid approach \citep[adapted and corrected from][tab.~4.1]{Llor05}. Normalized integral time, length, and viscosity scales, $\tau$, $\kappa$, and~$\sigma$, are defined in~\eqref{eq:tks} and~\eqref{eq:tksxy}. Notice the differences in these quantities for RT compared to RM and KH.}
\label{tab:0DSurrogate}
\end{table}\marginpar{Check numbers from \citet{Thornber17} with O.~Soulard! \citet{Youngs94}?}
Their tedious but straightforward derivation from the references will not be detailed here but the ensuing values are discussed in the following sections.
%
\subsection{Higher relative strength of directed energy in RT flows}
\label{ssec:SurrogateDirected}
	From early experimental works where only mean mass fraction profiles of fluids~$\overline{c^m}$ are considered, the mean fluid velocities can be reconstructed from the 1D mass conservation equations as \citep[§~4 and refs therein]{Llor05}
\begin{equation}
U^m_z = \frac{\overline{c^m u_z}}{\overline{c^m}}
	= \frac{-1}{\overline{c^m}} \int \partial_t \overline{c^m} \dd z .
\end{equation}
In this surrogate two-fluid approach, the drift velocity is found as $\delta U_z^\text{2Fluid} = U_z^2-U_z^1 \approx - L'/2$---consistently with~\eqref{eq:LasDrift},---the relationship being exact and uniform for vanishing Atwood number and linear profiles of~$\overline{c^m}$ \citep[eq.~4.5]{Llor05}. It is then found $\langle K_d\rangle^\text{2Fluid}=(L')^2/48$ regardless of the type of mixing layer. Now, because of the counter-flows of fluids by structure entrainment, this two-fluid value is actually a \emph{lower bound} of any two-structure-field value, \ie $\langle K_d\rangle^\text{2Fluid} \lesssim \langle K_d\rangle^\text{2Struct.}$. In the rest of this part~\ref{sec:QuantitativeArg}, all quantities will be considered as obtained from the surrogate two-fluid approach and label `2Fluid' will be omitted.

	The magnitude of the directed energy~$\langle K_d\rangle$ is to be estimated relative to the turbulent energy~$\langle K\rangle$ in~\eqref{eq:0DK}. Because published values of growth rate coefficients can vary by factors of up to two, both~$\langle K_d\rangle$ and~$\langle K\rangle$ must be obtained from the very \emph{same} experiment or simulation to ensure their mutual consistency. Results, gathered in table~\ref{tab:0DSurrogate} from three recently published works, show that the ratio $\langle K_d\rangle/\langle K\rangle$ in RT is almost six fold above its value in more common RM or KH flows. As expected, the strength of the directed energy also appears on the transfer terms of figure~\ref{fig:EnergyPaths}: in RT flows the otherwise dominant shear-drive terms in~$U_{i,j}$ and~$U^\pm_{i,j}$ are negligible compared to the pressure-drive terms in~$\bnabla P$ \citep[§~5.2.2]{Llor05}.

	In an early estimate \citep[tab.~4.1]{Llor05}, the $\langle K_d\rangle/\langle K\rangle$ ratio for RT appeared about three-fold above its present value. This was due to a combination of different effects, prominently the impact of typically two-fold higher~$\alpha$ values observed in experiments \citep[fig.~1]{Dimonte04}: the directed kinetic energy is then quadrupled. In contrast, the reference source for RT in table~\ref{tab:0DSurrogate} is a well controlled simulation with annular spectrum initialization yielding a smaller~$\alpha$ value.
%
\subsection{Non-standard integral turbulent scales in RT flows}
\label{ssec:NonStandard}
	The structure of turbulence in the TMZ is characterized by the large scale properties, related to the dissipation~$\langle \varepsilon \rangle$. In the self-similar limit, \eqref{eq:0DDissipation} is simplified by scaling all energies with respect to the energy input, thus yielding constant energy ratios. For vanishing Atwood number (hence negligible mean kinetic energy in RT)
\begin{subequations}
\begin{align}
\label{eq:SSDissipation}
\langle \varepsilon \rangle
	&= - \frac{\dd}{L\,\dd t}
		\big( L\langle E_p\rangle + L\langle K\rangle \big)
	= - \Big( 1 + \frac{\langle K\rangle}{\langle E_p\rangle} \Big)
		\frac{\dd \big( L\langle E_p\rangle \big)}{L\,\dd t}
\nonumber\\*
	&= - \big( \langle E_p\rangle + \langle K\rangle \big)
		\; \frac{\partial (L\langle E_p\rangle)}{\langle E_p\rangle\partial L}
			\; \frac{L'}{L} ,
\end{align}
where
\begin{align}
\label{eq:SSDissipationCoeff}
\frac{\partial (L\langle E_p\rangle)}{\langle E_p\rangle\partial L}
	= \left|\;\begin{aligned}
		& 2 ,
	\\*
		& 3-2/\theta_\text{RM} ,
	\\*
		& 1 ,
	\end{aligned}\right.
&&
\frac{L't}{L}
	= \left|\;\begin{aligned}
		& 2 && \text{Rayleigh--Taylor,}
	\\*
		& \theta_\text{RM} , && \text{Richtmyer--Meshkov,}
	\\*
		& 1 , && \text{Kelvin--Helmholtz.}
	\end{aligned}\right.
\end{align}
\end{subequations}
For the RT case this coefficient is readily obtained as $\langle E_p\rangle\propto L$. For the RM case, $\langle E_p\rangle=0$ but scaling by $\langle K\rangle\propto(L')^2$ instead of~$\langle E_p\rangle$ in~\eqref{eq:SSDissipation} eventually recovers~\eqref{eq:SSDissipationCoeff}---$\theta_\text{RM}\approx0.219$ being the self-similar growth exponent \citep{Thornber17}. For the KH case, \eqref{eq:SSDissipation} also holds, $\langle E_p\rangle\propto L^0$ being then the mean kinetic energy loss \citep[eq.~4.12]{Llor05}.

	In the spirit of the general analysis of turbulent scales in self-similar flows \citep[\eg][§~5.3, fig.~5.17 \&~5.18 for round jets]{Pope00},\marginpar{Cite also Tenekes and Lumley?} the following non-dimensional integral scales of time, length, and viscosity can be expressed from~\eqref{eq:SSDissipation}
\begin{subequations}
\label{eq:tks}
\begin{align}
\tau &= \frac{\langle K \rangle/\langle \varepsilon \rangle}{L/L'}
	= - \big( \langle E_p\rangle/\langle K\rangle + 1 \big)^{-1}
		\big( \tfrac{\partial (L\langle E_p\rangle)}
			{\langle E_p\rangle\partial L} \big)^{-1} ,
\\
\kappa &= \frac{\langle K \rangle^{3/2}/\langle \varepsilon \rangle}{L}
	= - \big( \langle E_p\rangle/\langle K\rangle + 1 \big)^{-1}
		\big( \tfrac{\partial (L\langle E_p\rangle)}
			{\langle E_p\rangle\partial L} \big)^{-1}
			\sqrt{\frac{\langle K \rangle}{48\langle K_d \rangle}} ,
\\
\sigma &= \frac{\langle K \rangle^2/\langle \varepsilon \rangle}{LL'}
	= - \big( \langle E_p\rangle/\langle K\rangle + 1 \big)^{-1}
		\big( \tfrac{\partial (L\langle E_p\rangle)}
			{\langle E_p\rangle\partial L} \big)^{-1}
			\frac{\langle K \rangle}{48\langle K_d \rangle} .
\end{align}
\end{subequations}
$\tau$, $\kappa$, and~$\sigma$ can be respectively identified with effective Stokes, Knudsen, and inverse Reynolds numbers of the turbulent transport in the TMZ---following \citet{Llor03, Llor05} $\kappa$~is also designated as a the `von Kármán number', not to be confused with the von Kármán \emph{constant}. Now, in order to correct the impact of the high anisotropy of RT-like flows, it also appears useful examine these quantities reconstructed from what would be the smallest expected isotropic part of the turbulence energy as given by $\tfrac{3}{2}\langle K_{xy}\rangle$
\begin{align}
\label{eq:tksxy}
\tau_{xy} = \tfrac{3\langle K_{xy}\rangle}{2\langle K\rangle} \tau ,
&&
\kappa_{xy} = \Big( \tfrac{3\langle K_{xy}\rangle}{2\langle K\rangle}
	\Big)^{3/2} \kappa ,
&&
\sigma_{xy} = \Big( \tfrac{3\langle K_{xy}\rangle}{2\langle K\rangle}
	\Big)^2 \sigma .
\end{align}

	Application of~\eqref{eq:tks} and \eqref{eq:tksxy} to up-to-date RT, RM, and KH published data yields the 0D estimates of the dimensionless turbulent scales in table~\ref{tab:0DSurrogate}. Significant differences appear between RM and KH on one side and RT on the other side: although the Stokes numbers are very similar, the von Kármán and inverse Reynolds numbers appear lower for RT, by up to a factor 5 for~$\sigma$ and even 20 for~$\sigma_{xy}$. The values of~$\tau$ around $\ensuremath{{\textstyle\frac{1}{3}}}$ confirm the significant persistence of the turbulent field within the TMZ and thus support the time-filtering approach discussed in section~\ref{ssec:Memory}.

	In an early estimate \citep[tab.~4.1]{Llor05}, $\kappa$~for RT appeared about three-fold below its present value, for similar reasons as for the overestimated $\langle K_d\rangle/\langle K\rangle$ ratio in section~\ref{ssec:SurrogateDirected}. \citet[fig.~2a]{Livescu09}; \citet[figs~13 \&~14]{Zhou20} later found higher values comparable with the present estimate. Yet, the impact analysis of low~$\kappa$ still holds as initially provided---especially for its consequences on modelling---and it was later paralleled by the implicit findings of low~$\sigma$ by \citet[p.~8]{Llor04}; \citet[§~4.2]{Schilling06}; \citet[§~4.4.1]{Livescu09}; \citet[eqs~11 \&~12]{Ristorcelli10}; \citet{Schilling09, Schilling17}. The contrast appears even more striking on~$\kappa_{xy}$ and~$\sigma_{xy}$.

	The value of the reduced large-eddy length scale~$\kappa_{xy}$ is indeed consistent with experimentally and numerically estimated sizes of bubbly structures at the edges of RT TMZs as reported or visible in \citet[fig.~2]{Linden94}; \citet[fig.~1]{Youngs94}; \citet[fig.~4]{Schneider98}; \citet[figs~9 to~10]{Dalziel99}; \citet[fig.~21]{Dimonte04}; \citet[fig.~12]{Burton11}. It is also consistent with the value of the Ozmidov length scale which separates the contributions to the turbulent energy spectrum \citep[§~3.3 \& fig.~6]{Griffond23}: inertia-dominated below and buoyancy-dominated (or directed) above.

	The values of~$\sigma$ or~$\sigma_{xy}$ provide a simple test for the Turbulent-Viscosity Hypothesis (TVH, \eg \citet[§~4.4]{Pope00}; \citet{Schmitt07} and refs therein). Simplifying the analysis of \citet{Ristorcelli10} and similarly to those of \citet[§~7.1]{Cook04}; \citet[§~5]{Epstein18}, the relative volume flux across the centre of a symmetric linear TMZ is trivially given by $L'/8$ and its closure would be $\approx C\ensuremath{{\textstyle\frac{3}{2}}}\sigma L'$ assuming a bell-shaped turbulent viscosity $\approx Ck^2/\varepsilon$: TVH is thus acceptable if $12C\sigma\approx1$ or $\sigma\approx0.65$ with the usual value of $C=C_\mu/\sigma_c=0.09/0.7$ from $k$--$\varepsilon$-like models. The value of $\sigma=0.1$ for RT TMZ in table~\ref{tab:0DSurrogate} makes transport clearly inconsistent with standard turbulent transport and thus supports a more direct non-Fickian description through the drift velocity and the associated directed energy. \emph{Understanding of this non-Fickian transport in RT flows is the \emph{core motivation} of the present work.}

	The low values of~$\kappa$ or~$\kappa_{xy}$ for RT---or equivalently of the integral length scale---may erroneously let believe that TVH could apply, just as usual Fickian microscopic transport holds at low Knudsen number. However, the associated velocity $\langle K\rangle^\shalf$ is too low to produce a high enough value of the reduced large eddy viscosity~$\sigma$.
%
\subsection{Mixing in RT flows}
\label{ssec:Mixing}
	At large Reynolds and moderate Schmidt numbers, experiments and simulations with miscible fluids at vanishing Atwood number have given uniform profiles of mixing fraction~$\theta$, with $\langle\theta\rangle\approx0.8$ for $\alpha\approx0.05$ \citep{Youngs91, Dimonte04, Livescu09, Schilling10, Dalziel99, Ramaprabhu04, Mueschke06, Banerjee10, Mueschke09}. Thus, assuming equivalent inter-structure and per-structure contributions to fluid segregation, as suggested by visual inspection of figure~\ref{fig:AlphaDNS} and in anticipation of results in section~\ref{ssec:0DResults}, the value of the effective mixing fraction $\langle\theta_e\rangle = \ensuremath{{\textstyle\frac{8}{9}}} \approx 0.89$ appears acceptable---as obtained for $\varkappa=\thalf$ in the expected volume fraction profiles~\eqref{eq:Prof}. This value should be independent of the miscible or immiscible nature of the fluids and a proper two-structure-field segmentation method should apply and behave equally in both cases.

	It must be noticed however, that $\varkappa=\thalf$ is the highest value giving \emph{monotonic} $\alpha^\pm(\zeta)$ as discussed in section~\ref{ssec:Profiles} and illustrated by the corresponding profiles in figure~\ref{fig:AlphaProfiles}. Setting $\varkappa=\ensuremath{{\textstyle\frac{1}{3}}}$ would yield $\langle\theta_e\rangle=\ensuremath{{\textstyle\frac{3}{4}}}$ which would assume negligible per-structure segregation, \ie $\langle\theta_e\rangle\approx\langle\theta\rangle$ as in model calibrations of \citet[§~4]{Youngs95} or \citet[§~2.1.2]{Kokkinakis15}. The range of acceptable~$\varkappa$ or~$\langle\theta_e\rangle$ values is thus somewhat restricted, unless non-monotonic volume-fraction profiles of structures $\alpha^\pm(\zeta)$ are accepted. This is not physically forbidden a priori but appears somewhat counter-intuitive.
%
\section{Present prescription for filtered two-structure-field segmentation in RT}
\label{sec:Prescription}
%
\subsection{General framework: structure-oriented Lagrangian space-time filtering}
\label{ssec:Filtering}
	As already introduced in section~\ref{ssec:Memory}, appropriate two-structure-field segmentation must involve a form of memory effect or time filtering in order to reduce distortions and be sensitive to the persistence of large-scale turbulent structures. A widely spread approach consists in generating a Lagrangian filtered field~$\widetilde{\beta}$ of some relevant variable~$\beta$ according to the general evolution equation \citep[\eg][eq.~1 \& fig.~2]{Lumley92}
\begin{equation}
\label{eq:Filtering}
\partial_t \widetilde{\beta} + u_j \, \widetilde{\beta}_{,j}
	= C_\omega \, \omega \, ( \beta - \widetilde{\beta})
	+ C_\omega C_\ell^2 \, \omega \langle\ell\rangle^2
		\widetilde{\beta}_{,jj} ,
\end{equation}
\marginpar{Switch to flux term for diffusion? Find reference for CDR equations?}%
which is a form of convection--diffusion--reaction equation. Segmentation of $\beta=\widetilde{\beta}$ is then performed according to generalized Otsu methods as detailed in appendix~\ref{app:Segmentation}. Filtering characteristics are controlled by a relaxation rate~$\omega$ and a diffusivity $\omega\langle\ell\rangle^2$ (possibly space or time dependent), both adapted from some characteristic energy-containing scales and tuned to specific needs with coefficients~$C_\omega$ and~$C_\ell$.

	Before specifying in section~\ref{ssec:UzFiltering} the filtering parameters adopted for the present study, it is worthwhile mentioning some important or critical features of~\eqref{eq:Filtering}:
\begin{itemize}
	\item\emph{Linear:} Beyond its convenience for numerical simulation, the linearity of~\eqref{eq:Filtering} provides a simple tool to combine different separator fields in predictable ways. For instance, no difference would appear between the final structure fields produced by applying~\eqref{eq:Filtering} to the full value or to the fluctuation of a given quantity.
	\item\emph{Autonomous:} None of the quantities appearing in~\eqref{eq:Filtering} is externally imposed by the observer (lengths, rates, \etc). When the underlying flow is self-similar, this ensures that~$\widetilde{\beta}$ is also self-similar and can be scaled by the characteristic yardsticks of the flow, $L$, $L'$, and their combinations.
	\item\emph{Lagrangian derivative:} The left hand side of~\eqref{eq:Filtering} is a Lagrangian or material derivative: for vanishing right hand side, it transports~$\widetilde{\beta}$ as an invariant and does not produce small scale mixing (other than numerical). It is thus the formal \emph{source of bimodality} by bringing into a TMZ the (different)~$\beta$ values of the laminar surroundings. When $C_\omega=0$, it provides an everlasting memory effect and, for non-miscible incompressible flows $u_{j,j}=0$, it coincides with the mass conservation equation~\eqref{eq:Dr} yielding $\widetilde{\beta}\propto\rho$.
	\item\emph{Relaxation towards local field value:} The first term of the right hand side of~\eqref{eq:Filtering} provides relaxation towards the local value of the non-filtered field~$\beta$: in the vanishing memory limit $C_\omega\rightarrow\infty$ with finite $C_\omega C_\ell^2$ it would yield $\widetilde{\beta}=\beta$.
	\item\emph{Diffusion of filtered field:} Fluctuations at small scales in~$\beta$ can produce fractal like interfaces after segmentation. The last term of the right hand side of~\eqref{eq:Filtering} controlled by~$C_\ell$ can thus provide diffusion and smooth~$\widetilde{\beta}$. In stationary conditions at zero velocity and with uniform~$\langle\ell\rangle$, the left hand side of~\eqref{eq:Filtering} cancels and the solution of the equation is provided by convolution with a Green kernel~$\mathcal{K}$
\begin{align}
\widetilde{\beta} = \mathcal{K} * \beta ,
&&
\text{with}
&&
\mathcal{K}(\bm{r})
	= \frac{e^{-\|\bm{r}\|/(C_\ell\langle\ell\rangle)}}
		{4\pi\,C_\ell^2\langle\ell\rangle^2\|\bm{r}\|} ,
\end{align}
akin to a screened-Coulomb or Yukawa potential. The width of the Kernel~$\mathcal{K}$ is uniform and given by a fraction~$C_\ell$ of the 0D-averaged integral length scale~$\langle\ell\rangle$.
	\item\emph{Persistence and bimodality control:} Following the last points above, the~$C_\omega$ and~$C_\ell$ values control the strength of memory effects and hence the bimodality level which can be maximized: for weak filtering, the bimodality is that of the raw $\beta$ field PDF, generally spread by turbulent fluctuations around an apparent single mode. For strong filtering, bimodality is that of a narrow PDF peaked at the mean of~$\beta$ over the TMZ. Both display poor bimodality and an optimum can be found in between.
	\item\emph{Non-conservative form of~\eqref{eq:Filtering}:} The filtering equation is a step in the (non-linear) signal processing leading to~$b^\pm$ and thus does not need to represent a physical conservation law. In particular, to reduce singularities produced by turbulent small-scale intermittency, the (possibly non-uniform) factor $C_\omega C_\ell^2 \, \omega$ applies to the flux gradient $\widetilde{\beta}_{,jj}$---instead of the flux $\widetilde{\beta}_{,j}$ for a conservative term. Similarly, filtering of the vertical velocity as introduced below in section~\ref{ssec:UzFiltering} would need to be carried out on its \emph{momentum} $\rho u_z$ in order to recover conservation.
	\item\emph{Initial condition:} At long enough times $t \gg 1/\omega$, results most often appear weakly sensitive to the initial condition which is then conveniently set to $\widetilde{\beta}(0,\bm{r}) = \beta(0,\bm{r})$.
	\item\emph{Consistency with previous approaches:} The present approach is akin to more convoluted Lagrangian filtering techniques developed in other contexts related to fluctuations \citep{Slooten98} or coherent structures \citep{Haller00, Samelson13, Haller15}.
\end{itemize}
%
\subsection{Present prescription: Lagrangian local space-time filtering of vertical velocity}
\label{ssec:UzFiltering}
	The general framework defined by~\eqref{eq:Filtering} leaves open the selection of the filtered field~$\beta$, the relaxation rate $C_\omega\omega$, and the diffusivity $C_\omega C_\ell^2 \, \omega \langle\ell\rangle^2$. A comprehensive exploration and optimization of such a wide space of parameters is of course out of question. Now, the principles provided in parts~\ref{sec:QualitativeArg} to~\ref{sec:QuantitativeArg} and preliminary tests \citep[app.~B]{Watteaux11} have hinted to a globally efficient combination of quantities, even if not fully optimal, given by
\begin{subequations}
\label{eq:Parameters}
\begin{align}
\label{eq:RawField}
\beta(t,\bm{r})
	&= u_z(t,\bm{r}) ,
\\
\label{eq:omegalim}
\omega(t,\bm{r})
	&= \min\Big\{\, \frac{\varepsilon}{3k_z}(t,\bm{r})
	\,,\, C_\wedge \frac{\langle\varepsilon\rangle}
		{3\langle K_z\rangle}(t) \,\Big\} ,
\\
\label{eq:Ell}
\langle\ell\rangle(t)
	&= \frac{(3\langle K_{xy}\rangle/2)^{3/2}}
		{\langle\varepsilon\rangle}(t) .
\end{align}
\end{subequations}
The integral length~$\langle\ell\rangle$ is \emph{global} (time dependent but position independent) and defined by the turbulent transverse kinetic energy and the turbulence dissipation, both averaged over the TMZ. In contrast, the integral rate~$\omega$ is \emph{local} and defined by the turbulent \emph{longitudinal} energy and the turbulence dissipation, both local
\begin{align}
\label{eq:omegaloc}
k_z(t,\bm{r}) &= \thalf (u_z'')^2
&&\text{and}&
\varepsilon(t,\bm{r}) &= 2 \nu s_{ij} s_{ij} ,
&&\text{with}&
s_{ij} = \thalf (u_{j,i}+u_{i,j}) .
\end{align}
In order to avoid overly stiff events---possibly divergent for $\Rey\rightarrow\infty$, with questionable physical relevance and annoying numerical impact---$\omega$ is also \emph{capped} by an adjustable multiple of the \emph{global} 0D-averaged integral rate obtained from~\eqref{eq:0DKpar} and~\eqref{eq:0DDissipation}.

	Before specifying in section~\ref{ssec:Optimization} the optimization criteria of the filtering coefficients~$C_\omega$, $C_\ell$, and~$C_\wedge$ adopted for the present study, it is worthwhile mentioning some important or critical features of~\eqref{eq:Parameters}:
\begin{itemize}
	\item \emph{Selection of raw field $\beta=u_z$~\eqref{eq:RawField}:} This choice is justified in section~\ref{ssec:Segmentation} and follows from the approximate but reasonable behaviour of the poor man's instantaneous approach.
	\item \emph{Dependence of relaxation rate~$\omega$ on \emph{local} turbulent conditions~\eqref{eq:omegalim}:} This has major impact because, as required, it tracks contrasts of turbulence and thus leaves a long-lasting memory effect in quasi-laminar zones. Preliminary tests showed a critical impact of the locality of~$\omega$ to capture the proper structure features at the TMZ edges (especially SP, SL, and EW in figure~\ref{fig:RTruc}).
	\item \emph{Dependence of relaxation rate~$\omega$ on \emph{vertical} turbulent kinetic energy~$k_z$~\eqref{eq:omegalim}:} $\omega$~is then divergent at~$u_z=0$, the threshold of the poor man's instantaneous approach. Loss of memory and diffusion are then maximal close to the two-structure-field boundary thus providing for a quick adjustment zone between~$\widetilde{\beta}$ on each side. Extensive preliminary explorations revealed this feature to be important to improve results compared to the a priori more intuitive closure based on the full turbulent kinetic energy~$k$.

	\item \emph{Uniform ratio of diffusion to relaxation~\eqref{eq:Ell}:} This ratio, defined by the global length scale~$\langle\ell\rangle$, makes~$\widetilde{\beta}$ to always evolve under the same diffusion--relaxation balance irrespective of local laminar or turbulent conditions (even if more or less rapidly). The uniformity of~$\langle\ell\rangle$ is consistent with the qualitative observation reported in part~\ref{sec:QualitativeArg} for RT at vanishing~$\Atw$, but it could possibly require adaptations at finite~$\Atw$ or in other flow configurations.
	\item \emph{Dependence of integral length scale~$\langle\ell\rangle$ on \emph{transverse} mean kinetic energy~$\langle K_{xy}\rangle$~\eqref{eq:Ell}:} This excludes the influence of directed energy effects which describe the relative motion of structures but not their internal dynamics based on Kolmogorov cascades. Factor~$3/2$ in~\eqref{eq:Ell} partly corrects $\langle K_{xy}\rangle$ for a missing $z$~contribution assuming an isotropic turbulent energy inside structures.
\end{itemize}
%
\subsection{Optimization criteria for space and time filtering coefficients}
\label{ssec:Optimization}
	The present prescription defined by~\eqref{eq:Filtering} and~\eqref{eq:Parameters} must be complemented with values for the filtering coefficients~$C_\omega$, $C_\ell$, and~$C_\wedge$. Coefficients $C_\omega$ and~$C_\ell$ primarily control the respective bimodal and fractal characters of the filtered field~$\widetilde{\beta}$ as commented in section~\ref{ssec:Filtering}, whereas, if large enough, $C_\wedge$~has a weaker effect and merely reduces the numerical stiffness of extreme turbulent events as commented in section~\ref{ssec:UzFiltering}. Coefficients can thus be optimized almost independently of each other, with cross adjustments being required for fine tuning only. In any case, this optimization requires quantitative definitions of bimodality, fractality, and segmentation threshold in the TMZ.

	The \emph{average bimodality coefficient}~$\langle\mathcal{B}\rangle$ is defined as\marginpar{Introduce factor $\alpha^2\alpha^1$ in $\langle\mathcal{B}\rangle$ definition.}
\begin{subequations}
\label{eq:FullBimodality}
\begin{align}
\label{eq:MeanBimodality}
\langle\mathcal{B}\rangle(t) &= \frac{1}{L}
		\int_{-L/2}^{L/2} \mathcal{B}(t,z) \dd z ,
\\
\label{eq:Bimodality}
\mathcal{B}(t,z) &= \frac{\mu_4\mu_2-\mu_3^2-\mu_2^3}{\tfourth\mu_3^2+\mu_2^3} ,
\\
\label{eq:Moments}
\mu_n(t,z) &= \overline{\big(\widetilde{\beta}-\mu_1\big)^n}
	= \int \big(\widetilde{\beta}-\mu_1\big)^n
		\, \mathcal{P}\big(t,z;\widetilde{\beta}\big)
			\, \dd \widetilde{\beta} .
\\
\label{eq:Mean}
\mu_1(t,z) &= \overline{\widetilde{\beta}}
	= \int \widetilde{\beta} \, \mathcal{P}\big(t,z;\widetilde{\beta}\big)
			\, \dd \widetilde{\beta} .
\end{align}
\end{subequations}
Selection of the bimodality coefficient~\eqref{eq:Bimodality} is briefly justified in appendix~\ref{app:Bimodality}. For the present study, suffice to mention the main properties of~$\mathcal{B}$:
\begin{itemize}
	\item $\mathcal{B} \geq 0$ always \citep[resp.\ p.~433 and eq.~iv]{Pearson16, Pearson29},
	\item $\mathcal{B} = 0$ if and only if~$\mathcal{P}$ is a purely bimodal (or Bernoulli) distribution,
	\item $\mathcal{B} \lesssim 1$ if the profile of~$\mathcal{P}$ displays a dip, as required for proper bimodal segmentation,
	\item $\mathcal{B} = 2$ if, but not only if, $\mathcal{P}$~is a single normal (or Gaussian) distribution.
\end{itemize}
Examples illustrating the behaviour of~\eqref{eq:Bimodality} on doubly-Gaussian PDFs are provided in figure~\ref{fig:BiModCoef}. An optimized bimodality is naturally expected to produce a least ambiguous segmentation of the PDF of~$\widetilde{\beta}$, yielding meaningful two-structure fields.

	The fractality of~$\widetilde{\beta}$ only needs to be monitored around the two-structure field threshold~$\widetilde{\beta}^\circ$ where it can be quantified by the \emph{normalized mean interfacial area}~$\langle\mathcal{A}\rangle$ defined as
\begin{equation}
\langle\mathcal{A}\rangle(t)
	= \int_{-\infty}^{\infty} \overline{\|\bnabla b^\pm\|} \dd z .
\end{equation}
$\langle\mathcal{A}\rangle$ must be kept non divergent for vanishing viscosity. In the present simulations an asymptotically constant value must thus be reached for the self-similar regime at late times: then $\langle\mathcal{A}\rangle\propto L/\langle\ell\rangle$ where~$\langle\ell\rangle$ is the integral length scale---which can be taken as a proxy for the size of large-scale structures.

	In order to obtain~$b^\pm$ (and then compute~$\langle\mathcal{A}\rangle$), the \emph{segmentation threshold}~$\widetilde{\beta}^\circ(t,z)$ of the PDF of~$\widetilde{\beta}$ must be defined. Here, it is given by the implicit equation in~$\widetilde{\beta}^\circ$
\begin{equation}
\label{eq:Threshold}
2\widetilde{\beta}^\circ = 2\overline{\widetilde{\beta}}
	+ (q-1) (\overline{b^+}-\overline{b^-})
		\bigg( \frac{\overline{b^+\widetilde{\beta}}}{\overline{b^+}}
			- \frac{\overline{b^-\widetilde{\beta}}}{\overline{b^-}} \bigg) ,
\end{equation}
with the $z$-independent weighting exponent $q=\thalf$.\marginpar{Romain, make sure this is the value used throughout.} Selection of this threshold is briefly justified in appendix~\ref{app:Segmentation}. It appears appropriate for the type of PDF asymmetry produced in the present RT simulations, with a dominant narrow mode and a weak wide mode at TMZ edges. The same threshold definition also appears appropriate for the (unfiltered) poor man's instantaneous approach defined in section~\ref{ssec:Segmentation}. Examples illustrating the behaviour of~\eqref{eq:Threshold} for $q=0$, $\thalf$, and~1 on doubly-Gaussian PDFs are provided in figure~\ref{fig:BiModCoef}.
%
\subsection{Optimized filtering for two-structure field segmentation in a turbulent RT flow}
\label{ssec:Coefficients}
\marginpar{Romain please approve section.}
	The optimization of coefficient according to the prescriptions in section~\ref{ssec:Optimization} was first carried out on a series of simulations of RT flows at $\Atw=0.01$ in a unit cubic box at $256^3$~resolution. This low resolution is convenient as numerous simulations can be computed on a limited number of processors ($\sim 4^3$) over a few hours. Coefficients were then finely adjusted at $512^3$ and eventually $1024^3$~resolutions. Initial conditions and numerical details are provided in appendix~\ref{app:Numerics}. Ensuing two-structure correlations will be examined in part~\ref{sec:RTStructures}.

	The following optimal coefficients were used at $1024^3$~resolution with the corresponding values of mean bimodality index and interfacial area
\begin{align}
\label{eq:bimod_sensi}
C_\omega = 1.7 ,
&& C_\ell = 0.15 ,
&& C_\wedge = 30 ,
&& \langle\widehat{\mathcal{B}}\rangle \approx 0.55 ,
&& \langle\widehat{\mathcal{A}}\rangle \approx 15 .
\end{align}
\marginpar{Correct values.}%
$\langle\widehat{\mathcal{B}}\rangle$ and $\langle\widehat{\mathcal{A}}\rangle$ are the weighted averages~\eqref{eq:SSRaveraging} of $\langle\mathcal{B}\rangle$ and $\langle\mathcal{A}\rangle$ over the estimated self-similar regime $\text{SSR}$---here chosen as $0.3\leq L\leq 0.7$ where the growth rate coefficient~$\alpha$ appears stabilized (see figure~\ref{fig:RT_results}). The value of ~$C_\wedge$ was somewhat lowered with respect to its optimum so as to significantly reduce the computational cost at a marginal loss of about 1\% on the bimodality coefficients---$C_\wedge$ bounds the stiffness of extreme turbulent events.\marginpar{Discuss descent to minimum here.}

\begin{figure}
\centerline{\begin{tabular}{ccc}
\hspace{\textwidth/3-1em}
	& \includegraphics[width=\textwidth/3-1em]{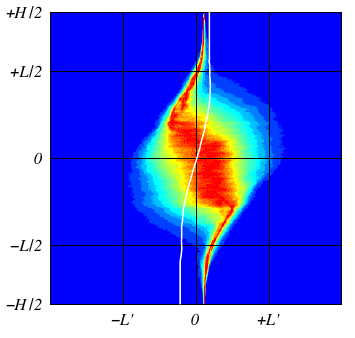}
	& \includegraphics[width=\textwidth/3-1em]{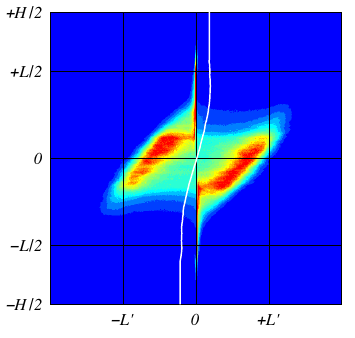}
\\*
(a) To be completed. & (b) Old. Update. & (c) Old. Update.
\\*
\includegraphics[width=\textwidth/3-1em]{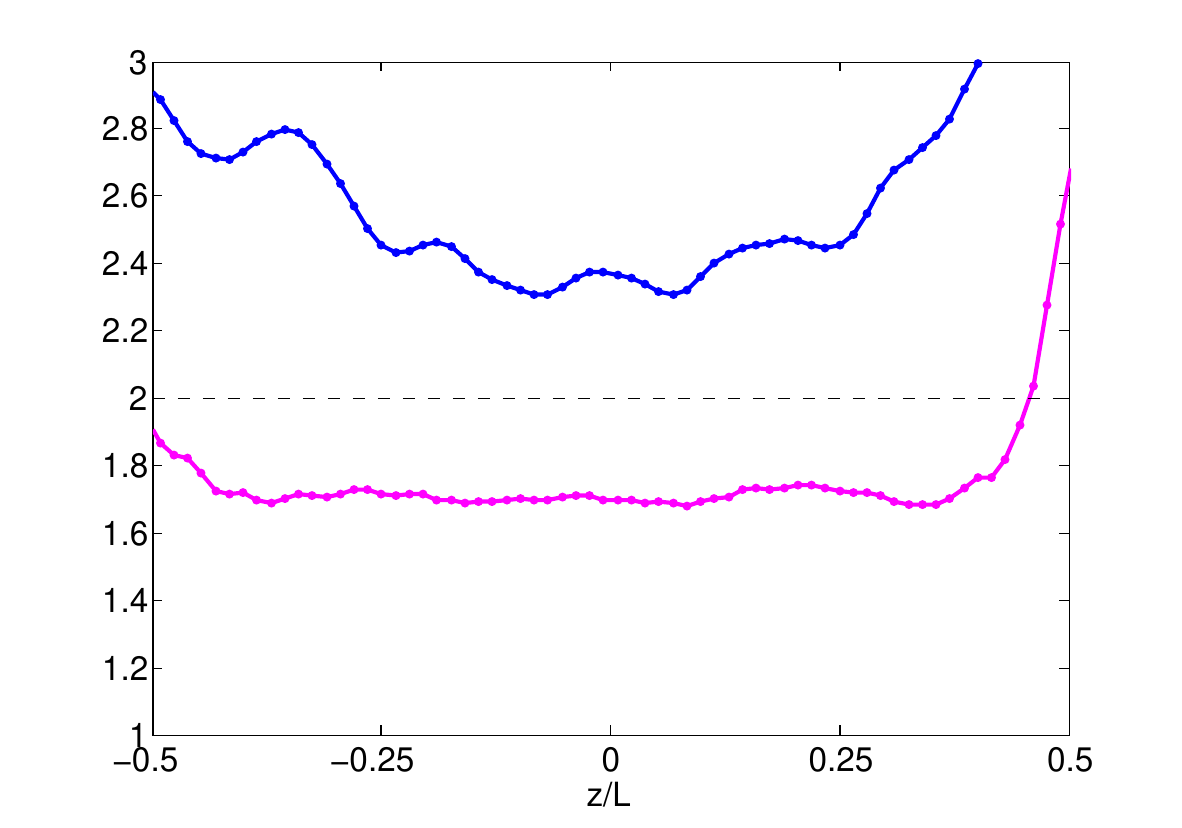}
	& \includegraphics[width=\textwidth/3-1em]{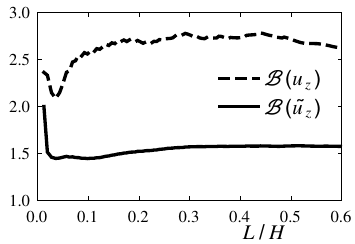}
	& \includegraphics[width=\textwidth/3-1em]{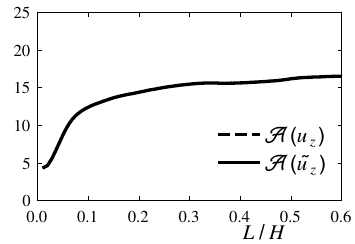}
\\*
(d) Old. $\mathcal{B}+1$. Update. & (e) Old. $\mathcal{B}+1$. Update. & (f) Old. Update.
\end{tabular}}
\caption{Illustration of relevant statistical properties of fields $\beta=\rho$, $u_z$, and $\widetilde{u_z}$---this last being filtered with optimized coefficients~\eqref{eq:bimod_sensi}---examined here for a two-structure-field segmentation in a $1024^3$ simulation of self-similar RT at $\Atw=0.01$: (a) to (c) PDF color maps of respectively~$\rho$, $u_z$, and $\widetilde{u_z}$ (scaled as $\mathcal{P}(\zeta,\beta)/\max_\beta\mathcal{P}(\zeta,\beta)$ for legibility) with threshold lines $\beta^\circ$ (white lines); (d) $\widehat{\mathcal{B}}(\zeta)$; (e) $\langle\mathcal{B}\rangle(L)$; (f) $\langle\mathcal{A}\rangle(L)$.
}
\label{fig:bimod_optim}
\end{figure}
\marginpar{Figure~\ref{fig:bimod_optim}: update. Include (scaled) profiles for $\beta=\rho$ and $u_z$?}
	The impact of filtering readily appears on the respective PDF maps of $\beta=\rho$, $u_z$, and $\widetilde{u_z}$ in figures~\ref{fig:bimod_optim}a to~c: a deep valley is visible on the filtered field $\widetilde{u_z}$ all across the TMZ height while mostly none appears on the unfiltered fields (except slightly at TMZ edges). As shown in figures~\ref{fig:bimod_optim}d and~e, this is quantitatively supported by significantly different values of 1.6 and 1.5 to 0.5\marginpar{Update values} for the bimodality coefficients---which are mostly uniform over both the TMZ height $\widehat{\mathcal{B}}(\zeta)$ and the TMZ growth $\langle\mathcal{B}\rangle(L)$, confirming self-similar behaviour. At the same time as shown in figure~\ref{fig:bimod_optim}f, the mean interfacial area $\langle\mathcal{A}\rangle(L)$ experiences an unbounded growth for $\beta=\rho$ and $u_z$ but appears to stabilize asymptotically around a consistent value of approximately 15\marginpar{Update values}. As visible at the edges of the TMZ in figure~\ref{fig:bimod_optim}d, $\widehat{\mathcal{B}}(\zeta)$ departs form its lower uniform level within the TMZ, where it becomes singular and statistically irrelevant for vanishing volume fractions $\alpha^\pm$. Also noticeable at the edges of the TMZ is the spreading of the velocity fluctuations $u_z$ beyond the $-\thalf\leq \zeta\leq +\thalf$ interval \citep[as analyzed by][]{Phillips55} due to long range pressure fluctuations: in contrast, because of the Lagrangian filtering which carries $u_z$ values from away of the TMZ, $\widetilde{u_z}$ is much closer to zero not only outside but also somewhat inside the $-\thalf\leq \zeta\leq +\thalf$ interval. As expected, the spreading of~$u_z$ and $\widetilde{u}_z$ values in figures~\ref{fig:bimod_optim}c and~d is consistent with~$L'$. The present results are reminiscent of previously explored bimodal distributions in turbulent flows as for instance the position-conditioned turbulent frequency observed by \citet[fig.~4]{Slooten98}.

\begin{figure}
\centerline{\begin{tabular}{ccc}
\hspace*{\textwidth/3-1em}
	& \hspace*{\textwidth/3-1em}
	& \includegraphics[width=\textwidth/3-1em]{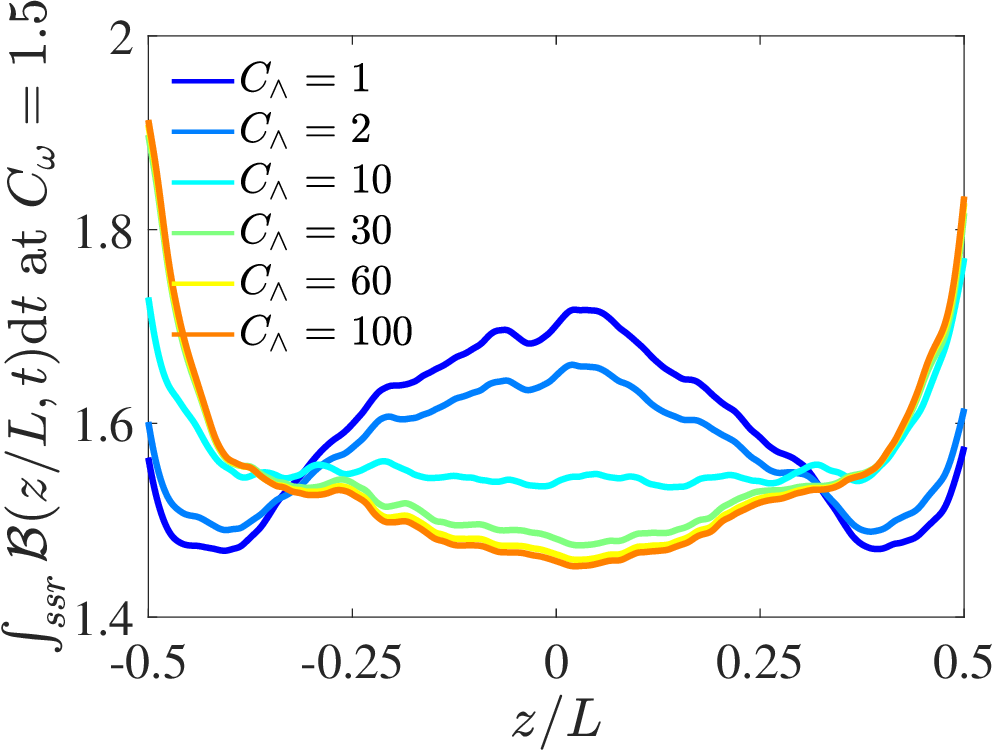}
\\*
(a) To be completed. & (b) To be completed. & (c) Old. Update.
\\*
\hspace*{\textwidth/3-1em}
	& \hspace*{\textwidth/3-1em}
	& \includegraphics[width=\textwidth/3-1em]{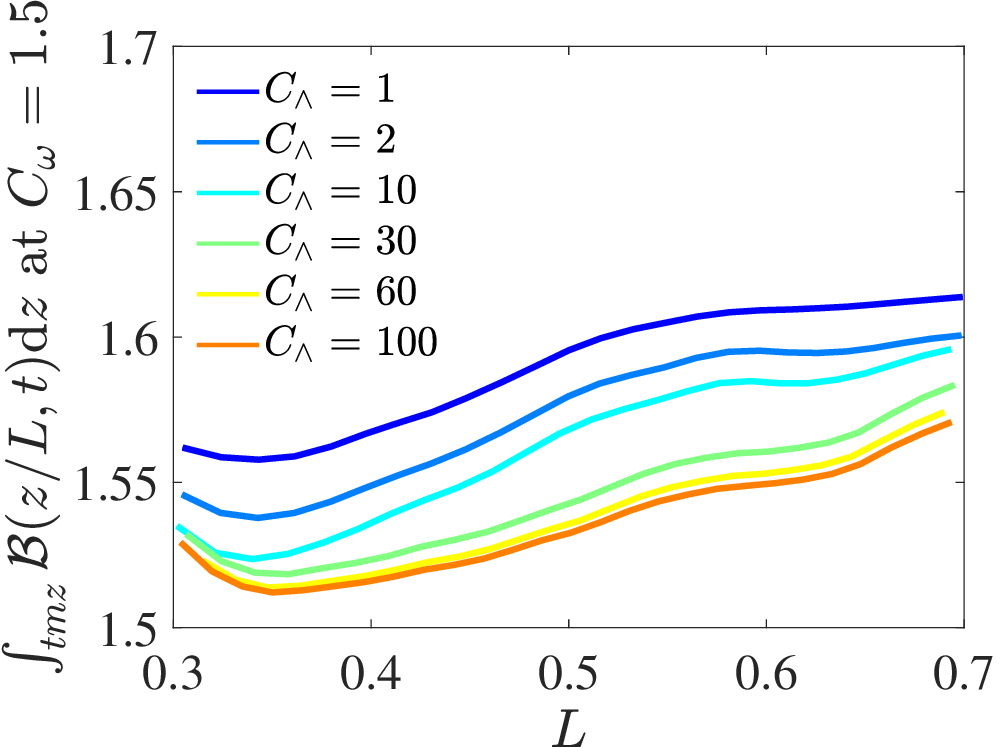}
\\*
(d) To be completed. & (e) To be completed. & (f) Old. Update.
\\*
\includegraphics[width=\textwidth/3-1em]{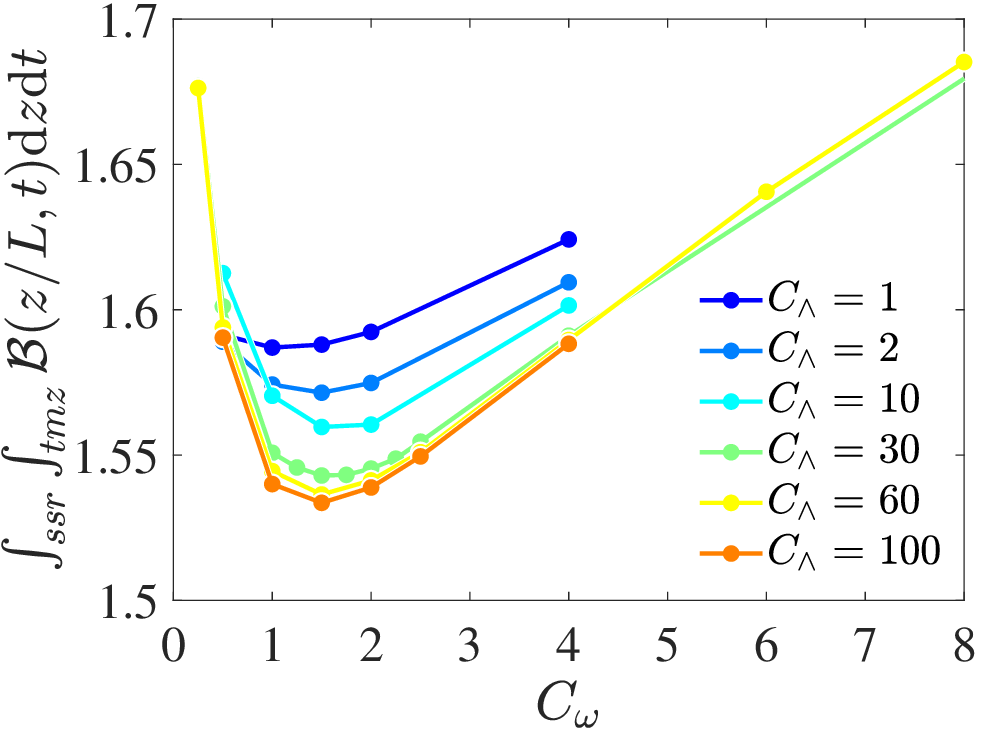}
	& \includegraphics[width=\textwidth/3-1em]{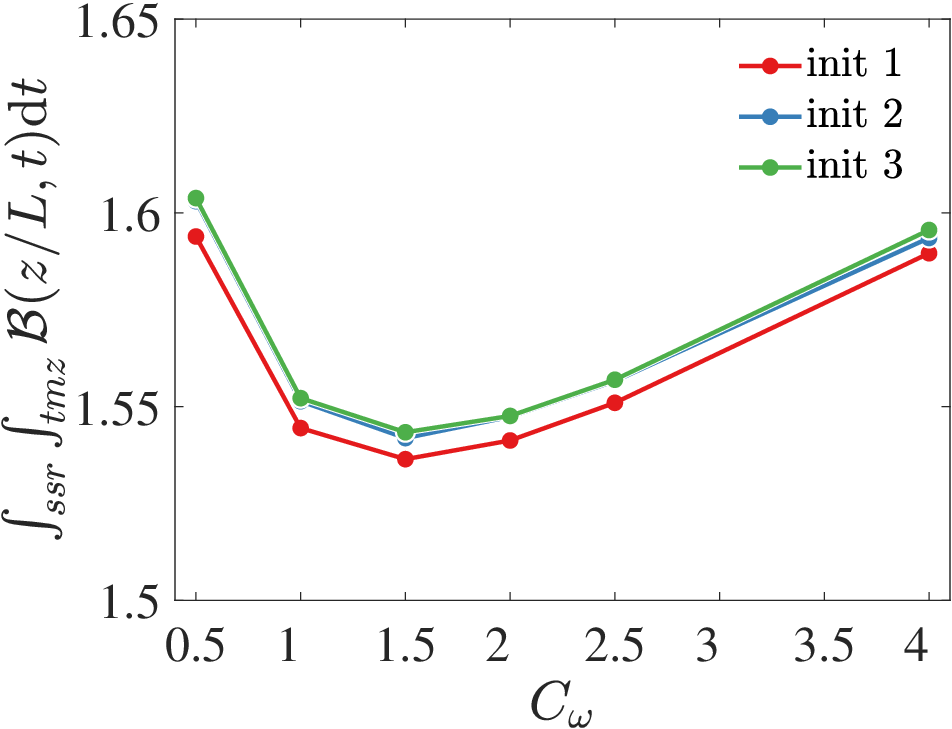}
	& \hspace*{\textwidth/3-1em}
\\*
(g) Old. Replace. & (h) Old. Replace. & (i) To be completed.
\end{tabular}}
\caption{Sensitivity towards coefficients~$C_\omega$, $C_\ell$, and $C_\wedge$ of $u_z$ filtering for two-structure-field segmentation in a $256^3$ simulations of self-similar RT at $\Atw=0.01$: (a) to~(c) Profiles of $\widehat{\mathcal{B}}(\zeta)$ for five variations of respectively~$C_\omega$, $C_\ell$, and $C_\wedge$ around their optimized values; (d) to~(f) Profiles of $\langle\mathcal{B}\rangle(L)$ for five variations of respectively~$C_\omega$, $C_\ell$, and $C_\wedge$ around their optimized values; (g) to~(i) Profiles of $\langle\mathcal{A}\rangle(L)$ for five variations of respectively~$C_\omega$, $C_\ell$, and $C_\wedge$ around their optimized values.
}
\label{fig:bimod_sensi}
\end{figure}
\marginpar{Figure~\ref{fig:bimod_sensi}: update.}
	The dependence of~$\widehat{\mathcal{B}}(\zeta)$, $\langle\mathcal{B}\rangle(L)$, and~$\langle\mathcal{A}\rangle(L)$ upon~$C_\omega$, $C_\ell$, and~$C_\wedge$ around their optimal values was explored on $256^3$~resolved simulations. As illustrated in figure~\ref{fig:bimod_sensi}, results globally confirm the optimal values in~\eqref{eq:bimod_sensi} up to slight variations attributed to the lower resolution. As expected in figures~\ref{fig:bimod_sensi}d to~e, an increase of~$C_\omega$, and marginally $C_\ell$ and $C_\wedge$, slows the convergence of $\widetilde{u_z}$ towards its self-similar behaviour. Less expected is the impact of~$C_\wedge$ on the~$\widehat{\mathcal{B}}(\zeta)$ profile: the concavity--convexity distortions observed in figure~\ref{eq:bimod_sensi}c can be explained by the different distribution of extreme turbulent events across the TMZ which are thus differently bounded when modifying~$C_\wedge$.

\begin{figure}
\centerline{\includegraphics[width=\textwidth]{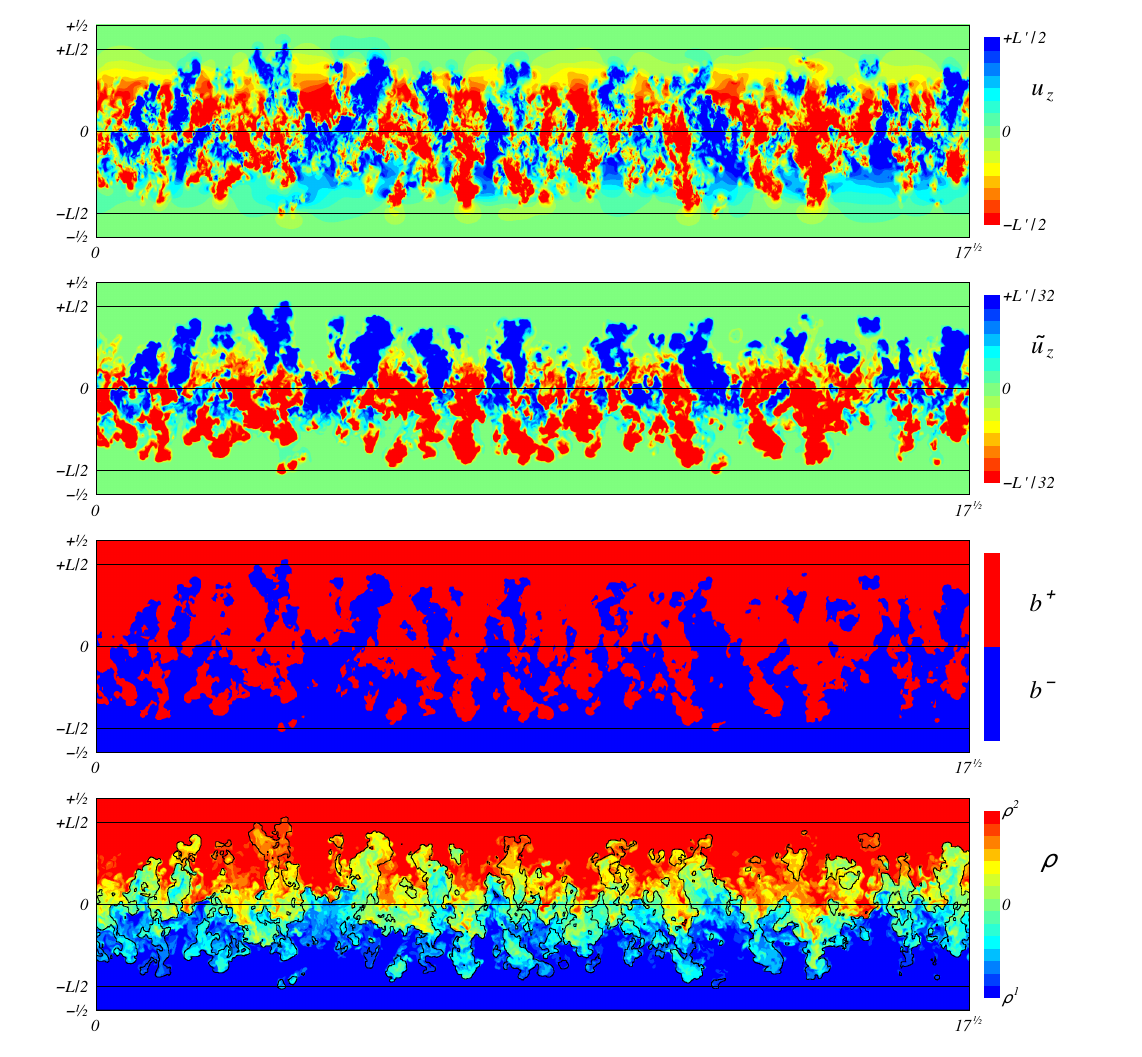}}
\caption{Colour maps at $L(t)\approx0.775$ in a vertical cross section along the $(x,y)=(4,1)$ direction of the calculation domain, of (top to bottom): vertical velocity~$u_z$, separator field $\beta=\widetilde{u_z}$ (filtered vertical velocity), corresponding structure fields~$b^\pm$ (segmented~$\beta$), and density~$\rho$ with superimposed structure boundaries (black contour).}
\label{fig:2DMaps}
\end{figure}
\marginpar{Figure~\ref{fig:2DMaps}: update. Make picture at $L=0.8$ exactly. Draw structure boundaries on $u_z$. Add poor man's instantaneous segmentation on~$b^\pm$?}
	In figure~\ref{fig:2DMaps}, maps of $u_z$, $\widetilde{u_z}$, $b^\pm$, and~$\rho$ in vertical cross sections of the TMZ illustrate the significant reduction of intermittency and fractality at small scales provided by filtering. Close inspection of the ensuing two-structure field boundary superposed to the~$u_z$ and~$\rho$ fields confirms the expected correlations: partial at large scales but weak at small scales, with significant corrections to the distortions discussed in section~\ref{ssec:Segmentation} and illustrated figure~\ref{fig:RTruc}.
%
\section{In progress: Two-structure field correlations from a simulated turbulent RT flow}
\label{sec:RTStructures}
%
\backsection[Supplementary data]{Supplementary material and movies are available at\dots
}

\backsection[Acknowledgements]{The authors benefited from numerous discussions on numerical aspects with {J.-M.}~Ghid\-aglia and F.~de~Vuyst, on bimodality with T.R.~Knapp, and on physical aspects with P.~Pailhoriès (who kindly provided the 1D simulations with the 2SFK model), O.~Poujade, A.~Briard, D.L.~Youngs, S.B.~Dalziel, B.~Aupoix, and the late and much missed P.~Comte.}

\backsection[Funding]{This work was partly supported by the Délégation Générale à l'Armement, France, under REI contract No.~2008.34.0026.0000.0000.}

\backsection[Declaration of interests]{The authors report no conflict of interest.}

\backsection[Data availability statement]{The data that support the findings of this study are openly available in [repository name] at http://doi.org/[doi], reference number [reference number].
}

\backsection[Author ORCIDs]{R.~Watteaux, \href{https://orcid.org/0000-0001-9905-9261}{https://orcid.org/0000-0001-9905-9261};\ J.A.~Redford, \href{https://orcid.org/0000-0002-6081-8790}{https://orcid.org/0000-0002-6081-8790}; A.~Llor, \href{https://orcid.org/0000-0002-4436-571X}{https://orcid.org/0000-0002-4436-571X}.}

%
\clearpage
\appendix
\section{Some previous works related to two-structure-field modelling}
\label{app:Previous}
\begin{itemize}
	\item \citet{Spiegel72} appears to have first applied two-fluid modelling to turbulence, using a variational principle in the spirit of Landau's pioneering work on liquid helium \citep{Landau41}. His aim was to retrieve the intermittency of turbulence in time or space which appears at either turbulent transitions or edges of free turbulent flows.
	\item To the authors' knowledge, the first publication of a truly conditional RANS modelling approach is due to \citet{Libby75, Libby76} also for capturing turbulence intermittency. The flow was again separated into laminar and turbulent regions, instead of upward and downward moving regions here, but the subsequent equations on conditionally averaged quantities appear formally identical in all these cases---even if closures may differ substantially. This general approach was further developed \citep{Byggstoyl86} and may be viewed as among the most appropriate and convenient to date for transition and intermittency modelling.
	\item For combustion applications, where strong density and pressure gradients can induce buoyancy driven instabilities and turbulence, various authors around 1980 also introduced Reynolds averaging conditioned by the presence of the different fluids \citep{Libby77, Libby81, Spalding85, Spalding86, Spalding87}. In the process, specific contributions to the turbulent fluxes (here designated as `directed') were identified as potentially dominant for transport. This approach has brought some success under the designation of BML model \citep[§§~7.3 \&~8.7 and refs therein]{Veynante02}.\marginpar{To finish? Check references from report of Soulard et al.} The phenomenology can be related to the `intermingling-fragments' concept introduced by \citet{Spalding00} for turbulence and combustion modelling.
	\item Shortly after, a two-fluid turbulent model was empirically developed by \citet{Youngs84, Youngs89, Youngs94} for RT type flows, and then extended into a two-structure-field model \citep{Youngs95, Kokkinakis15, Kokkinakis20}. Various complex 2D turbulent buoyancy-driven flows were successfully simulated by this model \citep{Youngs07, Youngs09, Kokkinakis15, Kokkinakis20}. It can also be interpreted and analysed from the rigorous standpoint of conditionally averaged statistical equations \citep{Llor05}.
	\item In a similar spirit, some single-fluid models were complemented with closures or evolution equations inspired by two-fluid approaches \citep{Polyonov89, Polionov91, Polionov93}. Despite some interesting properties and an appealing simplicity, these approaches do not seem to have been commonly used, probably for the significant distortions they introduce to directed effects \citep[§~7.4]{Llor05}. Other related two-fluid or two-structure-field models have also been reviewed by \citet[§~12.7]{Zhou17b}.
	\item Following the observations by \citet[fig.~17]{Brown74} of turbulent structures and the merging patterns of their trajectories, a model for RT type flows was designed by extending a laminar `bubble-competition' analysis due to \citet{Layzer55}. This approach does not rely explicitly on two-structure-field conditional averaging, but instead on population balance equations of dominant heavy and light bubble structures in the flow \citep{Alon94, Alon95, Shvarts95}. Beyond the usual comparison with available experimental data, the model was further calibrated using the observed dynamics of bubble tips and Voronoi tiling at the edges of numerically simulated RT \citep[figs~1, 3 \&~5]{Shvarts00, Oron01}. Despite some restrictions, this structure-like approach was successfully extended to the turbulent regime \citep[and refs therein]{Shvarts00} of mixed-fluid bubbles.
	\item Based on these previous works and on a theoretical analysis \citep{Llor03, Llor05} partly summarized in part~\ref{sec:Theory}, \citet{LlorBailly03, LlorPoujade04} proposed a novel two-structure-field- two-fluid- two-turbulent-field model designated as 2SFK, with consistent closures for the characteristic length scales.
	\item The relationship between the buoyancy--drag equation~\eqref{eq:BD} and statistical turbulence models is a significant source of insight, as shown for instance in the case of the simple $k$--$\varepsilon$ model \citep[§~4.2]{Grea15}. For two-structure-field approaches, systematic reduction procedures can be applied and lead to the buoyancy--drag equation in a natural manner (see part~\ref{sec:Theory} below).
	\item Severe difficulties are encountered in `simple' models (including buoyancy--drag) if~$g(t)$ increases too fast \citep{Llor03, Llor05, Grea16} or reverses \citep{Kucherenko93b, Kucherenko97, Dimonte07, Ramaprabhu13}. These stem from modelling limitations similar to those of counter-gradient fluxes in combustion \citep[and refs therein]{Sabelnikov17} which can be removed in two-structure-field models \citep[§~7.3.5]{Libby81, Veynante02}.
	\item In order to correct the response of Reynolds stress models to transients in Unstably Stratified Homogeneous Turbulence \citep{Grea16}, \citet[eq.~10]{Griffond23} mirrored the turbulent energy \emph{subtracted of the estimated directed energy} into the equation of dissipation~$\varepsilon$. As discussed in section~\ref{ssec:Directed}, the directed energy is a significant portion of turbulent energy and is associated with the relative motion of structures at large scales: as previously shown \citep[p.~9]{Llor04}, it therefore must not appear in the~$\varepsilon$ equation.
\end{itemize}
%
\section{On existing detection approaches of individual turbulent structures}
\label{app:Structures}
\marginpar{Romain please check section!}
	As previously shown, figure~\ref{fig:StructureApproaches} summarizes the two possible routes to the determination of the complementary structure fields $b^\pm$.\marginpar{Adapt.} The left path is undoubtedly more physically sound since a detailed detection of each individual coherent structure allows for a proper classification, ensuring each is entirely assigned to either~$b^+$ or~$b^-$. For this purpose, many pattern identification methods could be considered out of the numerous turbulent structure investigations carried out over the last forty years as reviewed for instance by \citet[Session~2]{Lumley90}, \citet{Metais91}, \citet{Holmes96}, \citet{Bonnet98}, \citet[§~7.4]{Pope00}, \citet{Kida06}.

	Historically, Eulerian methods using sets of velocity snapshots have first appeared, where structures are delineated using either: i)~threshold methods by means of velocity invariants \citep{Hunt88, Chong90, Soria93, Perry94, Ooi99} or vorticity \citep{Okubo70, Jeong95, Chakraborty05, Bremer16, Shivamoggi22}, ii)~modal decomposition methods such as POD \citep{Lumley81},\marginpar{Add possibly missing reference from  Lumley?} DMD \citep{Schmid10}, SPOD \citep{Towne18}, and their extensions \citep[\eg][for a partial review]{Taira17}, all based on the velocity fields to find the most energetic modes of the linearised system, or iii)~wavelet decomposition methods to separate scales and only keep non-noisy coherent vortices \citep{Okamoto07, Farge15}.\marginpar{Cite \citep{Lumley90}?} Regardless of their specificities, these methods stress the identification and classification of recurring but not necessarily persistent patterns which turn out to not being volume filling~\eqref{eq:Regroup}---coherent structures are detected but isolated from surrounding non-energetic fluid regions. An extra step of topological connection is therefore required, but it remains an open challenge which, to the authors knowledge, has not been properly tackled so far.

	\marginpar{Adapt §} The qualitative visual eduction in section~\ref{ssec:Visual} could let believe that the important feature for proper two-structure-field segmentation is the accurate reconstruction of boundaries or equivalently, of turbulence contrasts. At the edges of a TMZ this is akin to detecting the turbulent--non-turbulent transition around the cloud-like turbulent bubbles, an important and thoroughly explored topic \citep[and refs therein]{DaSilva14}. Although this approach appears relevant to understand the behaviour of clouds---with their buoyant, entrainment, and `nibbling' effects,---it is not retained here as it brings some serious issues: i)~the emphasis on small scales which is unnecessary as the conditional averages in~\eqref{eq:TSA} are dominated by bulk fields within structures, ii)~the convoluted reconstruction of two-structure fields starting from their boundaries, iii)~the difficulty of including persistence (see section~\ref{ssec:Persistence}), and iv)~the virtually hopeless task of producing sensible boundaries at the centre of the TMZ where both structures are highly turbulent.

	\marginpar{Adapt §} In order to include cohesion in structure detection, recent methods involve \emph{Lagrangian filtering} techniques, similar to those pioneered by Pope for modelling fluctuations \citep{Slooten98} or by Haller for separating Lagrangian Coherent Structures (LCS) \citep{Haller00, HallerY00, Haller01, Haller02, Peacock10}.

	Despite significant success of these methods into describing the dynamics of turbulent flows, intrinsic issues such as lack of persistence and of objectivity \citep{Haller05} motivated the development of new methods based on Lagrangian information in order to properly capture persistent regions in the flow. Central to this approach is the study of Lagrangian Coherent Structures \citep{Haller00, Haller15},\marginpar{Cite \citep{HallerY00} also?} specifically the hyperbolic and elliptic LCSs. The hyperbolic LCSs (detected using the finite-time Lyapunov exponent) highlight surfaces of the flow across which fluid transfers are minimal, and the elliptic LCSs \citep[captured by the Lagrangian-averaged vorticity deviation][]{Haller16} focus on the stable patterns within which entrainment and mixing are most efficient and correlated. Both are characterized by a timescale chosen by the user. In principle, the method could be used to generate two-structure-fields by regrouping the detected elliptic LCSs into upward and downward moving fields while creating boundaries following the hyperbolic LCSs, even in the turbulent core of the TMZ. However, hyperbolic LCS are not necessarily connected surfaces, which can lead to complicated processing in order to obtain space-filling tessellations of structure presence functions. Besides, detection of LCSs is significantly complex and, as a consequence, requires heavy post-processing calculations, especially when dealing with three-dimensional high-Reynolds-number turbulent flows. More recent works have opened the way to on-the-fly computation of the hyperbolic LCSs \citep{Finn13, Finn17} but work is still needed for the elliptic LCSs and, here again, the step of topological connection remains a daunting open challenge.

	As briefly reviewed and discussed in section~\ref{app:Structures}, identification algorithms of turbulent structures have been investigated for over forty years. For the present purposes they generally appear quite complex, computationally heavy, or with questionable time-persistence, and they seldom define explicit presence functions~$b^s$. Therefore, even if approximate for two-structure-field reconstruction, segmentation approaches such as~\eqref{eq:bpm_PM} remain convenient and relevant provided they include the effects of persistence: this can only be obtained by applying some form of \emph{preliminary filtering} to a separator field~$\beta$. This filtering is a proxy for the velocity averaging over individual structures~$\langle u_z\rangle_s$ appearing in~\eqref{eq:Regroup}, but without actual identification of structures.
%
\section{Single-fluid and two-structure-field statistical equations}
\label{app:2SF}
%
\subsection{Evolution equations of individual realizations}
	Individual realizations of the flow are described by the usual Navier--Stokes equation, complemented with mass conservations
\begin{subequations}
\begin{align}
\label{eq:Dc}
\partial_t ( \rho c^m ) + ( \rho c^m u_j )_{,j}
	&= - \vartheta^m_{j,j} ,
\\
\label{eq:Dr}
\big[\; \partial_t \rho + ( \rho u_j )_{,j}
	&= 0 \;\big] ,
\\
\label{eq:Du}
\partial_t ( \rho u_i ) + ( \rho u_i u_j )_{,j}
	&= - p_{,i} + \rho g_i + \tau_{ij,j} ,
\end{align}
\end{subequations}
where notations are those of section~\ref{ssec:TSCAE} with Einstein's implicit summation convention on repeated indices---notice that stress tensor $\tau_{ij}$ must not be confused with the normalized integral time $\tau$ defined in~\eqref{eq:tks}. Because $\bm{\vartheta}^2+\bm{\vartheta}^1=0$ the conservation equation of the total mass, here between brackets, is redundant with the equations on individual fluid masses but will be conserved for reference in all the following.

	Incompressibility does not reduce to $u_{j,j}=0$ in the the case of a fluid mixture: when the densities of the pure fluids~$\rho^2$ and~$\rho^1$ are constants, the mass conservation equations can be expressed with the fluid volume fractions $\alpha^m=\rho c^m/\rho^m$ as
\begin{equation}
\partial_t \alpha^m + ( \alpha^m u_j )_{,j} = - \vartheta^m_{j,j}/\rho^m .
\end{equation}
Adding equations over~$m$ then yields the incompressibility condition
\begin{align}
\label{eq:v}
v_{j,j} = 0 ,
&& \text{with} &&
\bm{v} = \bm{u} + \bm{\vartheta}^2/\rho^2 + \bm{\vartheta}^1/\rho^1 ,
\end{align}
where~$\bm{u}$ and~$\bm{v}$ represent respectively the mean mass and mean volume velocities. However, incompressibility will not be assumed in the following in order to preserve generality and simplicity.

	New (usually uppercase) variables have been introduced to describe the various single-fluid and two-fluid averages and correlations of all relevant quantities \emph{except} those involving the viscous stress~$\tau_{ij}$ and the diffusion flux~$\bm{\vartheta}$. Excluding dissipation, these terms are asymptotically small for large Reynolds and Péclet numbers, and they will be used as checks of the LES approximation.

	Reynolds stress evolution equations will not be considered because of their complexity in the two-structure-field case, and of their absence in two-structure-field models so far \citep{Youngs95, LlorBailly03, LlorPoujade04, Kokkinakis15, Kokkinakis20}.
%
\subsection{Single-fluid averaged equations (RANS)}
	For comparison with the two-structure-field approach, the well-known single-fluid statistical equations must be reminded here. They are obtained from ensemble averages of the mass and momentum conservation equations~\eqref{eq:Dc}--\eqref{eq:Du}, yielding
\begin{subequations}
\label{eq:D1F}
\begin{align}
\label{eq:DC}
\partial_t ( \overline{\rho} C^m )
	+ ( \overline{\rho} C^m U_j )_{,j}
	&= - \varPhi^m_{j,j}
		- \big(\overline{ \vartheta^m_j }\big)_{,j} ,
\\
\label{eq:DR}
\big[\; \partial_t ( \overline{\rho} )
	+ ( \overline{\rho} U_j )_{,j}
	&= 0 \;\big] ,
\\
\label{eq:DU}
\partial_t ( \overline{\rho} U_i )
	+ ( \overline{\rho} U_i U_j )_{,j}
	&= - R_{ij,j} - P_{,i}
		+ \overline{\rho} g_i
			+ \big(\overline{ \tau_{ij} }\big)_{,j} ,
\\
\label{eq:DK}
\partial_t ( \overline{\rho} K )
	+ ( \overline{\rho} K U_j )_{,j}
	&= - \varPhi^k_{j,j} - R_{ij} U_{i,j}
			- P_{,i} \overline{u''_i}
		- \varPhi^p_{j,j}
		+ \big(\overline{ \tau_{ij} u''_i }\big)_{,j}
		- \overline{\rho} \varepsilon ,
\end{align}
\end{subequations}
where the equation between brackets is redundant with the others but shown here for reference, and with
\begin{subequations}
\begin{align}
C^m &= \overline{\rho c^m}/\overline{\rho} ,
	& \bm{\varPhi}^m &= \overline{\rho c^m \bm{u}''} ,
\\
\bm{U} &= \overline{\rho \bm{u}}/\overline{\rho} ,
	& R_{ij} &= \overline{\rho u''_iu''_j} ,
\\
P &= \overline{p} ,
	& \bm{\varPhi}^p &= \overline{p' \bm{u}''} ,
\\
K &= \overline{\thalf\rho (\bm{u}'')^2}/\overline{\rho} ,
	& \bm{\varPhi}^k &= \overline{\thalf\rho (\bm{u}'')^2 \bm{u}''} ,
\end{align}
\end{subequations}
with velocity and pressure fluctuations $\bm{u}''=\bm{u}-\bm{U}$ and $p'=p-P$, and dissipation $\overline{\rho}\varepsilon = \overline{\tau_{ij}u''_{i,j}}$.
%
\subsection{Two-structure-field averaged equations}
	The generic two-structure-field equations~\eqref{eq:amean} complemented by definitions~\eqref{eq:all} can be applied to the quantities which describe the flow~\eqref{eq:Dc}--\eqref{eq:Du}: volume $a=1/\rho$, fluid concentrations $a=c^2$ and~$c^1$, density $a=1$, momentum $a=\bm{u}$, and turbulent kinetic energy components $a=\thalf(u^\pm_\alpha)^2$, with velocity fluctuations $\bm{u}^\pm=\bm{u}-\bm{U}^\pm$ and no sums on Greek indices. Lengthy but straightforward calculations eventually yield the two-structure-conditional evolution equations
\begin{subequations}
\label{eq:D2SF}
\begin{align}
\label{eq:DA_pm}
\partial_t ( \alpha^\pm )
	+ ( \alpha^\pm U^\pm_j )_{,j}
	&= - \varPhi^{\alpha\pm}_{j,j} \mp \varPsi^\alpha
			+ \overline{ b^\pm u_{j,j} } ,
\\
\label{eq:DC_pm}
\partial_t ( \alpha^\pm \rho^\pm C^{m\pm} )
	+ ( \alpha^\pm \rho^\pm C^{m\pm} U^\pm_j )_{,j}
	&= - \varPhi^{m\pm}_{j,j} \mp \varPsi^m
			- \big(\overline{ b^\pm \vartheta^m_j }\big)_{,j}
				+ \overline{ b^\pm_{,j} \vartheta^m_j } ,
\\
\label{eq:DR_pm}
\big[\; \partial_t ( \alpha^\pm \rho^\pm )
	+ ( \alpha^\pm \rho^\pm U^\pm_j )_{,j}
	&= \mp \varPsi^\rho \;\big] ,
\\
\label{eq:DU_pm}
\partial_t ( \alpha^\pm \rho^\pm U^\pm_i )
	+ ( \alpha^\pm \rho^\pm U^\pm_i U^\pm_j )_{,j}
	&= - R^\pm_{ij,j} \mp \varPsi^u_i - \alpha^\pm P_{,i}
		\mp \big(\alpha^+\alpha^- (P^+\!-P^-)\big)_{,i}
\nonumber\\*&\hspace{2em}
		\mp D_i + \alpha^\pm \rho^\pm g_i
			+ \big(\overline{ b^\pm \tau_{ij} }\big)_{,j}
				- \overline{ b^\pm_{,j} \tau_{ij} } ,
\\
\label{eq:DK_pm}
\partial_t ( \alpha^\pm \rho^\pm K^\pm_\alpha )
	+ ( \alpha^\pm \rho^\pm K^\pm_\alpha U^\pm_j )_{,j}
	&= - \varPhi^{k\pm}_{\alpha j,j} \mp \varPsi^{k\pm}_\alpha
		- R^\pm_{\alpha j} U^\pm_{\alpha,j}
			- P_{,\alpha} \varPhi^{\alpha\pm}_\alpha
\nonumber\\*&\hspace{2em}
		- \varPhi^{p\pm}_{\alpha j,j}
			+ \varPi^{d\pm}_\alpha + \varTheta^{p\pm}_\alpha
		+ \big(\overline{ b^\pm \tau_{\alpha j}
					u^\pm_\alpha }\big)_{,j}
\nonumber\\*&\hspace{3em}
		- \overline{ b^\pm_{,j} \tau_{\alpha j} u^\pm_\alpha }
		- \alpha^\pm \rho^\pm \varepsilon^\pm_\alpha ,
\end{align}
\end{subequations}
where equations between brackets are redundant with the others but shown here for reference, and with
\begin{subequations}
\begin{alignat}{3}
\label{eq:app:APP}
\alpha^\pm &= \overline{b^\pm} ,
	& \bm{\varPhi}^{\alpha\pm} &= \overline{b^\pm\bm{u}^\pm} ,
	& \varPsi^\alpha
	&= \overline{\mp (\dd_t b^\pm)} ,
\\\label{eq:app:CPP}
C^{m\pm} &= \overline{b^\pm\rho c^m}/\overline{b^\pm\rho} ,
	& \bm{\varPhi}^{m\pm} &= \overline{b^\pm\rho c^m \bm{u}^\pm} ,
	& \varPsi^m
	&= \overline{\mp (\dd_t b^\pm) \rho c^m} ,
\\
\rho^\pm &= \overline{b^\pm\rho}/\overline{b^\pm} ,
	& \big[\;\bm{\varPhi}^\rho &= 0 \;\big] ,
	& \varPsi^\rho
	&= \overline{\mp (\dd_t b^\pm) \rho} ,
\\\label{eq:app:URP}
\bm{U}^\pm &= \overline{b^\pm\rho \bm{u}}/\overline{b^\pm\rho} ,
	& R_{ij}^\pm &= \overline{b^\pm\rho u^\pm_iu^\pm_j} ,
	& \bm{\varPsi}^u
	&= \overline{\mp (\dd_t b^\pm) \rho\bm{u}} ,
\\
P^\pm &= \overline{b^\pm p}/\overline{b^\pm} ,
	& \bm{\varPhi}^{p\pm} &= \overline{b^\pm p' \bm{u}^\pm} ,
	& \bm{D}
	&= \overline{\mp (\bnabla b^\pm) p'} ,
\\
K^\pm_\alpha &= \overline{b^\pm\thalf\rho (u^\pm_\alpha)^2}
			/ \overline{b^\pm\rho} ,
	~& \bm{\varPhi}^{k\pm}_\alpha
		&= \overline{b^\pm\thalf\rho (u^\pm_\alpha)^2 \bm{u}^\pm} ,
	& \varPsi^{k\pm}_\alpha
	&= \overline{\mp (\dd_t b^\pm)
				\thalf\rho (u^\pm_\alpha)^2} ,
\\\label{eq:epsilon}
\varPi^{d\pm}_\alpha
	&= \overline{b^\pm_{,\alpha} p' u^\pm_\alpha} ,
	& \varTheta^{p\pm}_\alpha
		&= \overline{b^\pm p' u^\pm_{\alpha,\alpha}} ,
	~& \alpha^\pm \! \rho^\pm \! \varepsilon^\pm_\alpha
	&= - \overline{ b^\pm \tau_{\alpha j} u^\pm_{\alpha,j} } .
\end{alignat}
\end{subequations}

	It is to be noticed that for turbulent kinetic energies, exchange and production terms related to momentum exchange and drag are deeply intermingled. Their sums over the two structures give the total works due to momentum exchange and drag
\begin{subequations}
\label{eq:work_prod}
\begin{align}
- \varPsi^{k+} \!+ \varPsi^{k-}
	&= \big( \varPsi^u_i - \thalf\varPsi^\rho(U^+_i\!+\!U^-_i) \big)
			\delta U_i = \varPsi_i^{u \rho} \delta U_i ,
\\
\varPi^{d+} \!+ \varPi^{d-} &= D_i \delta U_i ,
\end{align}
\end{subequations}
where $\varPsi^{k\pm} = \sum_\alpha \varPsi^{k\pm}_\alpha$ and $\varPi^{d\pm} = \sum_\alpha \varPi^{d\pm}_\alpha$.
%
\subsection{Directed momentum and directed kinetic energy equations}
	To facilitate later calculations, it is useful to derive the following equations from~\eqref{eq:DR_pm}
\begin{equation}
\label{eq:Drr_pm}
\partial_t \big( \tfrac{\alpha^\pm\!\rho^\pm}{\overline{\rho}} \big)
	+ \big(\tfrac{\alpha^\pm\!\rho^\pm}{\overline{\rho}}\big)_{,j} U_j
	\pm \tfrac{1}{\overline{\rho}}
		\, \big(\tfrac{\alpha^+\!\alpha^-\!\rho^+\!\rho^-}{\overline{\rho}}
					\delta U_j \big)_{j}
	= \mp \tfrac{1}{\overline{\rho}} \, \varPsi^\rho .
\end{equation}
The per-structure momentum equations~\eqref{eq:DU_pm} can then be combined in order to produce the directed momentum equation according to
\begin{equation}
\tfrac{\alpha^-\!\rho^-}{\overline{\rho}}\text{\eqref{eq:DU_pm}}^+
		\!+ \text{\eqref{eq:Drr_pm}}^- \alpha^+\!\rho^+\bm{U}^+
	\!- \tfrac{\alpha^+\!\rho^+}{\overline{\rho}}\text{\eqref{eq:DU_pm}}^-
		\!- \text{\eqref{eq:Drr_pm}}^+ \alpha^-\!\rho^-\bm{U}^- .
\end{equation}
After rearrangements it is eventually found
\begin{align}
\label{eq:DUd}
\partial_t \big( \tfrac{\alpha^+\!\alpha^-\!\rho^+\!\rho^-}{\overline{\rho}}
				\delta U_i
	&\big) + \big( \tfrac{\alpha^+\!\alpha^-\!\rho^+\!\rho^-}{\overline{\rho}}
				\delta U_i \big[
				\tfrac{\alpha^-\!\rho^-}{\overline{\rho}} U^+_i
					+ \tfrac{\alpha^+\!\rho^+}{\overline{\rho}} U^-_i
				\big] \big)_{,j}
		\pm \tfrac{\alpha^+\!\alpha^-\!\rho^+\!\rho^-}{\overline{\rho}}
					\delta U_i \delta U_j
			\big( \tfrac{\alpha^\pm\!\rho^\pm}{\overline{\rho}}\big)_{,j}
\nonumber\\*&
	= - \big( \tfrac{\alpha^-\!\rho^-}{\overline{\rho}} R^+_{ij,j}
			\!- \tfrac{\alpha^+\!\rho^+}{\overline{\rho}} R^-_{ij,j}
				\big)
		- \tfrac{\alpha^+\!\alpha^-\!\rho^+\!\rho^-}{\overline{\rho}}
				\delta U_j U_{i,j}
		- \big( \varPsi^u_i \!- \varPsi^\rho U_i \big)
\nonumber\\*&\hspace{2em}
		- \tfrac{\alpha^+\!\alpha^-(\rho^--\rho^+)}{\overline{\rho}} P_{,i}
		- \big(\alpha^+\alpha^- (P^+\!-P^-)\big)_{,i} - D_i
\nonumber\\*&\hspace{4em}
		- \big( \tfrac{\alpha^-\!\rho^-}{\overline{\rho}}
					(\overline{ b^+\tau_{ij} })_{,j}
				- \tfrac{\alpha^+\!\rho^+}{\overline{\rho}}
					(\overline{ b^-\tau_{ij} })_{,j} \big)
				\mp \overline{ b^\pm_{,j} \tau_{ij} } .
\end{align}

	The per-structure momentum equations~\eqref{eq:DU_pm} can be combined in order to produce the equations of mean and directed kinetic energies according to
\begin{subequations}
\begin{align}
&[\text{\eqref{eq:DU_pm}}^+\!+\text{\eqref{eq:DU_pm}}^-] \bcdot \bm{U}
	- \thalf \;
	[\text{\eqref{eq:DR_pm}}^+\!+\text{\eqref{eq:DR_pm}}^-] \; \bm{U}^2 ,
\\
&\text{\eqref{eq:DU_pm}}^+ \bcdot \bm{U}^+
		\!- \thalf \; \text{\eqref{eq:DR_pm}}^+ \; (\bm{U}^+)^2
\nonumber\\*&\hspace*{8em}
	+ \text{\eqref{eq:DU_pm}}^- \bcdot \bm{U}^-
		\!- \thalf \; \text{\eqref{eq:DR_pm}}^- \; (\bm{U}^-)^2
	- \text{\eqref{eq:DKm}} .
\end{align}
\end{subequations}
After rearrangements it is eventually found
\begin{subequations}
\begin{align}
\label{eq:DKm}
\partial_t ( \thalf\overline{\rho}\bm{U}^2 )
	+ &( \thalf\overline{\rho}\bm{U}^2 U_j )_{,j}
\nonumber\\*&
	= - \big( R^+_{ij} + R^-_{ij}
		+ \tfrac{\alpha^+\!\alpha^-\!\rho^+\!\rho^-}{\overline{\rho}}
				\delta U_i\delta U_j \big)_{,j} U_i
	- P_{,j} U_j + \overline{\rho} g_i U_i + \overline{ \tau_{ij,j} } U_i ,
\\[\halflineskip]
\label{eq:DKd}
\partial_t \big( \tfrac{\alpha^+\!\alpha^-\!\rho^+\!\rho^-}{2\overline{\rho}}
				&\delta\bm{U}^2 \big)
	+ \big( \tfrac{\alpha^+\!\alpha^-\!\rho^+\!\rho^-}{2\overline{\rho}}
			\delta U_i\delta U_j \big[
				\tfrac{\alpha^-\!\rho^-}{\overline{\rho}} U^+_i
			+ \tfrac{\alpha^+\!\rho^+}{\overline{\rho}} U^-_i
				\big] \big)_{,j}
\nonumber\\*&
	= - \big( \tfrac{\alpha^-\!\rho^-}{\overline{\rho}}R^+_{ij,j}
		\!-\tfrac{\alpha^+\!\rho^+}{\overline{\rho}}R^-_{ij,j}
			\big) \delta U_i
		- \tfrac{\alpha^+\!\alpha^-\!\rho^+\!\rho^-}{\overline{\rho}}
				\delta U_i\delta U_j U_{i,j}
\nonumber\\*&\hspace{16em}
		- \big( \varPsi^u_i
			\!- \thalf\varPsi^\rho(U^+_i\!+U^-_i) \big) \delta U_i
\nonumber\\*&\hspace{2em}
		- \tfrac{\alpha^+\!\alpha^-(\rho^-\!-\rho^+)}{\overline{\rho}}
				P_{,j} \delta U_j
		- \big(\alpha^+\alpha^- (P^+\!-P^-)\big)_{,j} \delta U_j
			- D_i \delta U_i
\nonumber\\*&\hspace{4em}
			+ \big(\tfrac{\alpha^-\!\rho^-}{\overline{\rho}}
					(\overline{ b^+\tau_{ij} })_{,j}
				-\tfrac{\alpha^+\!\rho^+}{\overline{\rho}}
					(\overline{ b^-\tau_{ij} })_{,j} \big)
					\delta U_i
				\mp \overline{ b^\pm_{,j} \tau_{ij} } \delta U_i .
\end{align}
\end{subequations}

	It is to be noticed that the production terms of two-structure-field turbulent energies in~\eqref{eq:work_prod} are also identified as destruction terms of directed energy in~\eqref{eq:DKd}.
%
\subsection{Volume to mass reduction of exchange and flux terms in the incompressible case}
	In the incompressible limit, the flux and exchange terms of volume in~\eqref{eq:app:APP} can be related to their corresponding terms for mass in~\eqref{eq:app:CPP}. This is carried out by taking into account the volume filling relationship of the fluids
\begin{equation}
\frac{1}{\rho} = \frac{c^1}{\rho^1} + \frac{c^2}{\rho^2} ,
\end{equation}
where~$\rho^1$ and~$\rho^2$ are now constants, in order to transform the volume terms according to
\begin{align}
\bm{\varPhi}^{\alpha\pm} &= \overline{b^\pm\bm{u}^\pm}
&
\varPsi^\alpha &= \overline{\mp (\dd_t b^\pm)}
\nonumber\\*
	&= \overline{b^\pm\rho \big( \tfrac{c^1}{\rho^1}
		+ \tfrac{c^2}{\rho^2} \big) \bm{u}^\pm}
&
	&= \overline{\mp (\dd_t b^\pm) \rho \big( \tfrac{c^1}{\rho^1}
		+ \tfrac{c^2}{\rho^2} \big)}
\nonumber\\*
	&= \tfrac{1}{\rho^1} \overline{b^\pm\rho c^1 \bm{u}^\pm}
		+ \tfrac{1}{\rho^2}
			\overline{b^\pm\rho c^2 \bm{u}^\pm}
&
	&= \tfrac{1}{\rho^1} \overline{\mp (\dd_t b^\pm) \rho c^1}
		+ \tfrac{1}{\rho^2}
			\overline{\mp (\dd_t b^\pm) \rho c^2}
\nonumber\\*
	&= \tfrac{1}{\rho^1} \bm{\varPhi}^{1\pm}
		+ \tfrac{1}{\rho^2} \bm{\varPhi}^{2\pm}
&
	&= \tfrac{1}{\rho^1} \varPsi^1
		+ \tfrac{1}{\rho^2} \varPsi^2
\nonumber\\*
	&= \tfrac{\rho^2-\rho^1}{\rho^1\rho^2}
			\bm{\varPhi}^{1\pm} ,
&
	&= \tfrac{1}{\rho^2} \varPsi^\rho
		+ \tfrac{\rho^2-\rho^1}{\rho^1\rho^2}
			\varPsi^1 .
\end{align}
The last expression for~$\varPsi^\alpha$ is obtained from identities $\bm{\varPhi}^{1\pm} + \bm{\varPhi}^{2\pm} = 0$ and $\varPsi^1 + \varPsi^2 = \varPsi^\rho$. These expressions---which also hold under $1\leftrightarrow2$ permutation---show that the volume terms follow the mass terms scaled by the apparent Atwood number (\ie between fluid densities).
%
\section{Optimization tools for two-structure-field separation}
%
\subsection{Bimodality coefficient of a single-variable PDF}
\label{app:Bimodality}
\begin{figure}
\centerline{\begin{tabular}{@{}c*{5}{@{~}c}@{}}
$\mathcal{B}$
	& 1:1, 1:1 & 3:1, 1:1 & 3:1, 1:8 & 1:1, 1:8 & 1:3, 1:8
\\*
\raisebox{3in/4-1em}{0.5}
	& \includegraphics{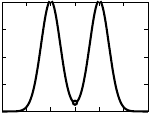}
	& \includegraphics{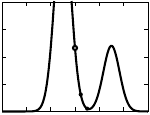}
	& \includegraphics{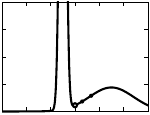}
	& \includegraphics{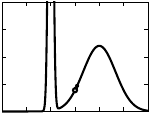}
	& \includegraphics{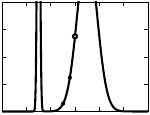}
\\*
\raisebox{3in/4-1em}{1.0}
	& \includegraphics{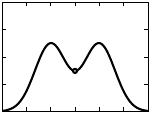}
	& \includegraphics{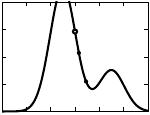}
	& \includegraphics{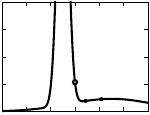}
	& \includegraphics{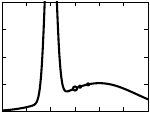}
	& \includegraphics{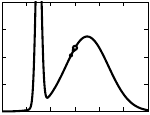}
\\*
\raisebox{3in/4-1em}{1.5}
	& \includegraphics{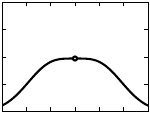}
	& \includegraphics{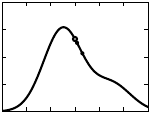}
	& \includegraphics{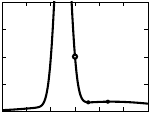}
	& \includegraphics{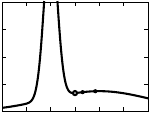}
	& \includegraphics{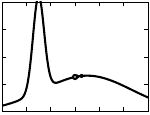}
\end{tabular}}
\caption{Graphical representations of doubly-Gaussian PDFs constrained by zero mean and unit variance, with all combinations of ratios of mode weights (1:1, 3:1, or~1:3), ratios of mode mean-square widths (1:1 or~1:8), and bimodality coefficients $\mathcal{B}=1.5$, 2.0, or~2.5. At $\mathcal{B}=0$ and~2 (not shown here) the PDFs reduce respectively to Bernoulli distributions (two Dirac functions) or to a single Gaussian. A trough between the peaks is generally observed when $\mathcal{B}\lesssim1$ irrespective of mode weights and widths. Thresholds given by the generalized Otsu procedure~\eqref{eq:OtsuEq} are represented for $q=0$, $\thalf$ (black points), and~1 (white points). PDFs of separator fields~$\widetilde{\beta}$ in a typical TMZ as described in part~\ref{sec:Prescription} are mostly of types \mbox{(1:1, 1:1)} and \mbox{(3:1, 1:8)}.}
\label{fig:BiModCoef}
\end{figure}
	Numerous measures of the bimodality of a single-variable PDF have been developed for statistical studies, with differing recommended uses \citep[and refs therein]{Knapp07} depending on specific applications. Here, consistency with Bernoulli (purely bimodal) and Gaussian (normal) distributions, robustness, and algebraic simplicity will be privileged. In this spirit, moment based approaches appear stable and weakly dependent on statistical noise because velocity fluctuations at large turbulent scales display weak extreme values (no long tails on PDFs).

	It was shown by \citet[resp.\ p.~433 and eq.~iv]{Pearson16, Pearson29} that the following inequality holds for any PDF of a single variable
\begin{equation}
\label{eq:Pearson}
\mu_4 \mu_2 - \mu_3^2 - \mu_2^3 \geq 0 ,
\end{equation}
where~$\mu_n$s are the~$n$th order centred moments as defined by~\eqref{eq:Moments}. It reduces to an equality only in the case of purely bimodal (or Bernoulli) distributions.

	The fourth reduced moment is known to depend on the `peakedness' of the probability distribution function. It is thus possible to build from~\eqref{eq:Pearson} various dimensionless coefficients~$\mathcal{B}$ which reflect bimodality, \ie so that $\mathcal{B}\geq0$ in general, equality holding only for Bernoulli distributions. It can be shown that, up to an irrelevant homography (positive and increasing over positive values), the set of such bimodality coefficients can be reduced to
\begin{equation}
\label{eq:Bc}
\mathcal{B}(c) = \frac{ \mu_4\mu_2-\mu_3^2-\mu_2^3 } { c\mu_3^2+\mu_2^3 } ,
\end{equation}
which displays the appealing complementary property $\mathcal{B}=2$ for normal (Gaussian) distributions. In order to ensure a non vanishing denominator for any distribution, $c$~must be positive, and the specific values $c=0$ and~1 provide the bimodality coefficients respectively recommended by \citet[and refs therein]{Knapp07} and Sarle \citep[p.~561]{Sarle89}---up to an inversion for the latter.

	From~\eqref{eq:Bc} a lengthy and tedious calculation provides an expression of $\mathcal{B}(c)$ for a doubly-Gaussian PDF defined by amplitudes $\alpha^\pm$ and mean square deviations~$\sigma^{2\pm}$ scaled by the peaks' square separation. The expansion of $\mathcal{B}(c)$ to lowest order at vanishing peak widths (thus in the limit of Bernoulli distributions) then yields
\begin{equation}
\label{eq:BcDG}
\mathcal{B}(c)
	= \frac{\alpha^+\sigma^{2+}\!+\alpha^-\sigma^{2-}}
		{\alpha^+\alpha^-\big(c+(1-4c)\alpha^+\alpha^-\big)} .
\end{equation}
This expression becomes singular with simple poles at $\alpha^\pm=0$ but it also displays two supplementary poles, imaginary or outside the $(0,1)$ interval, which depend on the value of $c$. In particular, $c=0$ turns the singularities into double poles. The most algebraically simple value thus appears to be $c=\tfourth$ which eliminates the supplementary poles.

	Value $c=\tfourth$ is retained in~\eqref{eq:Bimodality} because it provides a flatter response of~$\mathcal{B}$ as a function of height across the TMZ: when approaching the TMZ edges the dominant and vanishing modes are respectively the least and most turbulent, thus making products $\alpha^\pm\sigma^{2\pm}\sim\alpha^\pm\alpha^\mp$ match the $\alpha^+\alpha^-$ denominator in~\eqref{eq:BcDG}. The behaviour of $\mathcal{B}(\tfourth)$ is illustrated in figure~\ref{fig:BiModCoef} on the different cases of an asymmetric superposition of two Gaussians and summarized in section~\ref{ssec:Optimization} after~\eqref{eq:FullBimodality}.
%
\subsection{Segmentation threshold of a bimodal single-variable PDF}
\label{app:Segmentation}
	The segmentation of a single-variable bimodal PDF with a finite bimodality coefficient $\mathcal{B}>0$ involves some level of arbitrariness: whatever the threshold approach, the overlap between modes produces erroneous cross attributions ($+$ in~$-$ and $-$~in $+$) whose relative weights depend on the depth of the trough between modes and on the threshold value (see figure~\ref{fig:BiModCoef}). An intuitively reasonable segmentation threshold~$\beta^\circ$ could be defined by the minimal value of the PDF $\mathcal{P}'(\beta^\circ)=0$ (in the present section the random variable is denoted~$\beta$ instead of~$\widetilde{\beta}$ for a filtered field in part~\ref{sec:Prescription}). However, it is no more or less canonical than others and, by involving a derivative, it can be noisy and ill-conditioned. Numerous other approaches have thus been developed and adapted to specific applications in particular for image processing \citep{Pal93, Freixenet02, Sezgin04, Zhang08}.

	In the same spirit of the definition of~$\mathcal{B}$ in~\eqref{eq:Bc}, a moment-based definition of~$\beta^\circ$ is here privileged, akin for instance to the method of \citet{Otsu79}. It is one of the oldest, simplest and most widely used, and many variants have been developed in order to better capture some specific situations: a modification of marginal cost has been recently suggested to improve its performance on strongly asymmetric distributions \citep{Ng06, Hou06}.

	For a given PDF of~$\beta$, the here generalized Otsu method defines the threshold~$\beta^\circ$ so as to maximize the \emph{inter-structure contribution} to the variance, as weighted by some function of volume fractions. This is given by
\begin{equation}
\label{eq:Otsu}
\beta^\circ = \arg\max_{\beta^\circ}
		\bigg[\; \big(\overline{b^+}\;\overline{b^-}\big)^{1+q}
	\;\bigg( \frac{\overline{b^+\beta}}{\overline{b^+}}
		- \frac{\overline{b^-\beta}}{\overline{b^-}} \bigg)^2 \;\bigg] ,
\end{equation}
with~$\overline{b^\pm}$ and $\overline{b^\pm\beta}$ taken as functions of~$\beta^\circ$
\begin{subequations}
\label{eq:ThresholdDependence}
\begin{align}
\overline{b^\pm}(\beta^\circ)
	&= \int_{-\infty}^{+\infty} \rmH\big[
		\mp (\beta-\beta^\circ) \big] \mathcal{P}(\beta) \dd \beta ,
\\
\overline{b^\pm\beta}(\beta^\circ)
	&= \int_{-\infty}^{+\infty} \rmH\big[
		\mp (\beta-\beta^\circ) \big] \mathcal{P}(\beta) \beta \dd \beta ,
\end{align}
\end{subequations}
and where~$\mathcal{P}$ is the PDF of~$\beta$. $q$~is a predefined weighting exponent---with $q\geq-1$ to avoid singularities in the unimodal limit---for which values $q=0$ and~1 reduce equation~\eqref{eq:Otsu} to respectively the standard method of \citet[eqs~13 to~16]{Otsu79} and a symmetrized form of the method of \citet[eq.~8]{Ng06}.

	Assuming $\mathcal{P}(\beta)$ to be differentiable, the solution to~\eqref{eq:Otsu} is such that
\begin{equation}
\label{eq:OtsuDerivative}
0 = \frac{\dd}{\dd \beta^\circ}
	\big[ ( \alpha^+ \alpha^- )^{1+q} ( \beta^+\!-\beta^- )^2 \big] ,
\end{equation}
with the usual notations $\alpha^\pm = \overline{b^\pm}$ and $\beta^\pm = \overline{b^\pm\beta}/\overline{b^\pm}$. Derivatives of~\eqref{eq:ThresholdDependence} readily provide
\begin{align}
\label{eq:ThresholdDerivatives}
\frac{\dd}{\dd \beta^\circ} \alpha^\pm
	&= \mp \mathcal{P}(\beta^\circ) ,
&
\frac{\dd}{\dd \beta^\circ} (\alpha^\pm\beta^\pm)
	&= \mp \mathcal{P}(\beta^\circ) \beta^\circ ,
\end{align}
from which repeated application of the chain rule to~\eqref{eq:OtsuDerivative} eventually yields an implicit equation in~$\beta^\circ$ which takes either of the following forms
\begin{subequations}
\label{eq:OtsuEq}
\begin{align}
\label{eq:OtsuEq0}
2\beta^\circ &= 2\overline{\beta}
	+ (q-1) (\alpha^+\!-\alpha^-) (\beta^+\!-\beta^-) ,
\\\label{eq:OtsuEq1}
2\beta^\circ &= 2\beta^- \!+ ( 2q\alpha^+ \!+ 1 - q ) (\beta^+\!-\beta^-) .
\end{align}
\end{subequations}
These equations can be solved numerically by common algorithms (secant, dichotomy, \etc).

	For $q=1$ or on a distribution with equal median and mean $\overline{b^\pm}(\overline{\beta}) = \thalf$, \eqref{eq:OtsuEq0}~has the trivial solution $\beta^\circ = \overline{\beta}$. For~$\alpha^+\rightarrow0$ in~\eqref{eq:OtsuEq1}, the limit threshold~$\beta^\circ$ appears shifted away from $\beta^-$ by a fraction $(1-q)/2$ of the offset $\beta^+\!-\beta^-$. Adjusting~$q$ thus defines the behaviour of $\beta^\circ$ at the edges of the TMZ. $\beta^\circ$~must ideally fall at the edge of the dominant narrow peak: $q$~is then related to the width of the dominant mode as scaled by the separation between modes. For usual structure asymmetries at TMZ edges and with the presently optimized bimodality coefficients $\langle\mathcal{B}\rangle$ below 1, $\beta^\pm$~are good approximations of the means of individual modes and a value of $q\approx\thalf$ appears appropriate as visible in figure~\ref{fig:BiModCoef} columns~3:1 and~1:8.

	Because incompressibility forces $\overline{u_z} = 0$ in the RT flow, the poor man's instantaneous approach defined in section~\ref{ssec:Segmentation} merely appears as the special case of the method for $\beta = u_z$ and $q=1$ according to~\eqref{eq:OtsuEq0}: variants with $\beta = u_z$ but $q\neq1$ will then be designated as `optimized poor man's' methods. The poor man's method optimized at $q=0$ can be interpreted as maximizing the directed contribution to the Reynolds stress tensor~\eqref{eq:Rij}---though without the mass weighting by field~$\rho$.
%
\section{Numerical conditions for the of RT flow simulation with two-structure fields}
\label{app:Numerics}
%
\subsection{Numerical schemes and code}
\label{ssec:Schemes}
	All realizations of turbulent RT flows have been obtained from simulations carried out with a finite-volume 3D-parallelized numerical code \citep{Watteaux11} adapted from the incompressible multi-phase code \href{http://www.ida.upmc.fr/\~zaleski/codes/legacy_codes.html}{\textsc{Surfer}} \citep{Lafaurie94} as parallelized using \textsc{Mpich2} `Message Passing Interface'.

	The code evolves the Navier--Stokes and mass equations~\eqref{eq:Du} and~\eqref{eq:Dr} under the incompressibility constraint~\eqref{eq:v} on a `Marker-and-Cell' staggered mesh using a `Lagrange +~Remap' method with a Helmholtz--Hodge type decomposition \citep{Peyret83} and a `Red--Black Gauss--Seidel' pressure relaxation coupled with the `V-cycle' multigrid convergence method \citep{Brandt82, Briggs00, Press91, Press91b}. It is second order in space using an upwind TVD approach with Van Leer's flux limiters and second order in time using the \textsc{Ssprk2} `Strong Stabilization Preserved Runge--Kutta' approach \citep{Gottlieb01}. No interface reconstruction between fluids was used, hence inducing a numerical inter-diffusion of the fluids.

	The same scheme was used to evolve the passive filtering equation \eqref{eq:Filtering} supplemented with closures~\eqref{eq:Dc}, \eqref{eq:Parameters}, \eqref{eq:FullBimodality}, \eqref{eq:OtsuEq0} and~\eqref{eq:bpm_PM} to yield the two-structure-field segmentation. The 1D conditional averages~\eqref{eq:app:APP} were computed \emph{on the fly at all time steps}---all other correlations (such as single-fluid) being obtained from combinations or from residues of the evolution equations~\eqref{eq:D2SF}. Full 3D fields were only saved at occasional predefined times.

	Further details are provided by \citet{Watteaux11}. All simulations were carried out at the national cluster `\href{https://www-ccrt.cea.fr/}{Centre de Calcul Recherche et Technologie}' (CCRT) at Bruyères-le-Châtel, France.
%
\subsection{Initial state, interface perturbation}
\label{ssec:Conditions}
	The simulations were carried out in a cubic domain of size~1 with no-slip boundaries on top and bottom walls and periodic conditions on the sides. It was discretized with uniform cubic cells of size~$h$ with resolutions of $N = 1/h = 256$, $512$, and~$1024$ used at respectively exploratory, adjustment, and final stages.

	The Rayleigh--Taylor instability was initialized in a quiescent state, with `heavy' and `light' fluid above and below the interface at half-height of the domain where $z = 0$ with a gravity field $g_z=-1$. Respective fluid densities of $\rho_2 = 101/99$ and $\rho_1 = 1$ provided a low Atwood number of $\Atw=0.01$.

	The interface perturbation was generated using the now standard approach of \citet{Youngs94}: its spectrum was defined on the spectral annulus $[\kappa_{\min} ; \kappa_{\max}] = [2\pi/(8 h) ; 2\pi/(4 h)]$ with a uniform random amplitude such that the standard deviation was $\sigma = 0.02 \cdot \lambda_{\min} = 0.02 \cdot 2 \pi / \kappa_{\min}$. The locally computed interface displacements were then projected on homogeneously mixed cells on the two layers on each side of the $z=0$ plane. The influence of the random generator seed was found to be negligible on all averaged quantities, whether single-fluid like (such as growth rate $\alpha$) or two-structure-field conditioned (such as bimodality coefficients $\mathcal{B}$).
%
\subsection{Viscosity and diffusion}
\label{ssec:Dissipations}
	The fluids have an identical, uniform, but \emph{time-dependent} kinematic viscosity coefficient~$\nu(t)$ calibrated at every time-step so as to keep the mean Kolmogorov scale $\eta = \nu^{3/4} \langle\varepsilon\rangle^{-1/4}$ in the TMZ on the order of the grid mesh size. A common prescription for proper DNS simulations was provided by \citet[eq.~9.3]{Pope00} at $\eta \gtrsim h / 2.1$. The viscosity $\nu(t)$ was therefore defined as\marginpar{Check $8h$ threshold.}
\begin{equation}
\label{eq:nut}
\nu(t) = \left|\;\begin{aligned}
		& \nu_0 = 0.21 \sqrt{\Atw g h^3}
		& \text{if}\quad L(t) \leq 8h ,
	\\*
		& \max \big\{\, (h/2.1)^{4/3} \langle\varepsilon\rangle^{1/3}(t)
			\,,\, \nu_0 \,\big\}
		& \text{if}\quad L(t) > 8h ,
	\end{aligned}\right.
\end{equation}
updated at every time step. After the early linear to non-linear transition around $L\approx 8h$, this smoothly time-varying viscosity is proportional to the 0D-averaged dissipation $\langle\varepsilon\rangle(t)$ defined by the budget of turbulent kinetic energy in~\eqref{eq:SSDissipation}. It ensures a maximal extension of the Kolmogorov cascade within the available resolution, yielding significantly higher Reynolds numbers at late times than previously reported with identical mesh sizes. The value of $\nu_0$ was taken as given by~\citet[eq.~4]{Dimonte04} which preserves stability at early times while keeping short transients.

	As the main properties studied here concern large scales and agree with previously reported results, possible distortions induced by~\eqref{eq:nut} at small scales were not observed and were not further investigated. The existence of the Kolmogorov cascade---required for~\eqref{eq:nut} to hold---was explained theoretically \citep{Chertkov03, Soulard12} and observed empirically \citep{Cabot06} in RT flows at high enough Reynolds numbers. Because of the somewhat unusual time-dependence of viscosity, simulations with this approach could be considered as \textsc{Iles} \citep[§~13.6.4]{Boris92, Pope00} instead of DNS \citep[§~9]{Pope00}.

	Mass diffusion was locked to the viscosity through a constant Schmidt number $\Sch = \nu(t)/\kappa(t)$ equal to $1$ for the $1024^3$ resolution---higher values were used at lower resolutions due to the impact of numerical diffusion. The diffusion coefficient $\kappa(t)$ was therefore time dependent, as $\nu(t)$. Here again, small-scale effects of mass diffusion did not impact statistical results.
%
\subsection{Averaging over the self-similar regime}
\label{ssec:SSRaveraging}
	The mean quantities, profiles, and fields reported in the present study are obtained by averaging over sub-domains of the computational box (planes, mixing zone\dots) which sometimes restrict the sampling and introduce significant noise.

	Over the self-similar regime, it is possible to further improve the signal-to-noise ratio by performing a time averaging of the quantities as scaled by the proper functions of the TMZ width $L$ and its derivatives. Now, over a self-similar duration identified by some given prescription, the number of elementary statistical events of relevance can vary significantly and an adapted time weighting must thus be introduced for averaging.

	As the present study focuses on turbulent structures at integral scales, it appears consistent to define the weighting according to an estimate of the number and lifetime of such structures at any given time $t$. In order reduce the possible impact of the time-origin definition, averaging can be more efficiently performed over $L(t)$, with $\dd L = L' \dd t$. In the self-similar regime, the number of large scale structures over an $xy$ plane scales as $1/L^2$ and their inverse lifetime scales as $L'/L$. The weighted average of some dimensionless quantity $X(t,\zeta)=X(t,z/L(t))$ over a prescribed self-similar range of $L$ values labeled as $\text{SSR}$ is thus defined as
\begin{equation}
\label{eq:SSRaveraging}
\widehat{X}(t,\zeta) = \int_\text{SSR} X(t,\zeta) \frac{\dd L}{L^3}
		\Big/ \int_\text{SSR} \frac{\dd L}{L^3} .
\end{equation}
This definition shows that noise at large scales is actually lower at early times although the system's evolution may then be somewhat off self-similarity. Integration over $L$ instead of $t$ may reduce these artifacts as the internal structure of the TMZ is expected to stabilize before the growth rate of $L$---the later is a consequence of the internal dynamics of the TMZ.

	The weighting in \eqref{eq:SSRaveraging} has been seldom taken into account in previous works although it may have some impact. For instance, if the self-similar range spans a modest factor of two in $L$, weighting is reduced by a significant factor of eight between its beginning and its end.
%
\bibliographystyle{jfm}
\bibliography{2024_Watteaux_Redford_Llor}
%
%
%
%
\end{document}